\documentclass[11pt,a4paper]{article}
\pdfoutput=1

\usepackage{jcappub}

\usepackage{siunitx}
\DeclareSIUnit{\deg}{\text{deg}}
\DeclareSIUnit{\arcmin}{\text{arcmin}}
\DeclareSIUnit{\arcsec}{\text{arcsec}}
\DeclareSIUnit{\parsec }{\text{pc}}
\usepackage{mathrsfs}
\usepackage{bbm}
\usepackage{bm}
\usepackage{dsfont}
\usepackage{tensor}
\usepackage{xspace}
\usepackage{lmodern}
\usepackage{wasysym}
\usepackage{microtype}
\usepackage{empheq}
\usepackage{ifthen}
\usepackage{amsmath}
\usepackage{wrapfig}
\usepackage{cleveref}
\crefname{section}{sec.}{secs.}
\Crefname{section}{Section}{Sections}
\crefname{subsection}{subsec.}{subsecs.}
\Crefname{subsection}{Subsection}{Subsections}
\crefname{subsubsection}{subsubsec.}{subsubsecs.}
\Crefname{subsubsection}{Subsubsection}{Subsubsections}
\usepackage{xcolor}
\usepackage{import}
\usepackage{transparent}
\usepackage{orcidlink}
\usepackage{ntheorem} 
\theoremheaderfont{\bfseries}
\theorembodyfont{\normalfont}
\theoremseparator{\vspace*{3mm}}
\usepackage[framemethod=TikZ]{mdframed}
\newcommand\ph{\varphi}
\newcommand\thet{\vartheta}
\newcommand\eps{\varepsilon}

\newcommand\define{\equiv}

\newcommand\vect[1]{\boldsymbol{#1}}
\newcommand{\uvect}[1]{\hat{\vect{#1}}}
\newcommand\mat[1]{\boldsymbol{#1}}

\newcommand\ex[1]{\mathrm{e}^{#1}}
\newcommand\ii{\mathrm{i}}

\renewcommand\Re{\mathrm{Re}}
\renewcommand\Im{\mathrm{Im}}

\newcommand{\order}{\mathcal{O}}

\newcommand\e[1]{_{\mathrm{#1}}}
\newcommand\h[1]{^{\mathrm{#1}}}

\newcommand{\dd}{\mathrm{d}}

\newcommand{\delimiters}[4][]{
\ifthenelse{ \equal{#1}{1} }{  #2 #3 #4  }
					{ \ifthenelse{\equal{#1}{2}}{ \big#2 #3 \big#4 }
						{ \ifthenelse{\equal{#1}{3}}{ \Big#2 #3 \Big#4 }
							{ \ifthenelse{\equal{#1}{4}}{ \bigg#2 #3 \bigg#4 }
								{ \ifthenelse{\equal{#1}{5}}{ \Bigg#2 #3 \Bigg#4 }
									{ \left#2 #3 \right#4 }
								}
							}
						}
					}
													}
													
\newcommand{\pa}[2][]{\delimiters[#1]{(}{#2}{)}}
\newcommand{\pac}[2][]{\delimiters[#1]{[}{#2}{]}}
\newcommand{\paac}[2][]{\delimiters[#1]{\{}{#2}{\}}}

\newcommand{\ev}[2][]{\delimiters[#1]{\langle}{#2}{\rangle}}

\newenvironment{system}
{ \left\{ \begin{aligned} }
{ \end{aligned} \right. }

\newcommand{\amp}{\mathcal{A}}
\newcommand{\spinharmonic}[2]{{}_{#1}Y_{#2}}
\newcommand{\hLOS}{\gamma\h{LOS}}
\newcommand{\lLOS}{\gamma\e{LOS}}
\newcommand{\mhLOS}{\hat{\gamma}\h{LOS}}
\newcommand{\mlLOS}{\hat{\gamma}\e{LOS}}
\newcommand{\nlLOS}{\eta\e{LOS}}
\newcommand{\nhLOS}{\eta\h{LOS}}
\newcommand{\Dirac}{\delta\e{D}}

\newcommand{\galaxies}{\mathcal{G}}
\newcommand{\pairs}{\mathcal{P}}
\newcommand{\footprint}{\mathcal{F}}
\newcommand{\annulus}{\mathcal{A}}
\newcommand{\lens}{\mathrm{L}}
\newcommand{\shape}{\mathrm{E}}
\newcommand{\pol}{\mathrm{p}}
\newcommand{\position}{\mathrm{P}}
\newcommand{\observable}{\mathrm{O}}
\newcommand{\universe}{\mathcal{U}}
\newcommand{\vanishes}[1]{\textcolor{red4}{#1}}
\newcommand{\negligible}[1]{\textcolor{mygray}{#1}}
\newcommand{\wnoise}[1]{\widetilde{#1}}

\newcommand{\Euclid}{\textit{Euclid}\xspace}
\newcommand{\cev}[2][]{\delimiters[#1]{\langle}{#2}{\rangle}}
\newcommand{\gsev}[2][]{\delimiters[#1]{\lceil}{#2}{\rceil}}
\newcommand{\lsev}[2][]{\delimiters[#1]{\lfloor}{#2}{\rfloor}}
\newcommand{\Cov}{\mathrm{Cov}}
\newcommand{\Var}{\mathrm{Var}}
\definecolor{ccolour}{RGB}{128,0,0}
\definecolor{ncolour}{RGB}{0,128,0}
\definecolor{scolour}{RGB}{0,0,128}
\newcommand{\CCov}{{\color{ccolour}\mathrm{CCov}}}
\newcommand{\NCov}{{\color{ncolour}\mathrm{NCov}}}
\newcommand{\SCov}{{\color{scolour}\mathrm{SCov}}}
\newcommand{\deltag}{\delta\h{g}}
\newcommand{\SNR}{\text{SNR}}

\newcommand{\code}[2][]{%
    \if\relax\detokenize{#1}\relax
        \textsc{#2}%
    \else
        \href{#1}{\textsc{#2}}%
    \fi
}

\theoremstyle{nonumberbreak}
\newmdtheoremenv[%
outerlinewidth=0pt,%
roundcorner=5pt,%
backgroundcolor=black!3,%
ntheorem=true%
]{FAQ}{Frequently asked questions}

\newcommand{\modified}[1]{#1}
\definecolor{blue4}{RGB}{0,0,143}
\definecolor{red4}{RGB}{143,0,0}
\definecolor{orange}{RGB}{255,128,0}
\definecolor{darkcyan}{RGB}{0,128,128}
\definecolor{olive}{RGB}{0,128,0}
\definecolor{purple}{RGB}{128,0,128}
\definecolor{cyan2}{RGB}{0,255,255}
\definecolor{fushia}{RGB}{255,0,255}
\definecolor{mygray}{gray}{0.5}
\definecolor{lightgray}{gray}{0.85}

\newcommand{\EClink}{https://github.com/ELROND-project/encyclopaedia_covariancis/blob/main/Encyclopaedia_covariancis.pdf}
\newcommand{\EC}{\href{\EClink}{\emph{Encyclop{\ae}dia Covariancis}\xspace}}

\makeatletter
\g@addto@macro\bfseries{\boldmath}
\makeatother

\title{Cosmology with the line-of-sight shear of strong gravitational lenses}

\author[1]{Pierre~Fleury~\orcidlink{0000-0001-9292-3651},}
\author[1]{Daniel~Johnson~\orcidlink{0000-0002-0311-2513},}
\author[1]{Théo~Duboscq~\orcidlink{0009-0005-1300-3782},}
\author[2, 3]{Natalie~B.~Hogg~\orcidlink{0000-0001-9346-4477},}
\author[1]{Julien~Larena~\orcidlink{0000-0003-3301-0435}.}

\affiliation[1]{Laboratoire Univers et Particules de Montpellier, 
CNRS \& Université de Montpellier,\\
Parvis Alexander Grothendieck, F-34095 Montpellier Cedex 05, France}

\affiliation[2]{Institute of Astronomy, University of Cambridge,\\Madingley Road, Cambridge, CB3 0HA,~UK}

\affiliation[3]{Kavli Institute for Cosmology, University of Cambridge,\\Madingley Road, Cambridge, CB3 0HA,~UK}

\emailAdd{pierre.fleury@lupm.in2p3.fr}
\emailAdd{daniel.johnson@umontpellier.fr}
\emailAdd{theo.duboscq@umontpellier.fr}
\emailAdd{natalie.hogg@ast.cam.ac.uk}
\emailAdd{julien.larena@umontpellier.fr}

\abstract{
Stage-IV photometric galaxy surveys are designed to measure the position and shapes of billions of galaxies. Their aim is to characterise the large-scale distribution of matter in the Universe using galaxy clustering and weak gravitational lensing. As a byproduct, stage-IV surveys are expected to detect more than a hundred thousand strong gravitational lenses.
In this article, we propose the use of weak-lensing perturbations to strong lenses, specifically their line-of-sight (LOS) shear, as a cosmological probe.
This new observable allows us to define three new correlation functions: the LOS shear with itself, with galaxy positions, and with galaxy shapes, thereby promoting the standard $3\times 2$pt correlation method to a $6\times 2$pt scheme. We design estimators for these new correlation functions and determine their expectation values as a function of the matter power spectrum. We then derive the analytical expression for the full covariance matrix of the $6\times 2$pt correlation scheme.
Considering various scenarios for the stage-IV strong-lensing samples, we demonstrate that the cosmological information carried by the LOS shear of strong lenses will be detectable with a very high signal-to-noise ratio, even in the most pessimistic of cases. Strong lenses are thus extremely promising cosmological probes, whose synergy with galaxy positions and shapes should also contribute to mitigating systematics in stage-IV surveys.
}


\date{\today}

\begin{document}

\maketitle
\flushbottom

\section{Introduction}
\label{sec:introduction}

Gravitational lensing is the phenomenon resulting from the deflection of light rays by the gravitational fields between the source and the observer. Most of the time, gravitational lensing only leads to small effects -- a slight magnification, or demagnification, of light sources, accompanied by a slight distortion of their shapes. The latter manifestation of weak lensing, known as cosmic shear, produces alignment patterns of background galaxies -- tangentially around foreground matter overdensities, and radially around underdensities~\cite{2001PhR...340..291B}. The larger the density fluctuation, the more apparent the pattern, which is why cosmic shear has become a key probe of the large-scale distribution of matter in the Universe. It is particularly efficient at constraining the combination of cosmological parameters $S_8=\sigma_8\sqrt{\Omega\e{m}/0.3}$~\cite{1997ApJ...484..560J, 2021MNRAS.505.4935H}, where $\Omega\e{m}$ is the relative contribution of matter density to the dynamics of cosmic expansion, and $\sigma_8$ is the standard deviation of the matter density contrast smoothed on a scale of $8 h^{-1}\,\mathrm{Mpc}$.

Stage-III photometric galaxy surveys, namely the Dark Energy Survey (DES)~\cite{DES:2026fyc}, the Subaru Hyper Suprime-Cam (HSC) survey~\cite{Sugiyama:2023fzm, Miyatake:2023njf}, and the Kilo-Degree Survey (KiDS)~\cite{Wright:2025xka}, have constrained $S_8\approx 0.8$ with a precision ranging from $1.5\,\%$ to $5\,\%$. These constraints do not come from cosmic shear alone, but from the so-called $3\times 2$pt correlation scheme, in which the two-point correlation function of the apparent ellipticity of galaxies (cosmic shear) is combined with the two-point correlation function of galaxy positions (galaxy clustering) and their cross-correlation (galaxy--galaxy lensing). Stage-IV surveys, such as \Euclid~\cite{Euclid:2024yrr} and the Legacy Survey of Space and Time (LSST) of the Vera C. Rubin Observatory~\cite{LSST:2008ijt}, are expected to improve these constraints by about an order of magnitude~\cite{Euclid:2019clj}, thanks to unprecedented catalogues of billions of galaxies with excellent imaging quality. In the era of stage-IV galaxy surveys, our ability to extract cosmological information will no longer be limited by statistics, but by systematic uncertainties, such as the intrinsic alignment of galaxies, their redshift distribution, and the modelling of small-scale baryonic processes. Controlling systematics, is, therefore, the main challenge of stage-IV surveys.

Due to their sensitivity, resolution, and their wide sky coverage, stage-IV surveys will also observe an abundance of the rarer strong manifestations of gravitational lensing, namely multiple images, arcs and Einstein rings, which typically occur when a foreground galaxy happens to lie directly on line of sight to another, more distant, galaxy. As a rule of thumb, there is one strong-lensing event every ten thousand observed galaxies, which implies that stage-IV catalogues shall comprise over a hundred thousand strong lenses~\cite{2015ApJ...811...20C}, a number that is compatible with cosmological applications. This is the topic of the present article.

To first approximation, a strong gravitational lens may be modelled as some concentration of matter in an otherwise homogeneous and isotropic Universe. In this approach, light deflection only occurs in the vicinity of the main deflector, but on the scales relevant to strong lensing, the Universe is far from being homogeneous, and the propagation of light is perturbed by the matter structures between the source, the main deflector, and the observer. If the most prominent matter concentrations happen to lie far from the line of sight (LOS), then they may be treated in the tidal regime~\cite{1987ApJ...316...52K, Schneider:1997bq, 1996ApJ...468...17B, McCully:2013fga, Fleury:2020cal}, whereby their net effect is captured by a single complex parameter: the LOS shear~\cite{2021JCAP...08..024F}. Most importantly, ref.~\cite{Hogg:2022ycw} demonstrated on mocks that the LOS shear can be directly measured on individual strong-lensing images, thereby confirming the preliminary results of \cite{Birrer:2016xku}, which relied on a different parameterisation of the LOS effects. Put simply, this means that \emph{Einstein rings are standardisable shapes},\footnote{This does not mean that Einstein rings are or need to be perfectly circular in the absence of LOS perturbations, but rather that one can distinguish, in an image, between the morphological features due to the main deflector and those due to LOS perturbations.} whose distortions are a probe of cosmological perturbations~\cite{Birrer:2017sge}. Albeit still in their infancy, actual measurements of the LOS shear have already been performed~\cite{Hogg:2025wac, Hogg:2025asw} on a sample of 45 lenses from the Sloan Lens Advanced Camera for Surveys (SLACS) catalogue~\cite{2006ApJ...638..703B}.

In this article, we present a complete framework for the use of the LOS shear of strong lenses as a cosmological probe. In a nutshell, we propose to add the LOS shear to the standard list of observables, namely galaxy positions and shapes, so as to promote the $3\times 2$pt correlation method to a $6\times 2$pt one. This strategy is mainly motivated by the fact that galaxies and strong lenses are subject to different systematics, which should be significantly mitigated by their combination. We then evaluate the detectability of a cosmological signal from the LOS shear in the context of stage-IV surveys.

The remainder of the article is organised as follows. In \cref{sec:LOS_shear}, we summarise the main concepts associated with the LOS shear in strong lensing, and establish its expression in cosmological perturbation theory. In \cref{sec:cosmology_with_LOS}, we present the $6\times 2$pt correlation method, we define estimators for the three new correlation functions involving the LOS shear and we derive their expectation values in standard cosmology. In \cref{sec:covmat}, we analytically calculate the full covariance matrix of the $6\times 2$pt correlations, with a view for its use in a Fisher forecast. Finally, we apply these results in \cref{sec:forecasted_detectability} to evaluate the uncertainties on our new observables, and thus assess their detectability with stage-IV galaxy surveys.

\setlength{\textheight}{660pt} 
\paragraph{Conventions and notation} $\define$ indicates a definition. We use bold symbols for vectors and matrices. A hatted bold symbol ($\uvect{u}$) is a unit vector; other hatted quantities indicate either a noisy observable ($\mlLOS$) or an estimator ($\widehat{\observable_1\observable_2}$). There are many indices in this work ($a, A, i, \alpha, \mu, \ldots$) which are consistently associated with a specific meaning (see \cref{tab:convention_indices}). We work with units such that the speed of light is unity, $c=1$.

\begin{FAQ}
\begin{itemize}
\item\emph{What is line-of-sight (LOS) shear?} It is a complex number, $\lLOS$, which captures the effect of tidal perturbations along the LOS of a strong lens. The setup is depicted in \cref{fig:perturbed_ring}, where (o) observes a source~(s) strongly lensed by a deflector~(d), with LOS perturbations. This configuration actually involves three shears, $\gamma\e{od}, \gamma\e{os}, \gamma\e{ds}$, where $\gamma\e{ab}$ would be the complex ellipticity, as seen from (a), of a small circular source at (b), due to weak lensing. The combination $\lLOS=\gamma\e{od}+\gamma\e{os}-\gamma\e{ds}$ affects the image in a way that is maximally independent of the source and main lens. Thus, the LOS shear may be seen as the net weak-lensing distortion of a strong-lensing image. 
\item\emph{What is the difference between LOS shear and the usual weak-lensing shear?} The normal weak-lensing shear is only one of the three terms, namely $\gamma\e{os}$, in the expression of $\lLOS$. The contribution of some matter lump to $\gamma\e{os}$ peaks if it lies halfway between (o) and (s). For $\lLOS$, that contribution is enhanced between (o) and (d) and suppressed between (d) and (s); see \cref{eq:LOS_shear_potential,eq:LOS_potential,eq:lensing_kernel_LOS}.
\item\emph{How can we measure the LOS shear from strong-lensing data?} An image~$I(\vect{\theta})$ consists of surface brightness, $I$, in pixels, $\vect{\theta}$. If the image is produced by gravitational lensing of a source~$I\e{s}$, then $I(\vect{\theta})=I\e{s}[\vect{\theta}-\vect{\alpha}(\vect{\theta})]$, where $\vect{\alpha}$ is the lensing displacement field. When interpreting this image, one must model both the source profile $I\e{s}(\vect{\theta})$ and the lensing displacement $\vect{\alpha}(\vect{\theta})$. In that context, the LOS shear~$\lLOS$ is a parameter in $\vect{\alpha}(\vect{\theta})$ which can be inferred from the data. In \code[https://lenstronomy.readthedocs.io/en/latest/]{Lenstronomy}, this is implemented in the \code[https://github.com/lenstronomy/lenstronomy/tree/main/lenstronomy/LensModel/LineOfSight]{LineOfSight} module.
\item\emph{How can we use the LOS shear in cosmology?} \Cref{eq:LOS_shear_potential,eq:LOS_potential,eq:lensing_kernel_LOS} show that $\lLOS$ depends on the gravitational potential~$\Phi$, just like the density contrast~\eqref{eq:Poisson} or the weak-lensing shear~\eqref{eq:shear_potential}. Measuring correlation functions of these quantities then informs us about the matter power spectrum. Current galaxy surveys exploit three correlation functions built from galaxy positions ($\position$) and ellipticities ($\shape$): $\shape\shape$ (cosmic shear), $\position\position$ (galaxy clustering) and $\shape\position$ (galaxy--galaxy lensing). Adding the LOS shear ($\lens$) allows us to define three new correlation functions ($\lens\lens$, $\lens\shape$, $\lens\position$), thereby promoting the standard $3\times 2$pt method to $6\times 2$pt (see \cref{fig:6x2pt}).
\item\emph{Stage-IV surveys will observe $\order(10^9)$ galaxies and $\order(10^5)$ lenses. How can the latter compete with the former?} In cosmic shear, a very large number of galaxies are required for the cosmological signal to exceed their intrinsic shape noise. This is because, in the apparent ellipticity of a galaxy, $\eps=\eps_0+\gamma$, the intrinsic contribution, $\eps_0$, is typically $30$ times larger than the weak-lensing shear, $\gamma$. In other words, the signal-to-noise ratio (SNR) of weak lensing is about $3\,\%$ for one galaxy. In contrast, we should measure the LOS shear on individual strong lenses with an SNR of about $100\%$, so that comparatively few lenses are needed to extract a cosmological signal. Furthermore, we expect the LOS shear to be immune to cosmic-shear systematics like intrinsic alignment, which makes these probes highly complementary.
\item\emph{Why compute the $6\times 2$pt covariance matrix analytically instead of using simulations?} Analytical calculations have the advantage of producing results that are easier to interpret. Since the project to use the LOS shear as a cosmological observable is at an early stage, interpretation is essential to design and optimise the future observational strategy (which lenses to keep? how to bin them by redshift and angular separation?). Furthermore, analytical results are more flexible: it is easier to add new ingredients to a formula than to run an entirely new suite of simulations. Not to mention that simulating a cosmological sample of perturbed strong lenses is significantly more challenging than simulating a galaxy sample.
\item\emph{Will strong lenses improve the measurements of cosmological parameters?} In this article, we demonstrate that the new correlation functions, $\lens\lens$, $\lens\shape$ and $\lens\position$, are detectable from stage-IV surveys with high SNR, across a range of angular scales. We are currently conducting Fisher forecasts to quantify the expected gain in precision and accuracy for cosmological parameters such as $\Omega\e{m}$ and $\sigma_8$; the results will be presented in a subsequent paper. 
\end{itemize}
\end{FAQ}

\setlength{\textheight}{642pt} 
\section{Line-of-sight shear in strong lensing}
\label{sec:LOS_shear}

\subsection{Tidal perturbations to a strong lens}

\begin{figure}[t]
\centering
\begin{minipage}{0.5\columnwidth}
\includegraphics[width=\columnwidth]{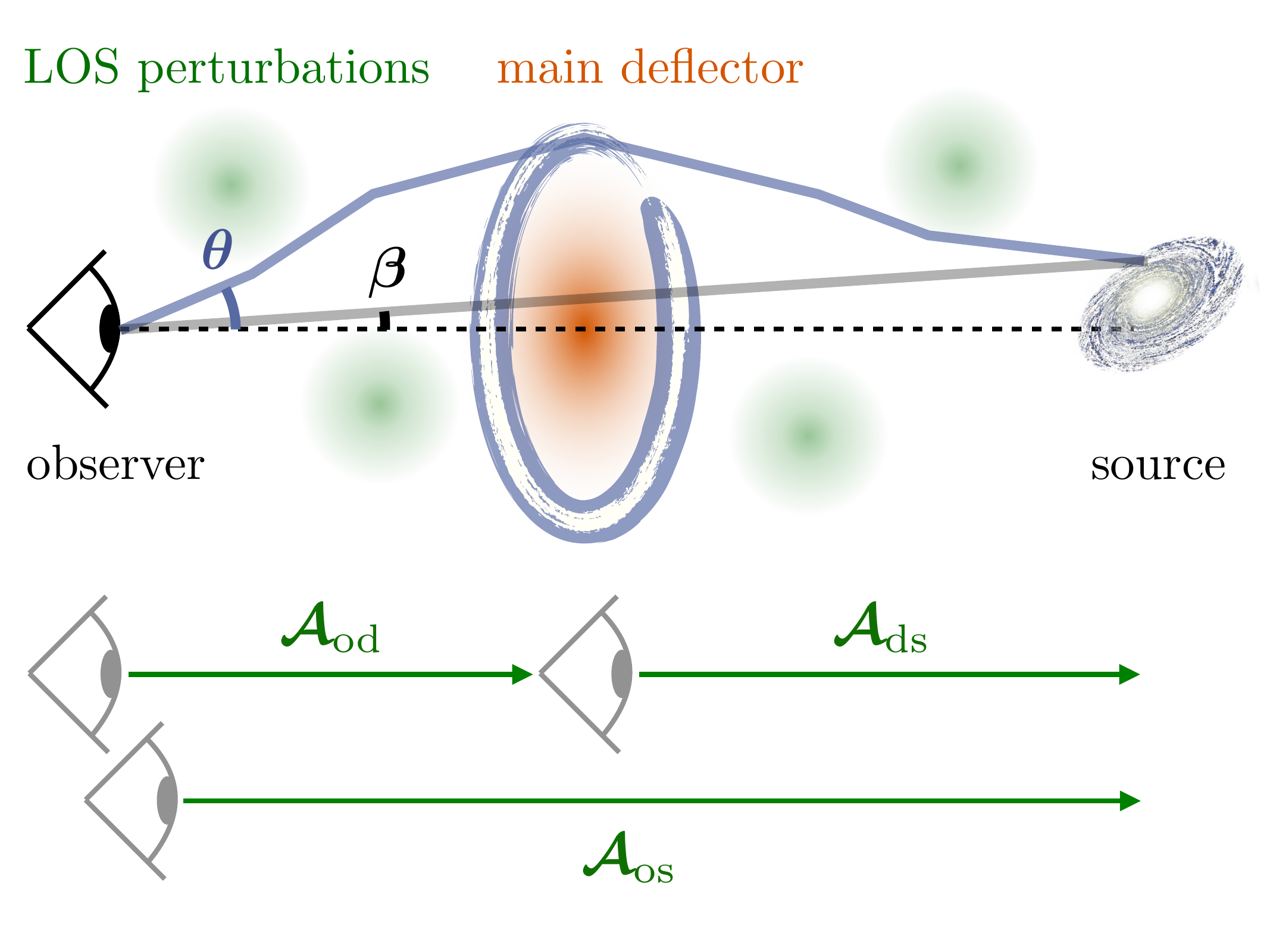}
\end{minipage}
\hfill
\begin{minipage}{0.46\columnwidth}
\caption{A strong-lensing system where a background galaxy (source), lies behind a foreground galaxy (main deflector), while inhomogeneities near the LOS are treated as tidal perturbations. In the absence of the main lens and LOS perturbations, a point in the source galaxy is observed at $\vect{\beta}$; $\vect{\theta}$ is an image of that point by the entire system $\{\text{main deflector} + \text{LOS perturbations}\}$. The $2\times 2$ distortion matrices $\mat{\amp}\e{ab}$ ($\mathrm{a, b}\in\{\mathrm{o, d, s}\}$) quantify the effect of the LOS perturbations when a source at (b) is observed from (a).}
\label{fig:perturbed_ring}
\end{minipage}
\end{figure}

Consider the scenario depicted in \cref{fig:perturbed_ring}, where a source galaxy is well aligned with a foreground galaxy acting as a strong gravitational lens. Let us denote with $\vect{\beta}$ the position at which some point within the source galaxy would be observed in the absence of the lens, through a homogeneous and isotropic Universe. We then denote with $\vect{\theta}$ the position of an image of that point, i.e. the direction in which it is observed in the presence of deflectors.

If the main deflector were isolated, that is, far from any other cosmic inhomogeneity, then the relationship between $\vect{\beta}$ and $\vect{\theta}$ -- known as the lens equation -- would read $\vect{\beta}=\vect{\theta}-\vect{\nabla}\Psi(\vect{\theta})$, where $\Psi(\vect{\theta})$ is the lensing potential of the main deflector~\cite{1992grle.book.....S}, and $\vect{\nabla}$ denotes the two-dimensional gradient with respect to $\vect{\theta}$. If, however, cosmological perturbations are present on the LOS, then the previous relation is modified to~\cite{1987ApJ...316...52K, 1996ApJ...468...17B, Schneider:1997bq, McCully:2013fga, Fleury:2020cal, 2021JCAP...08..024F}
\begin{equation}
\label{eq:lens_equation}
\vect{\beta}
=
\mat{\amp}\e{os}\vect{\theta}
- \mat{\amp}\e{ds} (\vect{\nabla}\Psi)(\mat{\amp}\e{od}\vect{\theta}) \ ,
\end{equation}
where $\mat{\amp}\e{os}, \mat{\amp}\e{ds}, \mat{\amp}\e{od}$ are called distortion matrices. Each one of the $\mat{\amp}\e{ab}$, with $\mathrm{a, b} \in \{\mathrm{o, d, s}\}$, is a $2\times 2$ matrix that encodes the effect of weak LOS perturbations, with respect to a homogeneous--isotropic Universe, when a source at (b) is observed from (a). For instance, in the absence of the main deflector ($\vect{\nabla}\Psi=\vect{0}$), the lens equation~\eqref{eq:lens_equation} reduces to $\vect{\beta}=\mat{\amp}\e{os}\vect{\theta}$, thereby illustrating how LOS perturbations alone affect the appearance of images.

Such a description of LOS effects is valid only if the secondary deflectors can be treated in the \emph{tidal regime}. This means that their gravitational field is well-approximated, at the scale of the strong-lensing image, by a second-order expansion of their gravitational potential around the LOS~\cite{McCully:2013fga, 2021JCAP...08..024F}. Physically speaking, this corresponds either to smooth distributions of matter lying near the LOS, or to concentrations of matter lying far from it. See refs.~\cite{2021JCAP...08..024F, Duboscq:2024asf} for a description of LOS effects beyond the tidal regimes, accounting in particular for flexion~\cite{2005ApJ...619..741G}.

If the LOS perturbations are weak, the distortion matrices~$\mat{\amp}\e{ab}$ are approximately symmetric~\cite{Fleury:2015hgz}. As such, they can be decomposed into a scalar component and symmetric trace-free component as
\begin{equation}
\mat{\amp}\e{ab}
=
\mat{1}
-
\begin{bmatrix}
\kappa\e{ab} & 0 \\
0 & \kappa\e{ab}
\end{bmatrix}
-
\begin{bmatrix}
\Re(\gamma\e{ab}) & \Im(\gamma\e{ab})
\\
\Im(\gamma\e{ab}) & -\Re(\gamma\e{ab})
\end{bmatrix} \ ,
\end{equation}
where the convergence $\kappa\e{ab}$ is caused by the smooth matter on the LOS, while the complex shear $\gamma\e{ab}$ is due to concentrations of matter around the LOS, between (a) and (b).
 
\subsection{Measuring the LOS shear with strong-lensing images}

From a strong-lensing image such as an Einstein ring, it is generally impossible to identify the effect of individual LOS perturbations $\mat{\amp}\e{os}, \mat{\amp}\e{od}, \mat{\amp}\e{ds}$, or distinguish them from the properties of the main deflector, $\Psi$. These degeneracies are partly due to our ignorance of the position and morphology of the source: in \cref{eq:lens_equation}, $\vect{\beta}$ is a parameter to be fitted just like $\Psi$ and $\mat{\amp}\e{ab}$.

This source-position degeneracy~\cite{2014A&A...564A.103S} can nonetheless be exploited so as to effectively reduce the number of parameters in the lens equation~\cite{2021JCAP...08..024F}. Multiplying \cref{eq:lens_equation} on the left with $\mat{\amp}\e{od}\mat{\amp}\e{ds}^{-1}$, and defining the displaced source position~$\vect{\beta}'=\mat{\amp}\e{od}\mat{\amp}\e{ds}^{-1}\vect{\beta}$, we have
\begin{equation}
\label{eq:minimal_model}
\vect{\beta}'
=
\mat{\amp}\e{LOS}\vect{\theta}
- (\vect{\nabla}\Psi\e{eff})(\vect{\theta}) \ ,
\qquad
\text{with}
\quad
\begin{system}
\mat{\amp}\e{LOS}
&\define \mat{\amp}\e{od}\mat{\amp}\e{ds}^{-1}\mat{\amp}\e{os}
\\
\Psi\e{eff}(\vect{\theta})
&\define
\Psi(\mat{\amp}\e{od}\vect{\theta})
\end{system} \ .
\end{equation}
\Cref{eq:minimal_model} is the \emph{minimal LOS model}~\cite{2021JCAP...08..024F}. It is formally equivalent to the description of a single-plane lens system, which would consist of a main deflector with potential~$\Psi\e{eff}(\vect{\theta})$, together with an ``external convergence'' and an ``external shear'' encoded in the distortion matrix~$\mat{\amp}\e{LOS}$. On the one hand, the expression of the effective potential~$\Psi\e{eff}$ shows that there is a strong degeneracy between the foreground perturbations~$\mat{\amp}\e{od}$ and the properties of the main deflector~\cite{2024JCAP...10..055J}. On the other hand, the expression of $\mat{\amp}\e{LOS}$ shows that it would be ill-advised to try to measure $\mat{\amp}\e{od}, \mat{\amp}\e{os}, \mat{\amp}\e{ds}$ independently from one another. It is, however, possible to measure the shear component of $\mat{\amp}\e{LOS}$, that is
\begin{empheq}[box=\fbox]{equation}
\label{eq:LOS_shear_def}
\gamma\e{LOS}
=
\gamma\e{od} + \gamma\e{os} - \gamma\e{ds}
\end{empheq}
in the weak-lensing limit ($\kappa\e{ab}, |\gamma\e{ab}|\ll 1$). We shall refer to $\gamma\e{LOS}$ as the \emph{LOS shear} throughout. From a theoretical point of view, one expects $\gamma\e{LOS}$ to be distinguishable from the properties of the main lens, such as its ellipticity. This is because the LOS shear represents a strictly homogeneous tidal field, while the main lens's tidal field always displays some radial gradient. See ref.~\cite{Birrer:2021iuz} for a comparison of the morphological features of shear and ellipticity.

The practical measurability of the LOS shear from realistic strong-lensing images was assessed in ref.~\cite{Hogg:2022ycw}, elaborating on a preliminary attempt~\cite{Birrer:2016xku} to infer $\gamma\e{os}-\gamma\e{ds}$, before $\gamma\e{LOS}$ was identified by ref.~\cite{2021JCAP...08..024F} as the least degenerate shear combination. This proof of concept relied on a sample of 64 mock Hubble Space Telescope (HST) images, produced by complex analytical lens models, in the presence of tidal LOS perturbations. It was shown that the LOS shear can be accurately measured, with a mean uncertainty on the order of 0.01, provided that the model for the main lens is elaborate enough. This corresponds to a signal-to-noise ratio of order unity for individual lenses. The first measurements of the LOS in real data were made~\cite{Hogg:2025wac, Hogg:2025asw} in 45 strong lenses from the SLACS catalogue~\cite{2006ApJ...638..703B}; this work also investigated possible systematics, such as the role of boxiness/diskiness in lens mass profiles~\cite{2022A&A...659A.127V}.

\subsubsection*{Caveats}

Albeit encouraging, these results also emphasise that LOS shear measurements are still in their infancy. In particular, it is unclear what ``elaborate enough'' means in practice for a lens model. In the context of Bayesian forward modelling of an image, if the actual lens possesses morphological features that are not accounted for in the model, then the sampler will tend to use the free shear in order to compensate for the missing features, thereby leading to spurious biases in the best-fit value of $\gamma\e{LOS}$. An eloquent example was found with the mock analysis of ref.~\cite{Hogg:2022ycw}: if a lens consists of a baryonic core and a dark-matter halo whose centres are offset, and if one tries to fit the image produced by that lens assuming that both components share the same centre, then the measured $\gamma\e{LOS}$ is clearly biased; the bias disappears if the model allows for the offset as a free parameter.

Whilst the necessary features of a good lens model -- as far as the LOS shear is concerned -- are still to be determined, we can already claim that the standard elliptical power-law~(EPL) model does not fulfil them. In addition to the biases observed in mocks~\cite{Hogg:2022ycw, Etherington:2023yyh}, studies on real data lead to the same conclusion. Analysing HST data using EPL plus external shear, ref.~\cite{Etherington:2023yyh} found suspicious correlations between the orientation of the external shear and the main lens's ellipticity, which suggests that the external-shear parameter may account for details of the lens's azimuthal structure that are not well-described by a mere ellipticity~\cite{VandeVyvere:2022gqa}. Reference~\cite{Hogg:2025wac}, besides, finds that the best-fit values of $\gamma\e{LOS}$ for SLACS lenses modelled as EPL plus LOS shear are typically ten times larger than expectations from cosmology.

\modified{Such large shears may be due to the lack of angular complexity of the EPL model, but they may also originate from the immediate environment of the main lens -- which contributes to $\lLOS$ via the $\gamma\e{os}$ term. Although it appears that lenses are not more likely to reside in group or clusters than non-lens galaxies with comparable properties~\cite{2000ApJ...545..129K, 2009ApJ...690..670T, 2018ApJ...867..107W}, those that do can experience a strong external shear from the host halo and from the neighbouring group members. There does not seem to be a consensus on the magnitude of this effect. From spectroscopic and photometric data in the field of nine lenses, ref.~\cite{2011ApJ...726...84W} estimated the average environmental shear magnitude to be $\ev{|\gamma\e{env}|}\approx 0.05$. This is twice larger than the result of ref.~\cite{2005tyad.confE..12D}, based on numerical simulations. The role of environmental shear in the much larger stage-IV lens samples remains, therefore, to be assessed.}

Another potential source of systematics in $\lLOS$ comes from neglecting higher-order LOS effects, such as the LOS flexion. Reference~\cite{Duboscq:2024asf} derived a minimal LOS flexion model, which features four additional complex parameters. In most realistic situations, the LOS flexion is too small to be measured, but it does produce a systematic bias on $\lLOS$ if not accounted for.

In this article, we shall assume that the LOS shear is strictly due to cosmological perturbations, and that it can be measured on strong-lensing images without systematic biases, albeit with a signal-to-noise ratio that may be smaller than unity on individual images. Our goal is to investigate its statistical measurability through correlation functions.

\subsection{LOS shear from cosmological perturbations}
\label{subsec:LOS_shear_perturbations}

Let us assume that the space-time geometry experienced by light on its way from the source to the main deflector, and then to the observer, is well approximated by a spatially flat Friedmann--Lemaître--Robertson--Walker model with perturbations. The line element reads
\begin{equation}
\dd s^2
=
- \pa{1+2\Phi} a^2(\eta) \, \dd\eta^2
+ \pa{1-2\Phi} a^2(\eta)
\pac{ \dd\chi^2 + \chi^2 \pa{\dd\theta^2+\sin^2\theta \, \dd\ph^2} }
 ,
\end{equation}
where $a(\eta)$ is the background scale factor, $\eta$ is conformal time, $\chi$ is the comoving distance to an arbitrary origin, $\theta$ and $\ph$ are polar coordinates on the two-sphere, and $\Phi(\eta, \chi, \theta, \ph)$ is the gravitational potential produced by matter inhomogeneities. The latter is connected to the density perturbations via the Poisson equation,
\begin{equation}
\Delta\Phi = 4\pi G a^{-1} \bar{\rho}_0 \delta \ ,
\label{eq:Poisson}
\end{equation}
where $\bar{\rho}_0$ is today's mean cosmic density and $\delta\define (\rho-\bar{\rho})/\bar{\rho}$ the dimensionless density contrast.

Solving the null geodesic deviation equation at leading order in $\Phi$, we find that the shear produced on an image at (b) observed from (a) can be expressed in terms of a potential as
\begin{equation}
\gamma\e{ab}(\uvect{u}) = \frac{1}{2} \, \eth^2 \Psi\e{ab}(\uvect{u}) \ ,
\end{equation}
where $\eth$ is the spin-raising operator (or eth\footnote{\ ``Eth'' is the name of the phonetic character $\eth$, although it seems to be mostly referred to as ``edth'' in the mathematical literature since ref.~\cite{1967JMP.....8.2155G}.} derivative); it acts on the angular variables~$(\theta, \ph)$, which we gather into the unit vector $\uvect{u}$ representing the corresponding position on the two-sphere; see e.g. ref.~\cite{2005PhRvD..72b3516C} and references therein for details on the $\eth$ operator. The potential~$\Psi\e{ab}$ is a weighted projection of the gravitational potential along the lightcone between (a) and (b),
\begin{equation}
\Psi\e{ab}(\uvect{u})
=
2\int_0^\infty \frac{\dd\chi}{\chi} \; K(\chi; \chi\e{a}, \chi\e{b}) \, \Phi(\eta_0-\chi, \chi, \uvect{u}) \ ,
\end{equation}
where $\eta_0$ is conformal time today, and the integration kernel reads
\begin{equation}
\label{eq:lensing_kernel}
K(\chi; \chi\e{a}, \chi\e{b})
=
\begin{cases}
\frac{(\chi\e{b}-\chi)(\chi-\chi\e{a})}{\chi(\chi\e{b}-\chi\e{a})}
& \text{if } \chi\e{a} \leq \chi \leq \chi\e{b} \ , \\
0 & \text{otherwise.}
\end{cases}
\end{equation}

The expression of the LOS shear then follows from \cref{eq:LOS_shear_def}, implying that $\gamma\e{LOS}$ derives from the potential $\Psi\e{LOS}=\Psi\e{od}+\Psi\e{os}-\Psi\e{ds}$; in other words,
\begin{empheq}[box=\fbox]{align}
\label{eq:LOS_shear_potential}
\gamma\e{LOS}
&=
\frac{1}{2} \, \eth^2\Psi\e{LOS} \ ,
\\
\label{eq:LOS_potential}
\Psi\e{LOS}(\uvect{u}, \chi\e{d}, \chi\e{s})
&=
2\int_0^\infty \frac{\dd\chi}{\chi} \; K\e{LOS}(\chi;\chi\e{d}, \chi\e{s}) \, \Phi(\eta_0 - \chi, \chi, \uvect{u}) \ ,
\\
\label{eq:lensing_kernel_LOS}
K\e{LOS}(\chi; \chi\e{d}, \chi\e{s})
&=
\begin{cases}
\frac{\chi\e{s}-\chi}{\chi\e{s}} + \frac{\chi\e{d}-\chi}{\chi\e{d}}
& \text{if } 0\leq\chi\leq\chi\e{d} \ , \\
\frac{\chi\e{s}-\chi}{\chi\e{s}} - \frac{(\chi-\chi\e{d})(\chi\e{s}-\chi)}{\chi(\chi\e{s}-\chi\e{d})}
& \text{if } \chi\e{d}\leq\chi\leq\chi\e{s} \ , \\
0 & \text{if } \chi\e{s} \leq \chi \ . 
\end{cases}
\end{empheq}
In the above, $\chi\e{d}$ and $\chi\e{s}$ respectively denote the comoving distances from the observer to the deflector and to the source, in a strong-lensing system.

For comparison, the usual weak-lensing shear~$\gamma$ affecting the apparent shapes of galaxies corresponds to the (os) part of the above only. In other words, if a galaxy at comoving distance~$\chi_*$ is observed in the direction~$\uvect{u}$, then its apparent (complex) ellipticity reads
\begin{equation}
\label{eq:apparent_ellipticity}
\eps = \eps_0 + \gamma(\uvect{u}, \chi_*) \ ,
\end{equation}
where $\eps_0$ denotes the intrinsic ellipticity that would be observed in the absence of lensing, and
\begin{equation}
\label{eq:shear_potential}
\gamma(\uvect{u}, \chi_*)
=
\frac{1}{2} \eth^2\Psi(\uvect{u}, \chi_*) \ ,
\qquad
\Psi(\uvect{u}, \chi_*)
=
2\int_0^\infty \frac{\dd\chi}{\chi} \; K(\chi; 0, \chi_*) \, \Phi(\eta_0-\chi, \chi, \uvect{u}) \ .
\end{equation}

\section{Cosmology with the line-of-sight shear}
\label{sec:cosmology_with_LOS}

As seen in \cref{subsec:LOS_shear_perturbations}, in particular with \cref{eq:LOS_potential}, the LOS shear directly depends on the gravitational field along the LOS of a strong-lens system. As such, it is a potential cosmological probe. In this section, we present a method to extract cosmological information from the LOS shear, inspired from the standard methods of current galaxy surveys.

\subsection{A six times two-point correlation scheme}

Stage-IV galaxy surveys, such as \Euclid~\cite{Euclid:2024yrr} or the Legacy Survey of Space and Time (LSST) of the Vera C. Rubin Observatory~\cite{LSST:2008ijt} will measure the position, redshift, and apparent ellipticity of $\order(10^9)$ galaxies. Cosmological information is usually extracted from the data with the $3\times 2$pt correlation method, which consists of: (1) the two-point correlation of galaxy positions, called \emph{galaxy clustering}, $\position\position\sim\cev{\delta\e{g}\times\delta\e{g}}$, where $\delta\e{g}$ denotes the density contrast of galaxy number counts; (2) the two-point correlation of galaxy ellipticities~$\eps$, called \emph{cosmic shear}, $\shape\shape\sim\cev{\eps\times\eps}$; and (3) the cross-correlation of galaxy shapes and positions, called \emph{galaxy--galaxy lensing}, $\shape\position\sim\cev{\eps\times\delta\e{g}}$.

\begin{figure}[t]
    \centering
    \includegraphics[width=0.75\columnwidth]{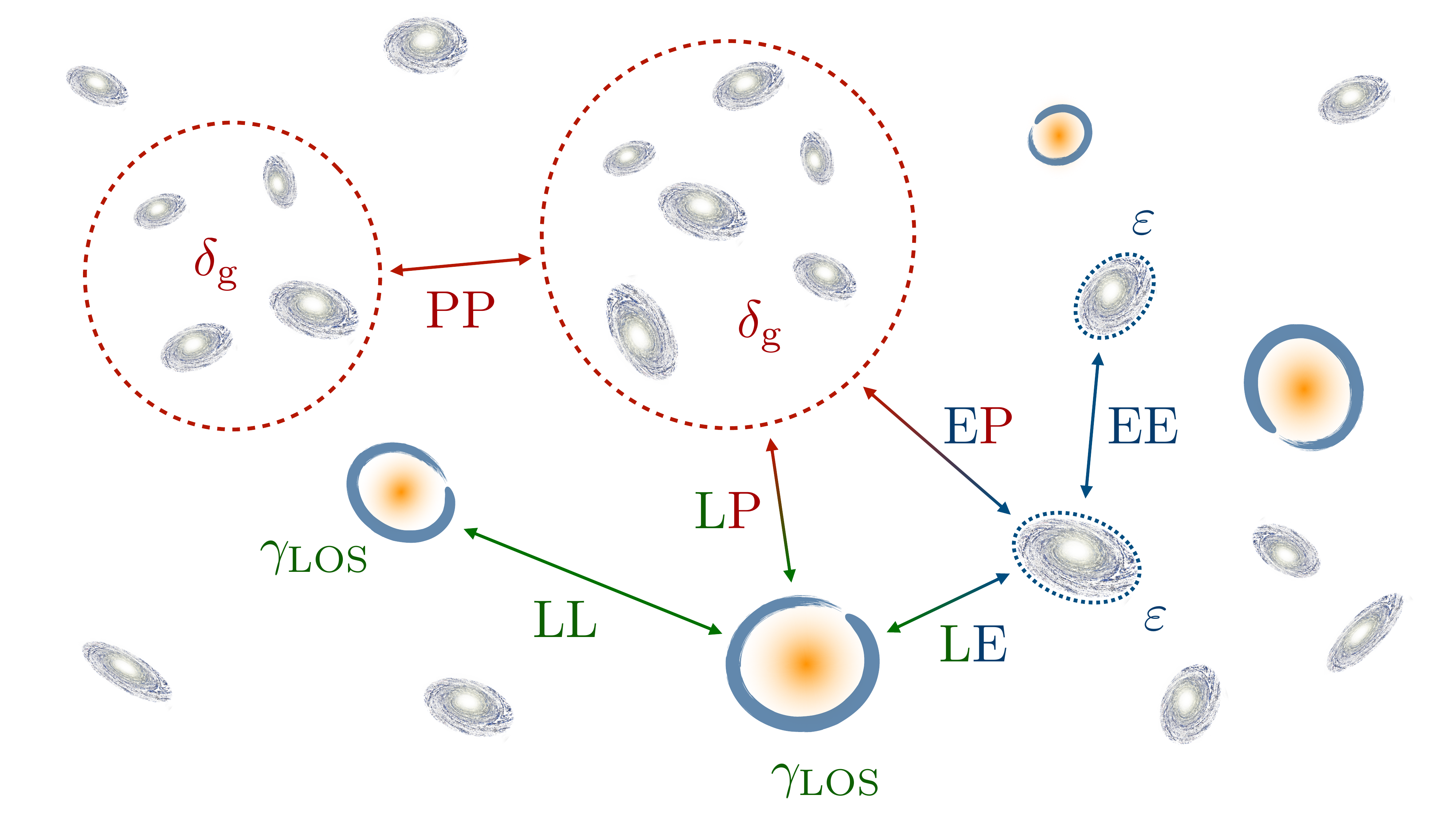}
    \caption{Stage-IV galaxy surveys will observe the position and apparent ellipticity~$\eps$ of more than a billion galaxies, and more than a hundred thousand strong lenses, from which we should be able to measure the LOS shear~$\gamma\e{LOS}$. Measuring all the two-point correlation functions of the galaxy position~(P), ellipticity~(E) and the LOS shear of lenses~(L) defines the $6\times 2$pt correlation scheme.}
    \label{fig:6x2pt}
\end{figure}

Thanks to their wide sky coverage and the quality of their imaging, these surveys are also expected to deliver an unprecedentedly large sample of $\order(10^5$) strong lenses~\cite{2015ApJ...811...20C, 2023MNRAS.525.2341H, Ferrami:2024obm}. If the LOS shear is measured from a significant fraction of this sample, then it could be used as a cosmological probe. By analogy with cosmic shear, we may naturally consider the two-point correlation of the LOS shear, $\lens\lens\sim\cev{\lLOS\times\lLOS}$; then, since the LOS shear has the same physical origin as the shear applied to galaxies, we may add the cross-correlation between LOS shear and galaxy ellipticities, $\lens\shape\sim\cev{\lLOS\times\eps}$; and finally, by analogy with galaxy--galaxy lensing, signal should be present in the cross-correlation between LOS shear and galaxy positions, $\lens\position\sim\cev{\lLOS\times \delta\e{g}}$, because galaxies trace the matter distribution whose gravitational effect is responsible for $\lLOS$. As summarised in \cref{fig:6x2pt}, the inclusion of strong lenses would promote the $3\times 2$pt method to a $6\times 2$pt method: $\shape\shape, \shape\position, \position\position, \lens\lens, \lens\shape, \lens\position$. 

Albeit ten thousand times less abundant than galaxies, strong lenses will result in much more precise measurements of the (LOS) shear. Indeed, following the results of \cite{Hogg:2022ycw}, the typical uncertainty on individual measurements of $\lLOS$ is expected to be $\Delta\lLOS\approx 0.01 \sim \lLOS$, leading to a signal-to-noise ratio (SNR) of order unity for individual measurements. For comparison, the intrinsic ellipticity of galaxies, responsible for shape noise in cosmic shear, is $\Delta\gamma=\eps_0\approx 0.3 \sim 30 \times \gamma$, which means an SNR of about $3\%$ for individual measurements~\cite{Wright:2025xka}.

In addition, the LOS shear of strong lenses is expected to be subject to very different systematics compared to the standard $3\times 2$pt observables. For instance, it should be immune to intrinsic alignments. For these reasons, the $6\times 2$pt correlation method has the potential to significantly enhance the cosmological power of stage-IV galaxy surveys by exploiting the synergy of weak-lensing and strong-lensing probes.

Let us mention for completeness that a first attempt to combine weak and strong lensing was performed by ref.~\cite{Kuhn:2020wpy} in the COSMOS field. The authors of that work considered a cross-correlation analogous to $\lens\shape$, but since they modelled only 3 out of the 67 lenses of the HST/COSMOS catalogue~\cite{2008ApJS..176...19F},\footnote{The more recent COSMOS-Web Lens Survey catalogue~\cite{Nightingale:2025mlk, Mahler:2025ajl, Hogg:2025plt} consists of over 100 strong lens candidates.} no signal could be measured. Note also that the concepts of minimal LOS model and LOS shear were not yet established at the time, which could explain the mismatch between the weak-lensing and strong-lensing shears observed by ref.~\cite{Kuhn:2020wpy}.

\subsection{Estimators of the LOS correlation functions}
\label{subsec:estimators_correlation_functions}

In this article, we shall focus on the three new correlation functions compared to the standard $3\times 2$pt method, namely $\lens\lens$, $\lens\shape$ and $\lens\position$. The first step is to define their estimators.

\paragraph{Setup} Consider a galaxy survey with footprint $\footprint\subset\mathcal{S}^2$, where $\mathcal{S}^2$ is the celestial two-sphere; we denote with $\Omega$ the survey area, that is the total solid angle covered by $\footprint$. We assume that the survey contains a total number of $G$ galaxies and $L$ lenses, and we denote with $n\e{G}\define G/\Omega, n\e{L}\define L/\Omega$ the corresponding angular densities.\footnote{For example, the \Euclid survey is expected to observe $G=2\times 10^9$ photometric galaxies and $L=1.7\times 10^5$ galaxy-scale strong lenses over $\Omega=\SI{15000}{\deg\squared}$; this implies $n\e{G}=\SI{40}{\per\arcmin\squared}$ and $n\e{L}=\SI{11}{\per\deg\squared}$.}

\begin{figure}[t]
\begin{minipage}{0.45\columnwidth}
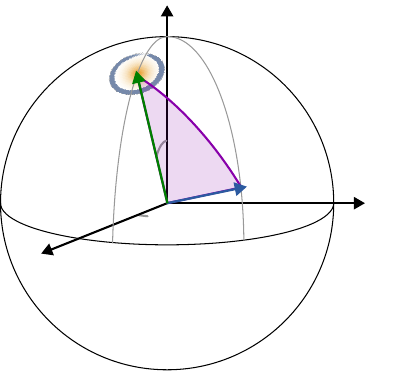
\end{minipage}
\hfill
\begin{minipage}{0.5\columnwidth}
\caption{Schematic summary of the galaxy survey setup and some of our notation. The survey footprint $\footprint\subset\mathcal{S}^2$ contains $G$ galaxies observed in directions $\uvect{u}_a$, and $L$ strong lenses observed in directions $\uvect{u}_i$. A direction $\uvect{u}$ is parameterised by its polar coordinates~$(\theta, \ph)$; the angle formed by the directions $\uvect{u}_i, \uvect{u}_a$ is denoted with $\thet_{ia}$.}
\label{fig:setup}
\end{minipage}  
\end{figure}

\paragraph{Notation} We denote with $\uvect{u}\in\mathcal{S}^2$ a direction in the celestial sphere, which may also be parameterised by its polar coordinates $(\theta, \ph)$ with respect to an arbitrary system of axes. In order to optimise the readability of the subsequent equations, indices will depend on the objects that they refer to, as summarised in \cref{tab:convention_indices}. We denote with $\thet_{\mu\nu}$ the angle formed by any two directions $\uvect{u}_\mu, \uvect{u}_\nu\in \mathcal{S}^2$. See \cref{fig:setup} for an illustrated summary.

\paragraph{Tomographic bins} If the galaxy sample is large enough, we may divide it into $T$ tomographic (redshift) bins, in which case $G_A$ will denote the number of galaxies in the $A$\textsuperscript{th} bin, and $n_A\define G_A/\Omega$. We shall order the set of galaxy indices per redshift bin as
\begin{equation}
\Big\{
\underbrace{1, 2, \ldots, G_1}_{\galaxies_1},
\underbrace{G_1 + 1, G_1 + 2, \ldots G_1 + G_2}_{\galaxies_2},
\ldots,
\underbrace{G-G_T+1, G-G_T+2, \ldots, G}_{\galaxies_T}
\Big\} ,
\end{equation}
so that $\galaxies_A$ denotes the set of galaxies in the $A$\textsuperscript{th} tomographic bin, with $|\galaxies_A|=G_A$.

\begin{table}[h]
    \centering
    \begin{tabular}{c|c|c|c|c}
    Alphabet & Position in alphabet & Letter case & Occurrences & Refers to \\
    \hline
    Latin & beginning & lowercase & $a, b, c, d$ & galaxy \\
    Latin & beginning & uppercase & $A, B, C, D$ & tomographic bin \\
    Latin & middle & lowercase & $i, j, k, l$ & lens \\
    Greek & beginning & lowercase &  $\alpha, \beta$ & angular bin \\
    Greek & middle & lowercase & $\mu, \nu, \rho, \sigma$ & generic index
    \end{tabular}
    \caption{Convention for indices throughout the article.}
    \label{tab:convention_indices}
\end{table}

\subsubsection{Autocorrelation of the LOS shear (LL)}
\label{subsubsec:estimator_autocorrelation_LOS_shear}

The estimators of the LOS shear two-point correlation functions are defined similarly to the standard cosmic shear estimators~\cite{2002A&A...396....1S}.

\paragraph{Noise} First of all, in order to account for the uncertainty of individual LOS shear measurements, we introduce an additive noise term~$\nlLOS$ which will encode both instrumental errors and systematics\modified{, including the environmental shear and the spurious shear due to inaccuracies of the lens model}. Specifically, we shall write the \emph{measured} value of the LOS shear on lens~$i$ as
\begin{equation}
\label{eq:LOS_shear_noise_def}
\mhLOS_i = \hLOS_i + \nhLOS_i \ ,
\end{equation}
where $\hLOS_i$ is the \emph{true} LOS shear field as defined in \cref{subsec:LOS_shear_perturbations} and evaluated at lens~$i$. Note the analogy between $\nlLOS$ and the intrinsic-ellipticity term~$\eps_0$ in cosmic shear; in that sense $\nlLOS$ may be considered as a kind of LOS shape noise. We shall assume throughout the article that LOS shear measurements are unbiased ($\nlLOS$ vanishes on average), and that the LOS noises of distinct lenses are independent.

\begin{figure}[t]
\centering
\begin{minipage}{0.5\columnwidth}
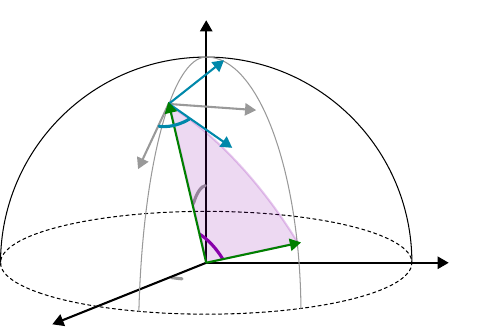
\end{minipage}
\hfill
\begin{minipage}{0.42\columnwidth}
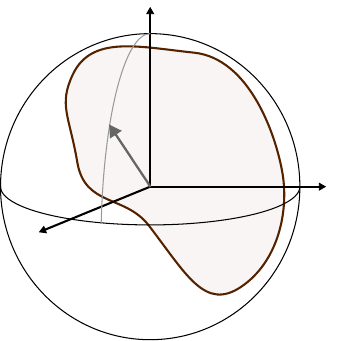
\end{minipage}
\caption{\textit{Left panel:} Two directions $\uvect{u}_i$ and $\uvect{u}_j$ on the two-sphere form an angle $\thet_{ij}$. The figure shows two orthogonal bases for the tangent space $\mathrm{T}_i\mathcal{S}^2$ at $\uvect{u}_i$: the coordinate basis~$(\uvect{e}_\theta, \uvect{e}_\ph)$ in grey; and the $(ij)$-basis $(\uvect{t}_{ij}, \uvect{n}_{ij})$ in turquoise. In the latter, $\uvect{t}_{ij}$ is tangent to the great circle connecting $\uvect{u}_i$ to $\uvect{u}_j$, and $\uvect{n}_{ij}\define \uvect{u}_i\times \uvect{t}_{ij}$. Both bases form an angle $\psi_{ij}$. The $(ij)$-basis is used to define the plus and cross polarisations of the shear for the couple of directions $(ij)$.
\textit{Right panel}: The annulus~$\annulus_\alpha(\uvect{u})$ is defined by the points of $\mathcal{S}^2$ that form an angle $\thet\in[\thet_\alpha, \thet_{\alpha+1})$ with $\uvect{u}$. Its intersection with the survey footprint, $\annulus_\alpha(\uvect{u})\cap\footprint$, represents the $\alpha$\textsuperscript{th} angular bin about $\uvect{u}$.}
\label{fig:geometry_sphere}
\end{figure}

\paragraph{Plus and cross polarisations} Correlation functions aim to identify alignment patterns in pairs of LOS shear measurements. For a pair~$(ij)$ of lenses, the relative orientation of their LOS shear is conveniently expressed in terms of a basis adapted to the pair. The construction goes as follows (see \cref{fig:geometry_sphere}, left panel). Consider the great arc that connects $\uvect{u}_i$ to $\uvect{u}_j$, and denote with $\uvect{t}_{ij}$ the unit vector, tangent to that curve, oriented from $i$ to $j$. A second unit vector tangent to $\mathcal{S}^2$ at $\uvect{u}_i$ may be defined as $\uvect{n}_{ij}\define \uvect{u}_i\times \uvect{t}_{ij}$. Together, these vectors form an orthonormal basis~$(\uvect{t}_{ij}, \uvect{n}_{ij})$ of $\mathrm{T}_i\mathcal{S}^2$ that is rotated by an angle $\psi_{ij}$ with respect to the coordinate basis~$(\uvect{e}_\theta, \uvect{e}_\ph)$. We then define the \emph{plus and cross polarisations} of the LOS shear of lens~$i$ as its real and imaginary parts when expressed in that basis; since $\lLOS$ is implicitly decomposed on the coordinate basis, we have
\begin{equation}
\label{eq:plus_cross_polarisations_definition}
\mhLOS_{+ij} + \ii\,\mhLOS_{\times ij}
=
\mhLOS_i \, \ex{-2\ii\psi_{ij}} \ .
\end{equation}
We stress that, in the notation $\hLOS_{\pol ij}$, $\pol\in\{+,\times\}$, the indices $i$ and $j$ are not interchangeable, because the former indicates the lens whose LOS shear is being measured. The same construction may be performed for lens $j$, thereby leading to
\begin{equation}
\mhLOS_{+ji} + \ii\,\mhLOS_{\times ji}
=
\mhLOS_j \, \ex{-2\ii\psi_{ji}} \ .
\end{equation}
The rotation angles $\psi_{ij}$ and $\psi_{ji}$ are different in general.\footnote{%
The explicit expression of the relevant phase, in terms of the polar coordinates of $\uvect{u}_i$ and $\uvect{u}_j$, reads
\begin{equation}
\ex{-2\ii\psi_{ij}}
=
\frac
{%
\pac{\cos\theta_i\sin\theta_j\cos(\ph_j-\ph_i) - \sin\theta_i\cos\theta_j
    - \ii \sin\theta_j \sin(\ph_j-\ph_i)
    }^2
}
{%
    \pac{\cos\theta_i\sin\theta_j\cos(\ph_j-\ph_i) - \sin\theta_i\cos\theta_j}^2
    + \sin^2\theta_j \sin^2(\ph_j-\ph_i)
} \ ,
\end{equation}
which is manifestly not symmetric under the exchange of $i$ and $j$.
}
However, when $\thet_{ij}\ll 1$ -- a condition typically satisfied in practice -- the tangent spaces $\mathrm{T}_i\mathcal{S}^2, \mathrm{T}_j\mathcal{S}^2$ coincide and we have $\psi_{ij}\approx\psi_{ji}$.

\paragraph{Simple estimator for the LOS shear autocorrelation} Consider all pairs~$(ij)$ of lenses, and sort them according to their angular separation~$\thet_{ij}$. In practice, we introduce angular bins~$(\pairs_\alpha)$ with limits $(\thet_\alpha)$; $\pairs_\alpha$ is defined as the set of pairs of lenses $(ij)$ such that $\thet_\alpha\leq\thet_{ij}<\thet_{\alpha+1}$. Since the LOS shear is a complex (or spin-two) quantity, two complementary correlation functions may be measured in the $\alpha$\textsuperscript{th} angular bin, namely the plus and minus~$(\pm)$ correlations. Estimators for these may be defined as
\begin{equation}
\label{eq:LL_estimator_initial}
(\widehat{\lens\lens})^\pm_\alpha
\define
\frac{1}{|\pairs_{\alpha}|}
\sum_{(ij)\in\pairs_{\alpha}}
\pa{
\mhLOS_{+ij} \mhLOS_{+ji} \pm \mhLOS_{\times ij} \mhLOS_{\times ji}
} ,
\end{equation}
where $|\pairs_\alpha|$ is the number of terms in the sum, that is the number of pairs $(ij)$ in $\pairs_\alpha$. A more general definition could involve different weights for the lenses in the sum.

\paragraph{Alternative expression} For later convenience, we shall rewrite \cref{eq:LL_estimator_initial} as follows, so as to make it formally closer to $\lens\shape$, $\lens\position$ estimators that we define in the next paragraphs,
\begin{empheq}[box=\fbox]{align}
\label{eq:LL_estimator}
(\widehat{\lens\lens})^\pm_\alpha
&=
(\widehat{\lens^+\lens^+})_\alpha \pm (\widehat{\lens^\times\lens^\times})_\alpha
\\
(\widehat{\lens^\pol\lens^\pol})_\alpha
&=
\frac{1}{L^2}
\sum_{i=1}^L\sum_{j=1}^L W^\alpha_{ij} \,
\mhLOS_{\pol ij} \mhLOS_{\pol ji}
\qquad
\text{for } \pol\in\{+,\times\}.
\end{empheq}
In the above, we have explicitly split $(\widehat{\lens\lens})^\pm_\alpha$ into their two elementary polarisation terms. Furthermore, we have turned the sum over pairs $(ij)\in\mathcal{P}_\alpha$ into a sum over all lenses in the sample by introducing the weighted window function
\begin{equation}
\label{eq:weight_function_estimators_1}
W^\alpha_{ij}
\define
\frac{L^2}{|\mathcal{P}_\alpha|}
\times
\begin{cases}
1 &\text{if } \thet_\alpha \leq \thet_{ij} < \thet_{\alpha+1} ,\\
0 &\text{otherwise, or if $i=j$.}
\end{cases}
\end{equation}

The prefactor in the above expression may be simplified by estimating the number of pairs~$|\pairs_\alpha|$ in the $\alpha$\textsuperscript{th} bin. Let $\annulus_\alpha(\uvect{u})$ denote the annulus spanned by the points in $\mathcal{S}^2$ forming, with $\uvect{u}$, an angle between $\thet_\alpha$ and $\thet_{\alpha+1}$ (see \cref{fig:geometry_sphere}, right panel). If a lens is observed at $\uvect{u}$, then $\annulus_\alpha(\uvect{u})$ is the region where other lenses can make, with the former one, a pair in $\pairs_\alpha$. Let $\Omega_\alpha(\uvect{u})$ be the solid angle covered by $\annulus_\alpha(\uvect{u})\cap\footprint$. For a full-sky survey ($\footprint=\mathcal{S}^2$), $\Omega_\alpha$ does not depend on $\uvect{u}$, in which case $|\pairs_\alpha| = L\times n\e{L}\Omega_\alpha=L^2\Omega_\alpha/\Omega$. In a survey that does not cover the entire celestial sphere, the exact result for $|\pairs_\alpha|$ would depend on the survey geometry. Since most of the cosmological signal is carried by small angular separations ($\thet\ll 1$), we shall neglect such boundary effects in the remainder of this article -- this is the \emph{pseudo-full-sky approximation}. When substituting the expression for $|\pairs_\alpha|$ in \cref{eq:weight_function_estimators_1}, we obtain
\begin{empheq}[box=\fbox]{equation}
\label{eq:weight_function_estimators_2}
W^\alpha_{\mu\nu}
=
\frac{\Omega}{\Omega_\alpha}
\times
\begin{cases}
1 &\text{if } \thet_\alpha \leq \thet_{\mu\nu} < \thet_{\alpha+1} ,\\
0 &\text{otherwise, or if $\mu=\nu$.}
\end{cases}
\end{empheq}
In this new expression~\eqref{eq:weight_function_estimators_2} of the weight function, the number of lenses~$L$ has disappeared, which makes it independent of the nature of the objects involved in the pairs; this is why we used generic indices $\mu, \nu$ which may refer to either lenses or galaxies.

\subsubsection{Cross-correlation of LOS shear and galaxy ellipticity (LE)}

The estimator of the cross-correlation of the LOS shear with the apparent ellipticity of galaxies is almost identical to that of the LOS shear autocorrelation. The difference is that we should now consider pairs of lenses~$i$ and galaxies $a$ rather than pairs of lenses. Averaging over a large number of pairs is expected to extract the relative alignment of the LOS shear and standard shear by removing the contribution of both noise and galaxy intrinsic ellipticities.

Just like $\lLOS$, the apparent (complex) ellipticity~$\eps_a$ of galaxy~$a$ may be decomposed into plus and cross polarisations when forming a pair with lens $i$,
\begin{equation}
\eps_{+ai} + \ii\,\eps_{\times ai}
=
\eps_a \, \ex{-2\ii\psi_{ai}} \ ,
\end{equation}
where $\psi_{ai}$ is defined as in \cref{fig:geometry_sphere}. Following the same construction as in \cref{subsubsec:estimator_autocorrelation_LOS_shear}, we get the following estimators for the two LE correlation functions,
\begin{empheq}[box=\fbox]{align}
\label{eq:LE_estimator}
(\widehat{\lens\shape_A})^\pm_\alpha
&=
(\widehat{\lens^+\shape^+_A})_\alpha \pm (\widehat{\lens^\times\shape^\times_A})_\alpha
\\
(\widehat{\lens^\pol\shape^\pol_A})_\alpha
&=
\frac{1}{L G_A}
\sum_{i=1}^L\sum_{a\in\galaxies_A} W_{ia}^\alpha \,
\mhLOS_{\pol ia} \eps_{\pol ai} \ ,
\qquad
\text{for } \pol\in\{+,\times\}.
\end{empheq}
In the above, we have assumed that the galaxy sample is large enough to be divided into tomographic redshift bins; $(\widehat{\lens\shape_A})^\pm_\alpha$ thus estimates the correlations between the strong lensing sample and the $A$\textsuperscript{th} tomographic sample, comprising $G_A$ galaxies. 

\subsubsection{Cross-correlation of LOS shear and galaxy positions (LP)}

The cross-correlation of LOS shear and galaxy positions is analogous to galaxy--galaxy lensing in the usual $3\times 2$pt analysis. It is based on the fact that galaxies, as matter overdensities, produce tangential shear around them. The LP estimator consists in extracting that information by summing the $+$ polarisation of the LOS shear (the $\times$ polarisation vanishes on average) for all lens--galaxy pairs $(ia)$ within a certain angular bin $\alpha$,
\begin{empheq}[box=\fbox]{equation}
\label{eq:LP_estimator}
(\widehat{\lens^+\position_A})_\alpha
=
\frac{1}{L G_A}
\sum_{i=1}^L \sum_{a\in\galaxies_A} W_{ia}^\alpha \,
\mhLOS_{+ia} \ .
\end{empheq}
Just like in the LE case, we have assumed that the galaxy sample is large enough to be divided into tomographic redshift bins. Note that a tangential shear corresponds to $\hLOS_+<0$, so \emph{the expectation value of $\lens\position$ will be negative}.

\subsection{Expectation values of the estimators in $\Lambda$CDM}
\label{subsec:expectation_values_estimators}

Let us now compute the expectation value of the estimators defined in \cref{subsec:estimators_correlation_functions}, that is, the signal that we aim to detect. Before we perform the actual calculation, we must specify what \emph{expectation value} really means in the present context.

\subsubsection{Three distinct notions of average}
\label{subsubsec:averages}

In our current understanding of cosmology, randomness in the sky originates from the quantum fluctuations of the physical fields that eventually lead to cosmic structures via gravitational collapse. Technically speaking, this concept is implemented by treating the gravitational potential~$\Phi$ -- and all fields that derive from it, such as the density contrast~$\delta$ and the peculiar velocity field -- as Gaussian random fields.

In this context, an expectation value means an average over realisations of those fields, i.e. a theoretical average over realisations of the Universe. For the observables defined in \cref{subsec:estimators_correlation_functions}, jumping from one realisation of the Universe to another has two consequences: on the one hand, the density and shear fields fluctuate; on the other hand, the positions and properties of their tracers (galaxies and lenses) change too. In practice, it is convenient to explicitly distinguish these two effects as \emph{cosmic averaging} and \emph{sample averaging}.

\paragraph{Cosmic averaging $\cev{\ldots}$} will refer to averages over realisations~$\universe$ of the \emph{macroscopic} properties of the Universe, i.e. the underlying fields. It is the usual notion of average involved in, e.g., the definition of the matter power spectrum.  

\paragraph{Galaxy-sample averaging $\gsev{\ldots}$} will be the average over the properties of all the galaxies in the sample, given a realisation~$\universe$ of the macroscopic properties of the Universe (in particular the density field). We shall parameterise a galaxy with (i) its redshift~$z$; (ii) its angular position~$\uvect{u}$; and (iii) its intrinsic complex ellipticity $\eps_0$, which we gather into the quintuplet~$\vect{\Gamma}=(z, \uvect{u}, \eps_0)$. For an observable $X$ that depends on the sample of $G$ galaxies, the galaxy-sample average then reads
\begin{equation}
\label{eq:galaxy-sample_average}
\gsev{X}(\universe)
\define
\pac{
\prod_{A=1}^T
\prod_{a\in\galaxies_A}
\int\dd^5\vect{\Gamma}_a \; p_A(\vect{\Gamma}_a|\universe)
}
X(\vect{\Gamma}_1, \ldots, \vect{\Gamma}_{G}) \ ,
\end{equation}
where $p_A(\vect{\Gamma}|\universe)$ is the probability density function (PDF) of $\vect{\Gamma}$ in the $A$\textsuperscript{th} tomographic bin, given the cosmic realisation~$\universe$. 

Let us be more explicit and decompose $p_A(\vect{\Gamma}|\universe)$ into elementary components:
\begin{equation}
\label{eq:decomposition_p_A_Gamma}
p_A(\vect{\Gamma}|\universe) \, \dd^5\vect{\Gamma}
=
f_A(z) \, \dd z
\times
\pac{1+b(z)\,\delta(z, \uvect{u};\universe)}
\frac{\dd^2\uvect{u}}{\Omega} 
\times
p(\eps_0; z) \, \dd^2\eps_0
\ .
\end{equation}
In \cref{eq:decomposition_p_A_Gamma}, $f_A(z)$ denotes the normalised redshift selection function of the survey in the $A$\textsuperscript{th} bin, which we assume to coincide with the \emph{observed} redshift distribution in that bin. For a given line of sight~$\uvect{u}$, this is modulated by the actual overdensity in galaxy number counts, which we assume to be linearly biased by $b(z)$ with respect to the total matter density contrast~$\delta$. Finally, $p(\eps_0; z)$ denotes the PDF of the intrinsic ellipticity of galaxies at $z$, which we assume to be independent of $\uvect{u}$; we shall also neglect intrinsic alignments in this work.

\paragraph{Lens-sample averaging $\lsev{\ldots}$} will be the average over the relevant properties of strong lenses in the sample given a macroscopic realisation~$\universe$. We shall parameterise a lens with (i) a pair of redshifts~$\vect{Z}=(z\e{d}, z\e{s})$, which denote respectively the redshift of the main deflector and source of the system; (ii) its position~$\uvect{u}$ in the sky; and (iii) the  error~$\nlLOS$ on the measurement of the LOS shear in that system, as introduced in \cref{eq:LOS_shear_noise_def}. We gather the aforementioned lens parameters into a sextuplet~$\vect{\Lambda}=(\vect{Z}, \uvect{u}, \nlLOS)$. For an observable $X$ that depends on a sample of $L$ lenses, the lens-sample average then reads
\begin{equation}
\lsev{X}(\universe)
\define
\pac{
\prod_{i=1}^{L}
\int\dd^6\vect{\Lambda}_i \; p(\vect{\Lambda}_i|\universe)
}
X(\vect{\Lambda}_1, \ldots, \vect{\Lambda}_{L}) \ ,
\end{equation}
where $p(\vect{\Lambda}|\universe)$ is the PDF of $\vect{\Lambda}$ in the cosmic realisation~$\universe$.
In the following, we shall neglect lens clustering in the expression of $p(\vect{\Lambda}|\universe)$, which implies that the angular positions of lenses are simply Poisson-distributed. We assume, besides, that the noise~$\nlLOS$ on LOS shear measurements is independent of the lens's position in the sky. Therefore,
\begin{equation}
p(\vect{\Lambda}) \, \dd^6\vect{\Lambda}
= p(\vect{Z}) \, \dd^2\vect{Z} \times \frac{\dd^2\uvect{u}}{\Omega} \times p(\nlLOS; \vect{Z}) \, \dd^2\nlLOS \ ,
\end{equation}
where the dependence on the cosmic macroscopic realisation~$\universe$ has dropped. As a consequence, lens-sample averaging and cosmic averaging commute in this framework.

This approximation is \modified{motivated by the fact that} none of the observables that we are interested in rely on lens clustering. In contrast, the LP cross-correlation directly depends on the fact that galaxies are inhomogeneously distributed in the sky. Here, neglecting lens clustering is analogous to neglecting source--lens clustering in cosmic shear, which leads to a higher-order bias proportional to the matter bispectrum~\cite{1998A&A...338..375B, 2002MNRAS.330..365H, 2009ApJ...702..593S, Yu:2014iea}.\footnote{PF thanks Calum Murray for bringing this up during the \emph{Rencontres de Moriond} in 2024.}

The impact of source--lens clustering on the \Euclid cosmic-shear power spectrum was comprehensively investigated in ref.~\cite{Euclid:2023wdq}, along with other higher-order corrections such as magnification bias and reduced shear~\cite{Deshpande:2019sdl}. The resulting bias was shown to be at most comparable to cosmic variance. \modified{Since a large fraction of strong lenses are massive ellipticals, which tend to be more clustered than the early-type \Euclid sources, lens clustering may have a stronger effect on the $\lens\lens$, $\lens\shape$, and $\lens\position$ correlations than on the usual $3\times 2$pt observables. However, since this effect is proportional to the galaxy bias, it cannot exceed the findings of ref.~\cite{Euclid:2023wdq} by more than a factor two or three~\cite{2005ApJ...630....1Z}. As we shall see in \cref{sec:forecasted_detectability}, the uncertainty budget of $\lens\lens$, $\lens\shape$, and $\lens\position$ is clearly dominated by noise and sparsity, so that higher-order biases should be negligible. This assumption should be confirmed, eventually, by a careful analysis of the selection function underlying stage-IV lens samples.}

\subsubsection{The LL signal}
\label{subsubsec:LL_signal}

Following the above, the expectation value of the estimator~\eqref{eq:LL_estimator} of the autocorrelation function of LOS shear reads
\begin{equation}
(\lens\lens)^\pm_\alpha
\define \cev{\lsev{(\widehat{\lens\lens})^\pm_\alpha}} 
.
\end{equation}
The averaging operator~$\gsev{\ldots}$ does not appear because the galaxy sample is not involved in $(\widehat{\lens\lens})^\pm_\alpha$. Since the latter is a combination of the $(++)$ and $(\times\times)$ correlation functions, we may work with those and start by evaluating their lens-sample average:
\begin{align}
\lsev{(\widehat{\lens^\pol\lens^\pol})_\alpha}(\universe)
&=
\pac{
\prod_{i'=1}^{L}
\int\dd^6\vect{\Lambda}_{i'} \; p(\vect{\Lambda}_{i'})
}
\frac{1}{L^2}
\sum_{i=1}^L\sum_{j=1}^L W^\alpha_{ij} \,
\mhLOS_{\pol ij} \mhLOS_{\pol ji}
\\
\label{eq:calculation_lsev_LpLp_2}
&=
\int \dd^6\vect{\Lambda}_1 \; p(\vect{\Lambda}_1)
\int \dd^6\vect{\Lambda}_2 \; p(\vect{\Lambda}_2) \,
W^\alpha_{12}\, \mhLOS_{\pol 12} \mhLOS_{\pol 21}
\\
\label{eq:calculation_lsev_LpLp_3}
&=
\int \dd^2\vect{Z}_1 \;  p(\vect{Z}_1)
\int \dd^2\vect{Z}_2 \;  p(\vect{Z}_2)
\int_\footprint\frac{\dd^2\uvect{u}_1}{\Omega}
\int_\footprint\frac{\dd^2\uvect{u}_2}{\Omega} \; W^\alpha_{12} \,
\hLOS_{\pol 12} \hLOS_{\pol 21} \ .
\end{align}
To obtain \cref{eq:calculation_lsev_LpLp_2}, we have noticed that all $L^2$ terms in the double sum over $i$ and $j$ were identical due to the integration over $\vect{\Lambda}_i, \vect{\Lambda}_j$, which can be freely renamed $\vect{\Lambda}_1, \vect{\Lambda}_2$. This is possible because $W^\alpha_{ij}$ forbids the case $i=j$, which would lead to different expressions. In \cref{eq:calculation_lsev_LpLp_3}, we integrated over the noise term~$\nlLOS$, which does not contribute if we assume that the noise of different lenses is uncorrelated and averages to zero:
\begin{equation}
\int \dd^2 \nlLOS \; p(\nlLOS; \vect{Z}) \, \nlLOS = 0 \ .
\end{equation}

We then apply the cosmic-averaging operator~$\cev{\ldots}$. Since the only $\universe$-dependence in the above lies in the LOS shear field~$\lLOS$, we may focus on
\begin{equation}
\label{eq:cev_LpLp}
\cev{\hLOS_{\pol 12} \hLOS_{\pol 21}}
=
\begin{cases}
\cev{\Re\pac{\lLOS(\uvect{u}_1)\,\ex{-2\ii\psi_{12}}} \Re\pac{\lLOS(\uvect{u}_2)\,\ex{-2\ii\psi_{21}}}}
& \text{if } \pol =+,
\\[2mm]
\cev{\Im\pac{\lLOS(\uvect{u}_1)\,\ex{-2\ii\psi_{12}}} \Im\pac{\lLOS(\uvect{u}_2)\,\ex{-2\ii\psi_{21}}}}
& \text{if } \pol =\times.
\end{cases}
\end{equation}
Now, because of the principles of statistical homogeneity and isotropy of our Universe, these expectation values only depend on the angle~$\thet_{12}$ formed by $\uvect{u}_1, \uvect{u}_2$. Besides, we shall work in the pseudo-full-sky approximation and neglect the boundary effects due to $\footprint\neq\mathcal{S}^2$. As a result, all values of $\uvect{u}_1$ in \cref{eq:cev_LpLp} are equivalent and we may choose to place it at the North pole~$\uvect{u}_1=\uvect{z}$. This allows us to remove the integral over $\uvect{u}_1$ in \cref{eq:calculation_lsev_LpLp_3}, while implying $\psi_{12}=\psi_{21}=0$ (see \cref{fig:geometry_sphere}). Putting everything together, we finally get
\begin{empheq}[box=\fbox]{equation}
\label{eq:ev_LL}
(\lens\lens)^\pm_\alpha
=
\frac{2\pi}{\Omega_\alpha} \int_{\thet_\alpha}^{\thet_{\alpha+1}}
\dd\thet \; \sin\thet
\int\dd^2\vect{Z}_1 \; p(\vect{Z}_1)
\int\dd^2\vect{Z}_2 \; p(\vect{Z}_2) \;
(\lens\lens)^\pm(\thet; \vect{Z}_1, \vect{Z}_2) \ ,
\end{empheq}
with
\begin{empheq}[box=\fbox]{align}
\label{eq:LL+_def}
(\lens\lens)^+(\thet; \vect{Z}_1, \vect{Z}_2)
&\define
\cev{\lLOS(\uvect{z}; \vect{Z}_1) \, \lLOS^*(\uvect{u}_\thet; \vect{Z}_2)} ,
\\
\label{eq:LL-_def}
(\lens\lens)^-(\thet; \vect{Z}_1, \vect{Z}_2)
&\define
\cev{\lLOS(\uvect{z}; \vect{Z}_1) \, \lLOS(\uvect{u}_\thet;\vect{Z}_2)} , 
\end{empheq} 
where $\uvect{u}_\thet=(\thet, 0)$ is the unit vector in the $(xz)$-plane forming an angle $\thet$ with the North pole. In the expression of $\lens\lens^\pm(\thet; \vect{Z}_1, \vect{Z}_2)$, we used the statistical invariance under parity, which implies that $\cev[2]{\hLOS_+\hLOS_\times}=0$, in order to get rid of the real- and imaginary-part operators.

\subsubsection{The LE signal}
\label{subsubsec:LE_signal}

In the case of the cross-correlation between the LOS shear and the apparent shapes of galaxies, since both the lens and galaxy samples are involved, the expectation value is obtained by applying all three averaging operators defined in \cref{subsubsec:averages},
\begin{equation}
(\lens\shape_A)^\pm_\alpha
\define \cev{\lsev{\gsev{(\widehat{\lens\shape_A})^\pm_\alpha}}}
.
\end{equation}
The rest of the calculation is quite similar to the LL case. Taking the lens-sample and galaxy-sample averages in a cosmic realisation~$\universe$, we find
\begin{align}
&\lsev[3]{\gsev[3]{(\widehat{\lens^\pol\shape_A^\pol})_\alpha}}(\universe)
\nonumber\\
&=
\pac{
\prod_{i'=1}^{L}
\int\dd^6\vect{\Lambda}_{i'} \; p(\vect{\Lambda}_{i'})
}
\pac{
\prod_{A'=1}^T
\prod_{a'\in\galaxies_{A'}}
\int\dd^5\vect{\Gamma}_{a'} \; p_{A'}(\vect{\Gamma}_{a'})
}
\frac{1}{L G_A}
\sum_{i=1}^L\sum_{a\in\galaxies_A} W^\alpha_{ia} \,
\mhLOS_{\pol ia} \eps_{\pol ai}
\\
\label{eq:calculation_sev_LpEp_2}
&=
\int \dd^6\vect{\Lambda}_1 \; p(\vect{\Lambda}_1) 
\int \dd^5\vect{\Gamma}_2 \; p_A(\vect{\Gamma}_2) \,
W^\alpha_{12} \mhLOS_{\pol 12} \eps_{\pol 21}
\\
\label{eq:calculation_sev_LpEp_3}
&=
\int \dd^2\vect{Z}_1 \;  p(\vect{Z}_1)
\int \dd z_2 \;  f_A(z_2)
\int_\footprint\frac{\dd^2\uvect{u}_1}{\Omega}
\int_\footprint\frac{\dd^2\uvect{u}_2}{\Omega} \; W^\alpha_{12}
\underbrace{
\pac{1+b(z_2) \delta(z_2, \uvect{u}_2)}
}_{\approx 1}
\hLOS_{\pol 12} \gamma_{\pol 21} \ .
\end{align}
When going from \cref{eq:calculation_sev_LpEp_2} to \cref{eq:calculation_sev_LpEp_3}, we substituted the expressions~\eqref{eq:LOS_shear_noise_def} and \eqref{eq:apparent_ellipticity} of $\eps$ and $\mlLOS$, and we used the assumption that $\eps_0$ and $\nlLOS$ average to zero and are uncorrelated. Furthermore, the galaxy overdensity~$b_A\delta$ can be neglected here because $\gamma$ and $\lLOS$ are already first-order terms in cosmological perturbations.

Averaging \cref{eq:calculation_sev_LpEp_3} over cosmic realisations, exploiting statistical homogeneity and isotropy just like the LL case, finally yields
\begin{empheq}[box=\fbox]{equation}
\label{eq:ev_LE}
(\lens\shape_A)^\pm_\alpha
=
\frac{2\pi}{\Omega_\alpha}
\int_{\thet_\alpha}^{\thet_{\alpha+1}} \dd\thet \; \sin\thet
\int\dd^2\vect{Z}_1 \; p(\vect{Z}_1)
\int\dd z_2 \; f_A(z_2) \,
(\lens\shape)^\pm(\thet; \vect{Z}_1, z_2) \ ,
\end{empheq}
with
\begin{empheq}[box=\fbox]{align}
\label{eq:LE+_def}
(\lens\shape)^+(\thet; \vect{Z}_1, z_2)
&=
\cev{\lLOS(\uvect{z}; \vect{Z}_1) \, \gamma^*(\uvect{u}_\thet; z_2)} ,
\\
\label{eq:LE-_def}
(\lens\shape)^-(\thet; \vect{Z}_1, z_2)
&\define
\cev{\lLOS(\uvect{z}; \vect{Z}_1) \, \gamma(\uvect{u}_\thet; z_2)} .
\end{empheq}

\subsubsection{The LP signal}
\label{subsubsec:LP_signal}

The expectation value of the cross-correlation between LOS shear and galaxy positions reads
\begin{equation}
(\lens^+\position_A)_\alpha
\define \cev{\lsev{\gsev{(\widehat{\lens^+\position_A})^\pm_\alpha}}}.
\end{equation}
Applying the sample-averaging operators yields
\begin{align}
&\lsev[3]{\gsev[3]{(\widehat{\lens^+\position_A})_\alpha}}(\universe)
\nonumber\\
&=
\pac{
\prod_{i'=1}^{L}
\int\dd^6\vect{\Lambda}_{i'} \; p(\vect{\Lambda}_{i'})
}
\pac{
\prod_{A'=1}^{T}
\prod_{a'=1}^{G_{A'}}
\int\dd^5\vect{\Gamma}_{a'} \; p_{A'}(\vect{\Gamma}_{a'})
}
\frac{1}{L G_A}
\sum_{i=1}^L\sum_{a=1}^{G_A} W^\alpha_{ia} \,
\mhLOS_{+ia}
\\
\label{eq:calculation_sev_L+P_2}
&=
\int \dd^6\vect{\Lambda}_1 \; p(\vect{\Lambda}_1) 
\int \dd^5\vect{\Gamma}_2 \; p_A(\vect{\Gamma}_2) \,
W^\alpha_{12} \mhLOS_{+12}
\\
\label{eq:calculation_sev_L+P_3}
&=
\int \dd^2\vect{Z}_1 \;  p(\vect{Z}_1)
\int \dd z_2 \;  f_A(z_2)
\int_\footprint\frac{\dd^2\uvect{u}_1}{\Omega}
\int_\footprint\frac{\dd^2\uvect{u}_2}{\Omega} \; W^\alpha_{12}
\pac{1+b(z_2) \delta(z_2, \uvect{u}_2)}
\hLOS_{+12} \ ,
\end{align}
where, contrary to the LE case, we must keep the $b\delta$ term. Indeed, when applying the cosmic-average operator, the first term $\cev[2]{[1+\ldots]\hLOS_{+12}}$ in the integral vanishes due to statistical isotropy, and we are left with
\begin{equation}
(\lens^+\position_A)_\alpha
=
\int \dd^2\vect{Z}_1 \;  p(\vect{Z}_1)
\int \dd z_2 \;  f_A(z_2) \, b(z_2)
\int_\footprint\frac{\dd^2\uvect{u}_1}{\Omega}
\int_\footprint\frac{\dd^2\uvect{u}_2}{\Omega} \; W^\alpha_{12}
\cev{\hLOS_{+12} \delta(z_2, \uvect{u}_2)}.
\end{equation}
Exploiting again statistical homogeneity and isotropy,
we obtain the final result
\begin{empheq}[box=\fbox]{equation}
\label{eq:ev_LP}
(\lens^+\position_A)_\alpha
=
\frac{2\pi}{\Omega_\alpha} \int_{\thet_\alpha}^{\thet_{\alpha+1}}
\dd\thet \; \sin\thet
\int\dd^2\vect{Z}_1 \; p(\vect{Z}_1)
\int\dd z_2 \; f_A(z_2) \,
\lens^+\position(\thet; \vect{Z}_1, z_2) \ ,
\end{empheq}
with
\begin{empheq}[box=\fbox]{equation}
\label{eq:LP_def}
\lens^+\position(\thet; \vect{Z}_1, z_2)
\define
\cev{\lLOS(\uvect{z}; \vect{Z}_1) \,  b(z_2)\delta(\uvect{u}_\thet, z_2)} .
\end{empheq}

\subsection{Angular power spectra}
\label{subsec:power_spectra_estimators}

In modern cosmology, the properties of the matter distribution are commonly expressed in Fourier space, with the matter power spectrum, rather than in real space with the two-point correlation function. In that context, it is useful to express the angular power spectra associated with the correlation functions $(\lens\lens)^\pm(\thet; \ldots)$, $(\lens\shape)^\pm(\thet; \ldots)$ and $\lens^+\position(\thet; \ldots)$ in order to make cosmological predictions.

\subsubsection{Spherical-harmonic decomposition}

The expectation values of the three correlation functions $\lens\lens$, $\lens\shape$, and $\lens\position$ considered here involve three distinct fields on the two-sphere: $\delta(\uvect{u}, z)$, which is a scalar (or spin-zero) field; and $\gamma(\uvect{u}; z), \lLOS(\uvect{u}; \vect{Z})$ which are spin-two fields. The former may be decomposed over the standard spherical-harmonic basis~$Y_{\ell m}(\uvect{u})$ according to
\begin{equation}
\delta(\uvect{u}, z)
=
\sum_{\ell=0}^\infty \sum_{m=-\ell}^\ell
\delta_{\ell m}(z) \,
Y_{\ell m}(\uvect{u}) \ ,
\qquad
\delta_{\ell m}(z)
=
\int_{\mathcal{S}^2} \dd^2\uvect{u} \;
Y_{\ell m}^*(\uvect{u}) \, \delta(\uvect{u}, z) \ ,
\end{equation}
and similarly for any other scalar field, such as the lensing potentials~$\Psi$ and $\Psi\e{LOS}$. As for the shear fields, we must use spin-weighted spherical harmonics\footnote{%
Spin-weighted spherical harmonics are obtained by successive eth differentiation of the usual spherical harmonics according to
\begin{equation}
\spinharmonic{s}{\ell m}(\uvect{u})
=
\sqrt{\frac{(\ell-s)!}{(\ell+s)}} \, \eth^s Y_{\ell m}(\uvect{u}) \ .
\end{equation}
}
\begin{equation}
\gamma(\uvect{u}; z)
=
\sum_{\ell=0}^\infty \sum_{m=-\ell}^\ell
\gamma_{\ell m}(z) \,
\spinharmonic{2}{\ell m}(\uvect{u}) \ ,
\qquad
\gamma_{\ell m}(z)
=
\int_{\mathcal{S}^2} \dd^2\uvect{u} \;
\spinharmonic{2}{\ell m}^*(\uvect{u}) \, \gamma(\uvect{u}; z) \ ;
\end{equation}
and similarly for $\lLOS(\uvect{u}; \vect{Z})$, thereby defining harmonic coefficients~$\hLOS_{\ell m}(\vect{Z})$.

Because $\gamma$ derives from the scalar lensing potential~$\Psi$ via \cref{eq:shear_potential}, their harmonic coefficients are related as
\begin{equation}
\label{eq:harmonic_coefficients_gamma_Psi}
\gamma_{\ell m}(z)
=
\frac{1}{2} \sqrt{\frac{(\ell+2)!}{(\ell-2)!}} \,
\Psi_{\ell m}(z) \ ,
\end{equation}
and similarly for the LOS shear and LOS potential.

We can then define the angular power spectrum for any two fields~$g, f$ (which may be identical) in our problem, regardless of their spin, as
\begin{empheq}[box=]{equation}
\cev{f_{\ell m} \, g^*_{\ell' m'}}
=
\Delta_{\ell\ell'} \Delta_{m m'} C^{fg}_\ell \ .
%
\end{empheq}
The Kr\"onecker symbols\footnote{We use a capital $\Delta$ to avoid confusion with the density contrast $\delta$.}~$\Delta_{\ell\ell'}, \Delta_{m m'}$ in the above are due to statistical isotropy, i.e. the fact that, in real space, the two-point correlation of $f, g$ only depends on the angle formed by the directions in which the fields are evaluated.

\subsubsection{Relation between correlation functions and power spectra}


\paragraph{Exact relations} Substituting the harmonic decomposition of $\delta$, $\gamma$ and $\lLOS$ in \cref{eq:LL+_def,eq:LL-_def,eq:LE+_def,eq:LE-_def,eq:LP_def} for the LL, LE and LP correlation functions, and exploiting the properties of the spin-weighted spherical harmonics, we find
\begin{align}
\label{eq:LL_C}
(\lens\lens)^\pm(\thet; \vect{Z}_1, \vect{Z}_2)
&=
\frac{1}{4\pi}
\sum_{\ell=2}^\infty (2\ell+1) d^\ell_{\pm 2 2}(\thet) \,
C^{\lens\lens}_\ell(\vect{Z}_1, \vect{Z}_2) \ ,
\\
\label{eq:LE_C}
(\lens\shape)^\pm(\thet; \vect{Z}_1, z_2)
&=
\frac{1}{4\pi}
\sum_{\ell=2}^\infty (2\ell+1) d^\ell_{\pm 2 2}(\thet) \,
C^{\lens\shape}_\ell(\vect{Z}_1, z_2) \ ,
\\
\label{eq:LP_C}
\lens^+\position(\thet; \vect{Z}_1, z_2)
&=
\frac{1}{4\pi}
\sum_{\ell=2}^\infty (2\ell+1) d^\ell_{20}(\thet) \,
C^{\lens\position}_\ell(\vect{Z}_1, z_2) \ ,
\end{align}
where $d^\ell_{s s'}(\thet)$ denote elements of the Wigner-$d$ matrices~\cite{1988qtam.book.....V}, and with the power spectra
\begin{align}
\label{eq:C_ell_LL_def}
C^{\lens\lens}_\ell(\vect{Z}_1, \vect{Z}_2)
&\define
\cev{\hLOS_{\ell m}(\vect{Z}_1) \, \pac{\hLOS_{\ell m}(\vect{Z}_2)}^*},
\\
\label{eq:C_ell_LE_def}
C^{\lens\shape}_\ell(\vect{Z}_1, z_2)
&\define
\cev{\hLOS_{\ell m}(\vect{Z}_1) \, \gamma_{\ell m}^*(z_2)} ,
\\
\label{eq:C_ell_LP_def}
C^{\lens\position}_\ell(\vect{Z}_1, z_2)
&\define
\cev{\hLOS_{\ell m}(\vect{Z}_1) \, b(z_2)\delta_{\ell m}^*(z_2)} .
\end{align}

\paragraph{Small-angle approximation} For $\thet\ll 1$, \cref{eq:LL_C,eq:LE_C,eq:LP_C} are dominated by the terms with $\ell\gg 1$. In such conditions, the Wigner-$d$ matrix elements may be approximated by Bessel functions~$J_n$ as~\cite{2018JMP....59b2102H}
\begin{equation}
d^\ell_{s s'}(\thet)
\approx
(-1)^{s-s'} J_{s-s'}(\ell\thet)
\qquad
\text{for} \quad \thet\ll 1 \quad s, s'\ll \ell ;
\end{equation}
in addition, the sums over $\ell$ may be turned into integrals to finally express the LL, LE, LP correlation functions as Hankel transformations of their power spectra:
\begin{empheq}[box=\fbox]{align}
\label{eq:correlations_as_Hankel}
(\lens\lens)^\pm(\thet; \vect{Z}_1, \vect{Z}_2)
&\approx
\int_0^\infty \frac{\ell \dd\ell}{2\pi} \, J_{0/4}(\ell\thet) \,
C^{\lens\lens}_\ell(\vect{Z}_1, \vect{Z}_2) \ ,
\\
(\lens\shape)^\pm(\thet; \vect{Z}_1, z_2)
&\approx
\int_0^\infty \frac{\ell \dd\ell}{2\pi} \, J_{0/4}(\ell\thet) \,
C^{\lens\shape}_\ell(\vect{Z}_1, z_2) \ ,
\\
(\lens^+\position)(\thet; \vect{Z}_1, z_2)
&\approx
\int_0^\infty \frac{\ell \dd\ell}{2\pi} \, J_{2}(\ell\thet) \,
C^{\lens\position}_\ell(\vect{Z}_1, z_2) \ .
\end{empheq}

\subsubsection{Expression of the LL, LE and LP power spectra}

The observables~$\delta, \gamma, \lLOS$ considered in the LL, LE and LP correlations are all linearly related to the gravitational potential~$\Phi$. Their angular power spectra can thus be expressed in terms of the power spectrum of $\Phi$, and hence of the matter power spectrum. In order to derive such expressions, the calculations are similar to the $3\times2$pt literature -- see, e.g., ref.~\cite{2017MNRAS.472.2126K} for the case of cosmic shear.

\paragraph{Harmonic coefficients} The first step consists in expressing $\delta_{\ell m}, \gamma_{\ell m}, \hLOS_{\ell m}$ in terms of the three-dimensional Fourier transform\footnote{%
We use the following convention for three-dimensional Fourier transforms,
\begin{equation}
\tilde{f}(\vect{k})
=
\int \dd^3\vect{x} \; \ex{-\ii \vect{k}\cdot\vect{x}} \, f(\vect{x}) \ ,
\qquad
f(\vect{x})
=
\int \frac{\dd^3\vect{x}}{(2\pi)^3} \;
\ex{\ii \vect{k}\cdot\vect{x}} \, \tilde{f}(\vect{k}) \ .
\end{equation}
} of the gravitational potential~$\tilde{\Phi}(\eta, \vect{k})$, or of the density contrast~$\tilde{\delta}(\eta, \vect{k})$. For that purpose, it is convenient to use \cref{eq:harmonic_coefficients_gamma_Psi} so as to work with spin-zero quantities only. In the definition of $\delta_{\ell m}, \Psi_{\ell m}, \Psi\h{LOS}_{\ell m}$, we introduce the Fourier transforms of $\delta, \Phi$, and substitute the Fourier--Bessel identity
\begin{equation}
\ex{\ii \vect{k}\cdot\vect{x}}
=
4\pi \sum_{\ell=0}^\infty \sum_{m=-\ell}^\ell
\ii^\ell \, j_\ell(kx) \,
Y_{\ell m}(\uvect{x}) \, Y^*_{\ell m}(\uvect{k}) \ ,
\end{equation}
where $j_\ell(x)\define \sqrt{\pi/(2x)} J_{\ell+1/2}(x)$ denotes the $\ell$\textsuperscript{th} spherical Bessel function. Using the orthogonality and normalisation of spherical harmonics,\footnote{%
We adopt the following convention for spherical harmonics,
\begin{equation}
\int \dd^2\uvect{u} \; Y^*_{\ell m}(\uvect{u}) \, Y_{\ell' m'}(\uvect{u})
=
\delta_{\ell \ell'} \delta_{m m'} \ .
\end{equation}
} we then obtain
\begin{align}
\delta_{\ell m}(z)
&=
4\pi \ii^\ell \int\frac{\dd^3\vect{k}}{(2\pi)^3} \;
j_\ell(k\chi) \, Y^*_{\ell m}(\uvect{k}) \, \tilde{\delta}(z, \vect{k})
\ ,
\\
\label{eq:Psi_ell_m}
\Psi_{\ell m}(z)
&=
8\pi \ii^\ell
\int_0^\infty \frac{\dd\chi}{\chi} \; K(\chi;z)
\int\frac{\dd^3\vect{k}}{(2\pi)^3} \; j_\ell(k\chi) \,
Y^*_{\ell m}(\uvect{k}) \, \tilde{\Phi}(\eta_0 - \chi, \vect{k})
\ ,
\\
\label{eq:Psi_LOS_ell_m}
\Psi\h{LOS}_{\ell m}(\vect{Z})
&=
8\pi \ii^\ell
\int_0^\infty \frac{\dd\chi}{\chi} \; K\e{LOS}(\chi; \vect{Z})
\int\frac{\dd^3\vect{k}}{(2\pi)^3} \; j_\ell(k\chi) \,
Y^*_{\ell m}(\uvect{k}) \, \tilde{\Phi}(\eta_0-\chi, \vect{k}) \ ,
\end{align}
with the lensing kernels~$K, K\e{LOS}$ introduced in \cref{eq:lensing_kernel,eq:lensing_kernel_LOS} and shorthand notation $K(\chi; z)\define K[\chi; 0, \chi(z)]$ and $K\e{LOS}(\chi; \vect{Z})\define K\e{LOS}[\chi; \chi(z\e{d}), \chi(z\e{s})]$. The Fourier transforms of $\Phi$ and $\delta$ are related by the Poisson equation~\eqref{eq:Poisson} in Fourier space,
\begin{equation}
\label{eq:Poisson_Fourier}
- k^2 \tilde{\Phi}(\eta, \vect{k})
= 4\pi G \bar{\rho}_0 [1+z(\eta)] \, \tilde{\delta}(\eta, \vect{k}) 
= \frac{3}{2} H_0^2 \Omega\e{m} \, [1+z(\eta)] \, \tilde{\delta}(\eta, \vect{k})  \ .
\end{equation}

\paragraph{Exact expressions of the power spectra} Having expressed the harmonic coefficients $\delta_{\ell m}(z), \Psi_{\ell m}(z), \Psi\h{LOS}_{\ell m}(\vect{Z})$, it is straightforward to obtain the following angular power spectra
\begin{align}
\label{eq:Psi_LOS_Psi_LOS_lm_exact}
\cev{\Psi\h{LOS}_{\ell m}(\vect{Z}_1) \pac{\Psi\h{LOS}_{\ell m}(\vect{Z}_2)}^*}
&=
\frac{8}{\pi}
    \int_0^\infty \frac{\dd\chi}{\chi} \; K\e{LOS}(\chi; \vect{Z}_1)
    \int_0^\infty \frac{\dd\chi'}{\chi'} \; K\e{LOS}(\chi'; \vect{Z}_2)
    \nonumber\\ &\quad \times
    \int_0^\infty \dd k \; k^2 j_\ell(k\chi) \, j_\ell(k\chi') \,
    P_\Phi(\eta_0-\chi, \eta_0-\chi', k) \ ,
\\
\label{eq:Psi_LOS_Psi_lm_exact}
\cev{\Psi\h{LOS}_{\ell m}(\vect{Z}_1) \Psi_{\ell m}^*(z_2)}
&=
\frac{8}{\pi}
    \int_0^\infty \frac{\dd\chi}{\chi} \; K\e{LOS}(\chi; \vect{Z}_1)
    \int_0^\infty \frac{\dd\chi'}{\chi'} \; K(\chi'; z_2)
    \nonumber\\ &\quad \times
    \int_0^\infty \dd k \; k^2 j_\ell(k\chi) \, j_\ell(k\chi') \,
    P_\Phi(\eta_0-\chi, \eta_0-\chi', k) \ ,
\\
\label{eq:Psi_LOS_delta_lm_exact}
\cev{\Psi\h{LOS}_{\ell m}(\vect{Z}_1) \delta^*_{\ell m}(z_2)}
&=
\frac{4}{\pi}
    \int_0^\infty \frac{\dd \chi}{\chi} \; K\e{LOS}(\chi; \vect{Z}_1)
    \nonumber\\ &\quad \times
    \int \dd k \; k^2 \,
    j_{\ell}(k \chi) \, j_{\ell}(k \chi_2) \,
    \frac{k^2 P_\Phi(\eta_0-\chi, \eta_0-\chi_2, k)}{-4\pi G \bar{\rho}_0 (1+z_2)} \ ,
\end{align}
with $\chi_2\define \chi(z_2)$. In the above, $P_\Phi(\eta, \eta', k)$ denotes the power spectrum of the gravitational potential, defined as
\begin{equation}
\cev{\tilde{\Phi}(\eta, \vect{k}) \, \tilde{\Phi}^*(\eta', \vect{k}')}
=
(2\pi)^3 \Dirac(\vect{k}-\vect{k}') \, P_\Phi(\eta, \eta', k) \ ,
\end{equation}
where the Dirac delta is due to statistical homogeneity, and $P_\Phi$ only depends on the magnitude $k=|\vect{k}|$ of $k$ due to statistical isotropy. The global factor of two between \cref{eq:Psi_LOS_Psi_LOS_lm_exact}, \eqref{eq:Psi_LOS_Psi_lm_exact}, and \cref{eq:Psi_LOS_delta_lm_exact} comes from the fact that $\Psi$ is \emph{twice} the line-of-sight projected potential.

The actual LL, LE, and LP power spectra defined in \cref{eq:C_ell_LL_def,eq:C_ell_LE_def,eq:C_ell_LP_def} are then obtained by multiplying \cref{eq:Psi_LOS_Psi_LOS_lm_exact,eq:Psi_LOS_Psi_lm_exact} by the relevant prefactors following \cref{eq:harmonic_coefficients_gamma_Psi}, and averaging over redshifts. For small angular scales ($\ell \gg 1$), we may further simplify these expressions using Limber's approximation.

\paragraph{Power spectra in Limber's approximation} For $\ell \gg 1$, Bessel functions~$J_n(x)$ are very small for $x<n$, they are at maximum at $x=n$, and then experience slowly decaying but rapid oscillations for $x>n$. When integrated with a slowly varying function, it implies that $J_n$ can be approximated as
\begin{equation}
J_n(x) \approx \Dirac(x-n) \ .
\end{equation}
In all the subsequent results, we shall assume the validity of Limber's approximation.

When substituted in \cref{eq:Psi_LOS_Psi_LOS_lm_exact,eq:Psi_LOS_Psi_lm_exact}, Limber's approximation reduces the number of integrals by two, with $\chi=\chi'$ and $k\chi=\ell+1/2$. The final angular power spectrum of any couple of observables $\observable_1, \observable_2\in\{\lens, \shape, \position\}$ then takes the following compact form,
\begin{empheq}[box=\fbox]{equation}
\label{eq:C_ell_final}
C_\ell^{\observable_1\observable_2}
=
\int_0^\infty \dd\chi \; Q_{\observable_1}(\chi) \, Q_{\observable_2}(\chi) \,
P\e{m}\pac{\eta_0-\chi, \frac{\ell+1/2}{\chi}} ,
\end{empheq}
where $P\e{m}(\eta, k)=[(3/2)\Omega\e{m} H_0^2 (1+z)]^2 P_\Phi(\eta, \eta, k)$ is the matter power spectrum, and the integration kernels $Q_{\observable}$ read
\begin{empheq}[box=\fbox]{align}
\label{eq:Q_L}
Q_\lens(\chi; \vect{Z})
&\define
-\frac{3}{2} \, \Omega\e{m} H_0^2 \, [1+z(\chi)] \, K\e{LOS}(\chi; \vect{Z}) \ ,
\\
\label{eq:Q_E}
Q_{\shape}(\chi; z_*)
&=
-\frac{3}{2} \, \Omega\e{m} H_0^2 \, [1+z(\chi)] \, K(\chi; z_*) \ ,
\\
\label{eq:Q_P}
Q_{\position}(\chi; z_*)
&=
\frac{b(z_*)}{\chi_*} \, \Dirac(\chi - \chi_*) \ ,
\end{empheq}
where $\chi_*$ is the comoving distance of a source at $z_*$. Note that the above equally applies to the usual correlation functions and power spectra of the $3\times 2$pt scheme, i.e. PP, PE and EE.

\begin{figure}[t]
    \centering
    \includegraphics[width=1\columnwidth]{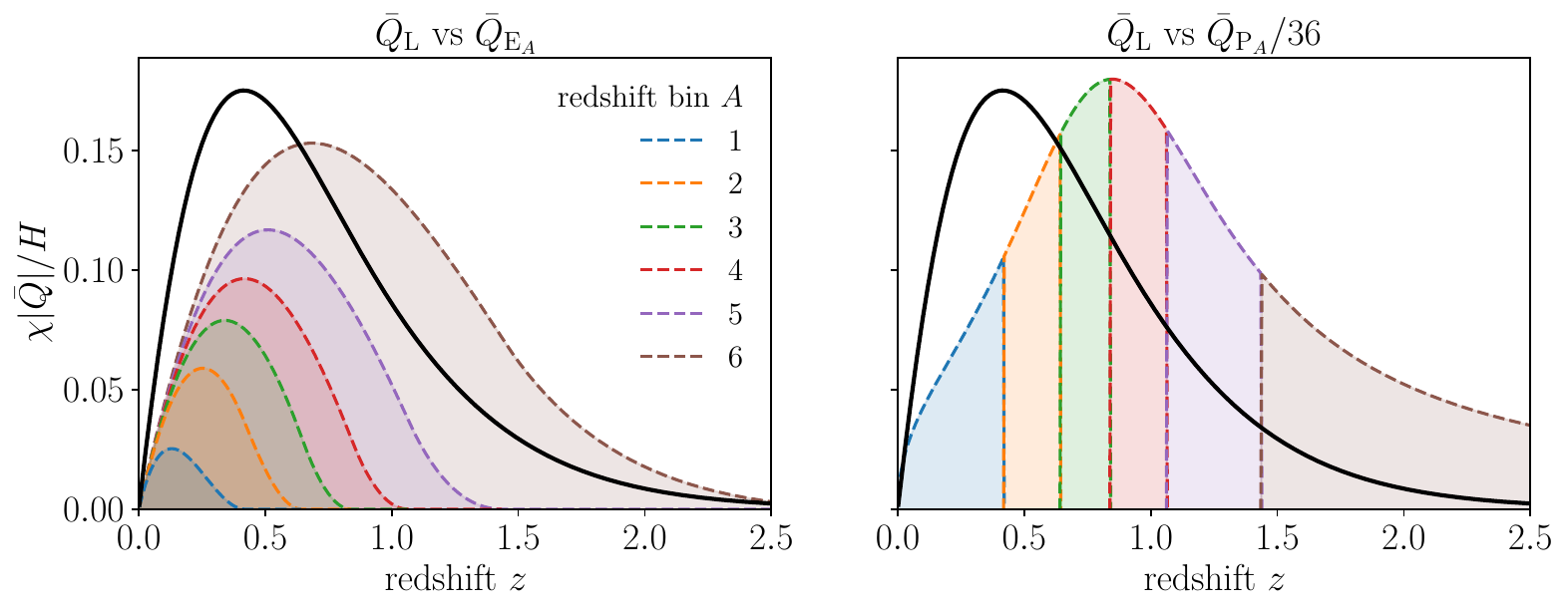}
    \caption{Integration kernels~$\bar{Q}_{\observable}[\chi(z)]$, with $\observable=\lens, \shape_A, \position_A$, for the angular power spectra of the $6\times 2\mathrm{pt}$ correlation scheme, as a function of redshift~$z$. The kernels, defined in \cref{eq:z-averaged_kernel_L,eq:z-averaged_kernel_E,eq:z-averaged_kernel_P}, are multiplied with $\chi(z)/H(z)$ to make them dimensionless. \textit{Left}: Comparison of the LOS-shear kernel~$\bar{Q}_\lens$ (black solid line) with the cosmic-shear kernels $\bar{Q}_{\mathrm{E}_A}$, for six equally-populated redshift bins~$A$ (dashed, coloured and shaded). \textit{Right}: Comparison of $\bar{Q}\e{L}$ with the galaxy-position kernel $\bar{Q}_{\mathrm{P}_A}$ for the same six redshift bins; $\bar{Q}_{\mathrm{P}_A}$ is scaled by an arbitrary factor of 1/36 for ease of comparison.}
    \label{fig:kernels_LEP}
\end{figure}

The integration kernels defined in \cref{eq:Q_L,eq:Q_E,eq:Q_P} all have the units of an inverse length squared. In the full expressions~\eqref{eq:ev_LL}, \eqref{eq:ev_LE} and \eqref{eq:ev_LP} of the expectation values of the LL, LE and LP correlation functions, the redshift integrals may be absorbed in the power spectra, thereby defining redshift-averaged spectra~$\bar{C}_{\ell}^{\observable_1\observable_2}$  for $\observable_1, \observable_2 \in \{\lens, \shape_A, \position_A\}$; these have the same formal expression as in \cref{eq:C_ell_final} but with redshift-averaged kernels
\begin{align}
\label{eq:z-averaged_kernel_L}
\bar{Q}_\lens(\chi)
&\define
\int \dd^2\vect{Z} \; p(\vect{Z}) \, Q_\lens(\chi; \vect{Z}) \ ,
\\
\label{eq:z-averaged_kernel_E}
\bar{Q}_{\shape_A}(\chi)
&\define
\int \dd z_* \; f_A(z_*) \, Q_\shape(\chi; z_*) \ ,
\\
\label{eq:z-averaged_kernel_P}
\bar{Q}_{\position_A}(\chi)
&\define
\int \dd z_* \; f_A(z_*) \, Q_\position(\chi; z_*) \ .
\end{align}
These three kernels are plotted in \cref{fig:kernels_LEP} as a function of redshift for illustration. Since the integration kernels have the units of an inverse length squared, we plot the dimensionless combination $\chi(z) \bar{Q}(z)/H(z)$.\footnote{This specific combination is the integration kernel for the effective fields involved in our observables, expressed as a redshift-integral of the density contrast. For instance, $\bar{C}_\ell^{\lens\lens}=\cev{\bar{\gamma}\h{LOS}_{\ell m}(\bar{\gamma}\h{LOS}_{\ell m})^*}$ may be seen as the angular power spectrum of the redshift-integrated LOS shear~$\bar{\gamma}\e{LOS}(\uvect{u})$, whose expression is
\begin{equation}
\bar{\gamma}\e{LOS}(\uvect{u}) 
\define \int \dd^2\vect{Z} \; \lLOS(\uvect{u}; \vect{Z})
= \int_0^\infty \dd z \;
\underbrace{
\paac{\frac{\chi(z)\bar{Q}_\lens[\chi(z)]}{H(z)}}
}_{\text{plotted in \cref{fig:kernels_LEP}}}
\underbrace{
\pac{\frac{\chi^{-2}(z)\eth^2\Phi(\uvect{u}, z)}{4\pi G (1+z) \bar{\rho}_0}} 
}_{\sim \text{density contrast }\delta}
.
\end{equation}
The same goes for the effective fields associated with $\shape, \position$.
}
We used strong-lens and galaxy redshift distributions as expected from the \Euclid survey (see \cref{subsec:optimistic} for details).

In \cref{fig:kernels_LEP}, we can see that the shape of $\bar{Q}_\lens(z)$ is similar to the shape of the standard cosmic-shear kernels $\bar{Q}_{\shape_A}(z)$; it peaks for cosmological perturbations around $z=0.5$. On the left panel, the amplitude of $\bar{Q}_{\shape_A}$ is seen to grow with the galaxy redshift bin~$A$, which reflects that the more distant a galaxy is, the more its image is sheared. Since the overlap of $\bar{Q}_{\shape_A}$ and $\bar{Q}_{\lens}$ grows with $A$, we expect their cross correlation $\lens\shape_A$ to do the same. The opposite is true for $\lens\position_A$, as the overlap of $\bar{Q}_{\position_A}$ and $\bar{Q}_{\lens}$ decreases with $A$; this is due to the fact that foreground galaxies have a stronger impact on the LOS shear of strong lenses than background galaxies.

\subsection{Standard $3\times 2$pt observables with our notation}

Let us mention, for completeness, that the formalism and notation used in this section equally applies to the usual $3\times 2$pt correlation scheme. The two-point correlations of cosmic shear~($\shape\shape$), galaxy--galaxy lensing~($\shape\position$) and galaxy clustering~($\position\position$) read, respectively
\begin{align}
\label{eq:EE_C}
(\shape\shape)^\pm(\thet; z_1, z_2)
&=
\frac{1}{4\pi}
\sum_{\ell=2}^\infty (2\ell+1) d^\ell_{\pm 2 2}(\thet) \,
C^{\shape\shape}_\ell(z_1, z_2)
\approx
\int_0^\infty \frac{\ell \dd\ell}{2\pi} \, J_{0/4}(\ell\thet) \,
C^{\shape\shape}_\ell(z_1, z_2) ,
\\
\label{eq:EP_C}
\shape^+\position(\thet; z_1, z_2)
&=
\frac{1}{4\pi}
\sum_{\ell=2}^\infty (2\ell+1) d^\ell_{20}(\thet) \,
C^{\shape\position}_\ell(z_1, z_2)
\approx
\int_0^\infty \frac{\ell \dd\ell}{2\pi} \, J_{2}(\ell\thet) \,
C^{\shape\position}_\ell(z_1, z_2) ,
\\
\label{eq:PP_C}
\position\position(\thet; z_1, z_2)
&=
\frac{1}{4\pi}
\sum_{\ell=2}^\infty (2\ell+1) d^\ell_{00}(\thet) \,
C^{\position\position}_\ell(z_1, z_2) 
\approx
\int_0^\infty \frac{\ell \dd\ell}{2\pi} \, J_{0}(\ell\thet) \,
C^{\position\position}_\ell(z_1, z_2) ,
\end{align}
and the angular power spectra $C^{\shape\shape}_\ell, C^{\shape\position}_\ell, C^{\position\position}_\ell$ are given by \cref{eq:C_ell_final} in Limber's approximation, with the kernels provided in \cref{eq:Q_E,eq:Q_P}.

\section{Uncertainties and covariance matrix}
\label{sec:covmat}

In order to assess the detectability of the LL, LE and LP correlations, we must evaluate the expected uncertainty on their measurements, i.e. the variance of the estimators defined in \cref{eq:LL_estimator,eq:LE_estimator,eq:LP_estimator}. Since the ultimate goal will be to perform Fisher forecasts, we shall go one step further and calculate the full covariance matrix of our new observables, whose diagonal elements are precisely the desired uncertainties.

In this work, we have chosen to conduct an analytical, rather than numerical, evaluation of the covariance matrix. Albeit technically challenging, this approach has the advantage of producing results that are more easily adapted to changes in the cosmological and observational setup. In addition, analytical results are more directly interpretable, which is particularly advantageous for optimising the data-analysis strategy.

\subsection{Definition of the covariance matrix}

We are interested here in estimators of the two-point correlation functions of three classes of observables: LOS shear, apparent ellipticity and positions of galaxies. These estimators all take the form $(\widehat{\observable_1 \observable_2})_\alpha$, where $\observable_1, \observable_2\in\{\lens^\pol, \shape_A^\pol, \position_A\}$ and $\alpha$ denotes an angular bin. The covariance of any two such measurements is defined as
\begin{equation}
\label{eq:covariance_definition}
\Cov\pac{(\widehat{\observable_1 \observable_2})_\alpha,
(\widehat{\observable_3 \observable_4})_\beta}
\define
\cev{\gsev{\lsev{
(\widehat{\observable_1 \observable_2})_\alpha
(\widehat{\observable_3 \observable_4})_\beta
}}}
-
(\observable_1 \observable_2)_\alpha
(\observable_3 \observable_4)_\beta \ ,
\end{equation}
with $(\observable_1 \observable_2)_\alpha$ the expectation value of the estimator $(\widehat{\observable_1 \observable_2})_\alpha$. The covariance \emph{matrix} is the symmetric matrix formed by the covariance of every pair of estimators. The diagonal elements of the covariance matrix then correspond to the variance of the estimators,
\begin{equation}
\Var\pac{(\widehat{\observable_1 \observable_2})_\alpha}
=
\Cov\pac{(\widehat{\observable_1 \observable_2})_\alpha,
(\widehat{\observable_1 \observable_2})_\alpha} .
\end{equation}
We shall consider this variance as the (squared) uncertainty on measurements of the correlation function $\observable_1\observable_2$ in the $\alpha$\textsuperscript{th} angular bin.

The remainder of this subsection is dedicated to the actual calculation of the covariance matrix of the $6\times 2$pt correlation scheme.

\subsection{Technical subtlety: coincident indices}
\label{subsec:coincident_indices}

There is an important technical subtlety that explains why the calculation of covariances is more complicated than that of the expectation value of the correlation functions. Readers who are only interested in a ready-to-use recipe may directly skip to \cref{subsec:recipe_covariance_matrix}.

We have already calculated, in \cref{subsec:expectation_values_estimators,subsec:power_spectra_estimators}, the expectation values of our estimators. The remaining quantity to be evaluated in \cref{eq:covariance_definition} is, therefore, the first term on the right-hand side, which generally takes the form
\begin{equation}
\label{eq:ev_four-point_term}
\cev{\gsev{\lsev{
(\widehat{\observable_1 \observable_2})_\alpha
(\widehat{\observable_3 \observable_4})_\beta
}}}
=
\frac{1}{N_1 N_2 N_3 N_4}
\sum_{\mu=1}^{N_1}
\sum_{\nu=1}^{N_2}
\sum_{\rho=1}^{N_3}
\sum_{\sigma=1}^{N_4}
W^\alpha_{\mu\nu} W^\beta_{\rho\sigma}
\cev{\gsev{\lsev{m^1_{\mu\nu} m^2_{\nu\mu} m^3_{\rho\sigma} m^4_{\sigma\rho}}}} ,
\end{equation}
where $N_I$ is the number of objects (galaxies or lenses) relevant to observable~$\observable_I\in\{\lens, \shape_A, \position_A\}$, while $m^I_{\mu\nu}\in\{\mhLOS_{\mu\nu}, \eps_{\mu\nu}, 1\}$ is the measurement associated with $\observable_I$, made in the direction~$\uvect{u}_\mu$ and paired with a second direction~$\uvect{u}_\nu$ to form the local basis. When $\observable_I=\position$, galaxy-sample averaging $\gsev{\ldots}$ eventually replaces $m^I_{\mu\nu}=1$ with $m^I_{\mu\nu}=b\delta_\mu$.

The core of the calculation then consists in evaluating $\cev[2]{\gsev[2]{\lsev[2]{m^1_{\mu\nu} m^2_{\nu\mu} m^3_{\rho\sigma} m^4_{\sigma\rho}}}}$. At first glance, this is a similar problem to the calculation of the LL, LE and LP expectation values performed in \cref{subsubsec:LL_signal,subsubsec:LE_signal,subsubsec:LP_signal}. The first step is to apply the sample-averaging operators so as to get
\begin{multline}
\label{eq:sev_four-point_term}
\cev{\gsev{\lsev{m^1_{\mu\nu} m^2_{\nu\mu} m^3_{\rho\sigma} m^4_{\sigma\rho}}}}\\
=
\pac{
\prod_{i=1}^{L}
\int\dd^6\vect{\Lambda}_{i} \; p(\vect{\Lambda}_{i})
}
\pac{
\prod_{A=1}^T
\prod_{a\in\galaxies_A}
\int\dd^5\vect{\Gamma}_{a} \; p_{A}(\vect{\Gamma}_{a})
}
\cev{m^1_{\mu\nu} m^2_{\nu\mu} m^3_{\rho\sigma} m^4_{\sigma\rho}} .
\end{multline}
Now, in contrast with \cref{subsubsec:LL_signal,subsubsec:LE_signal,subsubsec:LP_signal}, not all the terms of the form~\eqref{eq:sev_four-point_term}, involved in the quadruple sum of \cref{eq:ev_four-point_term}, are equivalent. This is because the window functions $W^\alpha_{\mu\nu}, W^\beta_{\rho\sigma}$ ensure that $\mu\neq\nu$ and $\rho\neq\sigma$, but they do not prevent the cases $\mu=\rho$, $\mu=\sigma$, etc. As a result, several indices in \cref{eq:sev_four-point_term} may coincide, which affects the marginalisation scheme.

For example, when addressing $\Var\pac[2]{(\widehat{\lens^+ \shape_A^+})_\alpha}$ the calculation will feature the term $\cev[2]{\lsev[2]{\gsev[2]{\mhLOS_{+ia} \eps_{+ai} \mhLOS_{+jb} \eps_{+bj}}}}$. When all indices are distinct, sample averaging leads to
\begin{align}
&\cev{\lsev{\gsev{\mhLOS_{+ia} \eps_{+ai} \mhLOS_{+jb} \eps_{+bj}}}}
\nonumber\\
&=
\int \dd^6\vect{\Lambda}_i \; p(\vect{\Lambda}_i)
\int \dd^6\vect{\Lambda}_j \; p(\vect{\Lambda}_j)
\int \dd^5\vect{\Gamma}_a \; p_A(\vect{\Gamma}_a)
\int \dd^5\vect{\Gamma}_b \; p_A(\vect{\Gamma}_b)
\cev{\mhLOS_{+ia} \eps_{+ai} \mhLOS_{+jb} \eps_{+bj}} .
\label{eq:lele_4_indices}
\end{align}
But if, say, $i=j$, then we end up with only three integrals, because the corresponding term only features three indices ($i, a, b$),
\begin{align}
&\cev{\lsev{\gsev{\mhLOS_{+ia} \eps_{+ai} \mhLOS_{+ib} \eps_{+bi}}}}
\nonumber\\
&=
\int \dd^6\vect{\Lambda}_i \; p(\vect{\Lambda}_i)
\int \dd^5\vect{\Gamma}_a \; p_A(\vect{\Gamma}_a)
\int \dd^5\vect{\Gamma}_b \; p_A(\vect{\Gamma}_b)
\cev{\mhLOS_{+ia} \eps_{+ai} \mhLOS_{+ib} \eps_{+bi}} .
\label{eq:lele_3_indices}
\end{align}
A key difference between \cref{eq:lele_4_indices} and \cref{eq:lele_3_indices} is that the latter has a non-zero contribution from the LOS noise~$\nhLOS_i$, because the LOS shear is measured on the same lens~$i$.

In summary, the evaluation of covariances is made more complicated by the presence of coincident indices in \cref{eq:ev_four-point_term}. The calculation requires one to count the number of terms in the quadruple sum that have identical indices, and to treat those separately. Eventually, the terms whose indices are all different lead to \emph{cosmic covariance}, while those with coinciding indices will lead to the contribution of \emph{noise} and \emph{sparsity covariance}~\cite{2025OJAp....8E..57F}, which may be seen as a generalisation of shot noise.

\subsection{Recipe to calculate the covariance matrix}
\label{subsec:recipe_covariance_matrix}

The above discussion emphasises that the analytical calculation of covariances is generally tedious, not to mention the fact that it must be repeated for all the combinations of $\observable_1\observable_2$ and $\observable_3\observable_4$, with $\observable_I\in\{\lens^\pol, \shape_A^\pol, \position_A\}$.

Fortunately, various shortcuts and patterns can be identified in the calculations, which eventually lead to a general recipe that is quite straightforward to use. We only present the recipe here; as a justification of its relevance, we present the explicit calculation of $\Cov\pac[2]{(\widehat{\lens^+\position_A})_\alpha,(\widehat{\lens^+\position_B})_\beta}$ in appendix~\ref{appendix:detailed_calculation_LPLP}.

\subsubsection{The formula}

The covariance of any pair of correlation functions reads
\begin{empheq}[box=\fbox]{multline}
\label{eq:covariance_matrix_recipe}
\Cov\pac{(\widehat{\observable_1 \observable_2})_\alpha,
(\widehat{\observable_3 \observable_4})_\beta}
=
\int_\footprint \frac{\dd^2\uvect{u}_\mu}{\Omega}
\int_\footprint \frac{\dd^2\uvect{u}_\nu}{\Omega} \; W^\alpha_{\mu\nu}
\int_\footprint \frac{\dd^2\uvect{u}_\rho}{\Omega}
\int_\footprint \frac{\dd^2\uvect{u}_\sigma}{\Omega} \; W^\beta_{\rho\sigma}
\\
\times
\pac{
(\wnoise{\observable_1 \observable_3})_{\mu\nu, \rho\sigma} \,
(\wnoise{\observable_2 \observable_4})_{\nu\mu, \sigma\rho}
+
(\wnoise{\observable_1 \observable_4})_{\mu\nu, \sigma\rho} \,
(\wnoise{\observable_2 \observable_3})_{\nu\mu, \rho\sigma}
} ,
\end{empheq}
where the tilded correlation functions such as $\wnoise{(\observable_1\observable_3)}_{\mu\nu, \rho\sigma}\sim\cev[2]{m^1_{\mu\nu}m^3_{\rho\sigma}}$ correspond to those computed in \cref{subsec:expectation_values_estimators}, with an \emph{additional noise term} if $\observable_1=\observable_3$. The weight function~$W^\alpha_{\mu\nu}$ is defined in \cref{eq:weight_function_estimators_2}.

\paragraph{Domain of validity} \Cref{eq:covariance_matrix_recipe} is valid under the following assumptions:
\begin{itemize}
\item \textit{Gaussianity.} We assume that the metric perturbation~$\Phi$ can be treated as a Gaussian random field. All the observables considered here inherit that property, because they are linearly related to $\Phi$. We may then apply Isserlis' theorem~\cite{Isserlis_1918} to express four-point correlation functions in terms of two-point functions; this is why \cref{eq:covariance_matrix_recipe} takes the general form
\begin{align}
\Cov[(\observable_1 \observable_2), (\observable_3 \observable_4)]
&\sim
\cev{m_1 m_2 m_3 m_4}
- \cev{m_1 m_2} \cev{m_3 m_4}
\\
&\sim (\observable_1\observable_3)(\observable_2\observable_4)
    + (\observable_1\observable_4)(\observable_2\observable_3) \ .
\end{align}
In reality, the non-linear formation of structures on small scales causes departures from Gaussianity. Such non-Gaussianities are known to play an important role in the cosmic covariance of cosmic shear~\cite{Takada:2013wfa}. However, since we focus here on the detectability of the LL, LE and LP signals, whose variance turns out to be dominated by noise and sparsity (see \cref{sec:forecasted_detectability}), it is justified to neglect the non-Gaussian covariance at this stage.
\item \textit{Pseudo full sky.} In the calculation of the covariance matrix, some terms involve integrals over directions~$\uvect{u}_\mu, \uvect{u}_\nu, \ldots$ that vanish in full sky due to the presence of trigonometric functions of their relative azimuthal angles. For $\footprint\neq\mathcal{S}^2$, such terms are not exactly zero near the boundaries of $\footprint$. We choose to neglect such boundary terms.
\item \textit{Large samples.} The full calculation of the covariance matrix features terms that are integrals of, e.g., $(\observable_1\observable_3)(\observable_2\observable_4)$ -- without noise -- divided by the lens or galaxy sample size, $L, G \gg 1$. It is safe to neglect such terms. 
\end{itemize}

\subsubsection{Using the formula}
\label{subsubsec:using_covariance_formula}

Let us now explain how to use \cref{eq:covariance_matrix_recipe} in three steps.

\paragraph{Step 1: Set indices up} Pick the required observables $\observable_1, \ldots, \observable_4$ among $\{\position_A, \shape_A^\pol, \lens^\pol\}$, where $\pol\in\{+,\times\}$ denotes polarisation. Replace the generic indices $\mu, \nu, \rho, \sigma$ in \cref{eq:covariance_matrix_recipe} with $a, b, c, d$ or $i, j, k, l$ depending on the nature (galaxy or lens) of the object that it refers to. Pairs of indices separated by a comma, like in $\wnoise{(\observable_1\observable_3)}_{\mu\nu, \rho\sigma}$, are associated with the observables following their order: $(\mu\nu)$ is associated with $\observable_1$ and $(\rho\sigma)$ with $\observable_3$. These pairs of indices appear due to the construction of local bases for spin-two observables. When $\observable=\position$, keep only the first index of the pair. For example,
\begin{multline}
\Cov\pac{(\widehat{\lens^\times \shape^\times_B})_\alpha,
(\widehat{\lens^+ \position_D})_\beta}
=
\int_\footprint \frac{\dd^2\uvect{u}_i}{\Omega}
\int_\footprint \frac{\dd^2\uvect{u}_b}{\Omega} \; W^\alpha_{ib}
\int_\footprint \frac{\dd^2\uvect{u}_k}{\Omega}
\int_\footprint \frac{\dd^2\uvect{u}_d}{\Omega} \; W^\beta_{kd}
\\
\times
\pac{
(\wnoise{\lens^+ \lens^+})_{ib, kd} \,
(\wnoise{\shape^\times_B \position_D})_{bi, d}
+
(\wnoise{\lens^+ \position_D})_{ib, d} \,
(\wnoise{\shape^\times_B \lens^+})_{bi, kd}
} .
\end{multline}

\paragraph{Step 2: Unfold pairs of indices} The correlation functions with pairs of indices are then expressed in terms of the regular correlation functions according to the following dictionary (we drop the tildes to alleviate notation here).

\begin{enumerate}
\item \textit{Two spin-two observables:}
\begin{align}
(\observable^+_1 \observable^+_3)_{\mu\nu, \rho\sigma}
&=
\observable_1^+\observable_3^+(\uvect{u}_\mu, \uvect{u}_\rho)
\cos 2(\psi_{\mu\rho} - \psi_{\mu\nu})
\cos 2(\psi_{\rho\mu} - \psi_{\rho\sigma})
\nonumber\\ &\quad
+
\observable_1^\times\observable_3^\times(\uvect{u}_\mu, \uvect{u}_\rho)
\sin 2(\psi_{\mu\rho} - \psi_{\mu\nu})
\sin 2(\psi_{\rho\mu} - \psi_{\rho\sigma}) \ ,
\\
(\observable^+_1 \observable^\times_3)_{\mu\nu, \rho\sigma}
&=
\observable_1^+\observable_3^+(\uvect{u}_\mu, \uvect{u}_\rho)
\cos 2(\psi_{\mu\rho} - \psi_{\mu\nu})
\sin 2(\psi_{\rho\mu} - \psi_{\rho\sigma})
\nonumber\\ &\quad
-
\observable_1^\times\observable_3^\times(\uvect{u}_\mu, \uvect{u}_\rho)
\sin 2(\psi_{\mu\rho} - \psi_{\mu\nu})
\cos 2(\psi_{\rho\mu} - \psi_{\rho\sigma}) \ ,
\\
(\observable^\times_1 \observable^\times_3)_{\mu\nu, \rho\sigma}
&=
\observable_1^+\observable_3^+(\uvect{u}_\mu, \uvect{u}_\rho)
\sin 2(\psi_{\mu\rho} - \psi_{\mu\nu})
\sin 2(\psi_{\rho\mu} - \psi_{\rho\sigma})
\nonumber\\ &\quad
+
\observable_1^\times\observable_3^\times(\uvect{u}_\mu, \uvect{u}_\rho)
\cos 2(\psi_{\mu\rho} - \psi_{\mu\nu})
\cos 2(\psi_{\rho\mu} - \psi_{\rho\sigma}) \ .
\end{align}

\item \textit{One spin-two observable with one spin-zero observable:}
\begin{align}
(\observable_1^+ \observable_3)_{\mu\nu, \rho}
&=
\observable_1^+ \observable_3(\uvect{u}_\mu, \uvect{u}_\rho) \,
\cos2(\psi_{\mu\rho} - \psi_{\mu\nu}) \ ,
\\
(\observable_1^\times \observable_3)_{\mu\nu, \rho}
&=
\observable_1^+ \observable_3(\uvect{u}_\mu, \uvect{u}_\rho) \,
\sin 2(\psi_{\mu\rho} - \psi_{\mu\nu}) \ .
\end{align}

\item \textit{Two spin-zero observables:} $(\observable_1 \observable_3)_{\mu, \rho} = \observable_1\observable_3(\uvect{u}_\mu, \uvect{u}_\rho)$.
\end{enumerate}

\paragraph{Step 3: Substitute the elementary building blocks} After step~2, the covariance is fully expressed in terms of trigonometric functions and the following building blocks, which are redshift-integrated correlation functions, possibly with an additional noise term. The final result is obtained by substituting in their explicit expressions:
\begin{align}
\wnoise{\lens^\pol\lens^\pol}(\uvect{u}_i, \uvect{u}_j)
&\define
\int \dd^2 \vect{Z}_i \; p(\vect{Z}_i)
\int \dd^2 \vect{Z}_j \; p(\vect{Z}_j) \,
\lens^\pol\lens^\pol(\thet_{ij}; \vect{Z}_i, \vect{Z}_j)
+ \frac{\sigma\e{LOS}^2}{2n\e{L}} \, \Dirac(\uvect{u}_i - \uvect{u}_j) \ ,
\\
\wnoise{\lens^\pol\shape^\pol_B}(\uvect{u}_i, \uvect{u}_b)
&\define
\int \dd^2 \vect{Z}_i \; p(\vect{Z}_i)
\int \dd z_b \; p_B(z_b) \,
\lens^\pol\shape^\pol(\thet_{ib}; \vect{Z}_i, z_b) \ ,
\\
\wnoise{\lens^+\position_B}(\uvect{u}_i, \uvect{u}_b)
&\define
\int \dd^2 \vect{Z}_i \; p(\vect{Z}_i)
\int \dd z_b \; p_B(z_b) \,
\lens\position(\thet_{ib}; \vect{Z}_i, z_b) \ ,
\\
\wnoise{\shape^\pol_A\shape^\pol_B}(\uvect{u}_a, \uvect{u}_b)
&\define
\int \dd z_a \; p_A(z_a) \int \dd z_b \; p_B(z_b) \,
\shape^\pol\shape^\pol(\thet_{ab}; z_a, z_b) 
+ \frac{\delta_{AB}\sigma_A^2}{2n_A} \, \Dirac(\uvect{u}_a - \uvect{u}_b) \ ,
\\
\wnoise{\shape^+_A\position_B}(\uvect{u}_a, \uvect{u}_b)
&\define
\int \dd z_a \; p_A(z_a) \int \dd z_b \; p_B(z_b) \,
\shape^+\position(\thet_{ab}; z_a, z_b) \ ,
\\
\wnoise{\position_A\position_B}(\uvect{u}_a, \uvect{u}_b)
&\define
\int \dd z_a \; p_A(z_a) \int \dd z_b \; p_B(z_b) \,
\position\position(\thet_{ab}; z_a, z_b)
+ \frac{\delta_{AB}}{n_A} \, \Dirac(\uvect{u}_a - \uvect{u}_b) \ .
\end{align}

Each of the three tilded autocorrelation functions ($\wnoise{\lens\lens}, \wnoise{\shape\shape}, \wnoise{\position\position}$) exhibits a Dirac-delta term. Those \emph{noise terms} effectively account for the cases of coincident indices in the sum of \cref{eq:ev_four-point_term}, which were discussed in \cref{subsec:coincident_indices}.

In $\wnoise{\position_A\position_A}$, this is the usual shot noise, or Poisson noise, which is inversely proportional to the surface density~$n_A$ of galaxies in the $A$\textsuperscript{th} tomographic bin. In $\wnoise{\shape^\pol_A\shape^\pol_A}$, the noise term is proportional to $\sigma_A^2$, which is the variance of the apparent ellipticity of galaxies in that bin,
\begin{align}
\label{eq:ellipticity_variance}
\sigma_A^2
&\define
\int \dd^5\vect{\Gamma} \; p_A(\vect{\Gamma}) \, \cev[2]{|\eps|^2}
\\
&=
\int \dd^5\vect{\Gamma} \; p_A(\vect{\Gamma})
\pac{|\eps_0|^2 + \cev[2]{|\gamma|^2}}
\\
\label{eq:ellipticity_variance_2_terms}
&=
\underbrace{%
\int \dd z \; f_A(z)
\int \dd^2\eps_0 \; p(\eps_0; z) \, |\eps_0|^2
}_{\text{shape noise}~\sigma_{\eps_0, A}^2}
+
\underbrace{%
\int \dd z \; f_A(z) \, (\shape\shape)^+(0; z, z)
}_{\text{shear variance}} .
\end{align}
We see that $\sigma_A^2$ receives a contribution from intrinsic ellipticities, usually referred to as shape noise, as well as a contribution from the variance of the weak-lensing shear. The latter contribution, which is absent from the standard weak-lensing literature, has been recently introduced in ref.~\cite{2025OJAp....8E..57F} as \emph{sparsity variance}. It comes from the fact that cosmic-shear measurements are performed on a finite sample of galaxies that are randomly distributed in the sky; this leads to an uncertainty on estimates of the $\shape\shape$ correlation function that would persist even if we could perfectly measure the shear on individual galaxies.

In cosmic shear, sparsity variance leads to a sub-percent contribution to the covariance matrix, because $|\gamma|\ll|\eps_0|$. The situation is quantitatively different for the LOS shear. The noise term in $\wnoise{\lens\lens}$ is proportional to the variance of the \emph{measured} LOS shear,
\begin{align}
\label{eq:LOS_variance}
\sigma\e{LOS}^2
&\define
\int\dd^6\vect{\Lambda} \; p(\vect{\Lambda}) \cev[2]{|\mlLOS|^2}
\\
&=
\int\dd^6\vect{\Lambda} \; p(\vect{\Lambda})
\pac{|\nlLOS|^2 + \cev[2]{|\lLOS|^2}}
\\
&=
\underbrace{%
\int \dd^2\vect{Z} \; p(\vect{Z})
\int \dd^2\nlLOS \; p(\nlLOS; \vect{Z}) \, |\nlLOS|^2
}_{\text{LOS shear measurement noise}~\sigma^2_{\nlLOS}}
+
\underbrace{%
\int \dd^2\vect{Z} \; p(\vect{Z}) \,
(\lens\lens)^+(0; \vect{Z}, \vect{Z})
}_{\text{LOS shear variance}~\sigma^2_{\lLOS}}
\ ,
\end{align}
which combines the variance of the noise on individual LOS shear measurements, $\sigma_{\nlLOS}$, with the variance of the LOS signal, $\sigma_{\lLOS}$. Since we expect $\sigma_{\nlLOS} \sim \sigma_{\lLOS}$, sparsity variance must absolutely be taken into account in the covariance matrix of the $6\times2$pt correlation scheme.

When dealing with forecasts, $\sigma_A^2$ and $\sigma\e{LOS}^2$ must be estimated by assuming the noise properties and a cosmological model. Eventually, these quantities will be measured directly from the data.

\subsection{Result and comments on the sources of uncertainty}

We calculated the full covariance matrix of the $6\times2$pt scheme using the recipe presented in \cref{subsec:recipe_covariance_matrix}. Due to the lengthiness of the resulting expressions, we do not report them here. Instead, they are presented in the \EC, a separate document stored in a dedicated \href{https://github.com/ELROND-project}{public Github repository}.

The covariance of two correlation-function estimators is generally composed of three types of contributions: $\Cov=\CCov+\NCov+\SCov$.
\begin{itemize}
\item \emph{Cosmic covariance}, $\CCov$, is caused by the intrinsically correlated nature of the physical fields at play. It is the limit of $\Cov$ if we could observe an infinite number of  galaxies and lenses ($L, G\to\infty$), i.e. \cref{eq:covariance_matrix_recipe} without tildes on the right-hand side. The associated variance, $\CCov(\widehat{\observable_1\observable_2}, \widehat{\observable_1\observable_2})$, is an irreducible uncertainty on $\observable_1\observable_2$, due to the fact that we observe a single realisation of the cosmic fields on a finite sky.
\item \emph{Noise covariance}, $\NCov$, is due to (i) the uncertainty~$\nlLOS$ on individual measurements of the LOS shear; and (ii) the intrinsic ellipticity~$\eps_0$ of galaxies (shape noise). 
\item \emph{Sparsity covariance}, $\SCov$, comes from the fact that we observe a finite number of galaxies and strong lenses with random positions. It may be seen as a generalisation of shot noise~\cite{2025OJAp....8E..57F}. This contribution would persist even if galaxies were intrinsically spherical and if we could measure $\lLOS$ with an arbitrarily large precision.
\end{itemize}

In practice, $\NCov$ and $\SCov$ always come together, via $\sigma_A^2$ and $\sigma\e{LOS}^2$ defined in \cref{eq:ellipticity_variance,eq:LOS_variance}.\footnote{Some variances, such as $\Var(\widehat{\lens^+\lens^+})_\alpha$ -- see eq.~(11) in the \EC~-- contain terms proportional to $\sigma\e{LOS}^4=(\sigma_{\nlLOS}^2+\sigma_{\lLOS}^2)^2$. We choose to count the cross term, $2\sigma_{\nlLOS}^2\sigma_{\lLOS}^2$, as a part of $\NCov$, so that $\SCov$ represents the contribution that survives in the absence of noise ($\sigma_{\nlLOS}=0$). The same is true for terms proportional to $\sigma\e{LOS}^2\sigma_A^2$ or $\sigma_A^4$. This convention will tend to boost the contribution of noise with respect to that of sparsity.}
The relative importance of $\SCov$ and $\NCov$ is driven by the signal-to-noise ratio of individual observations. For standard cosmic shear, $\SCov\ll\NCov$ because the intrinsic ellipticity of a galaxy is typically larger than the weak-lensing shear; in other words, the second term on the right-hand side of \cref{eq:ellipticity_variance_2_terms} may be neglected. But for the LOS shear, $\SCov\sim\NCov$ because the uncertainty on $\lLOS$ for a single strong lens is comparable to the expected signal. Sparsity covariance is therefore a crucial concept in this work.

\subsection{Numerical implementation}

The correlation functions and their covariance are computed using the Python code \code[https://github.com/ELROND-project/loscov]{loscov}, which we made publicly available. The code assumes, as input parameters, a background cosmology and the survey specifications. These include sky coverage, $\Omega$; the limits of the tomographic redshift bins, $z_A$; the properties of the galaxy sample -- $G$, $f_A(z)$, $\sigma_{\eps_0, A}$ -- and those of the strong-lens sample -- $L$, $p(\vect{Z})$, $\sigma_{\nlLOS}$. The limits of the angular bins, $\thet_\alpha$, can be either set by hand or computed automatically according to a chosen mathematical prescription. \code[https://github.com/ELROND-project/loscov]{loscov} then returns the binned correlation functions and associated covariance for those specifications, examples of which are presented in \cref{sec:forecasted_detectability}. 

The matter power spectrum, $P\e{m}(z, k)$,\footnote{In practice, we use the Weyl power spectrum as computed from \code{camb},
\begin{equation}
P_W\h{\code{camb}}(z, k)
\define k^4 P_\Phi(z, k)
= \pa{\frac{3}{2} H_0^2\Omega\e{m}}^2 (1+z)^2 P\e{m}(z, k) \ .
\end{equation}
} is computed using \code[https://camb.readthedocs.io/en/latest/]{camb}~\cite{Lewis_2000,Howlett_2012} for $z\in[0, 7]$. Following the \code{camb} default setting, non-linear scales are modelled with \code[https://www.roe.ac.uk/~jap/haloes/]{halofit}~\cite{Peacock:2000qk, Smith:2002dz, Mead:2020vgs}; we use this prescription up to $k=\SI{500}{\per\mega\parsec}$, and further extrapolate the result as a power law to $k=\SI{e10}{\per\mega\parsec}$. The angular power spectra, $C_\ell^{\observable_1\observable_2}$, of \cref{eq:C_ell_final} are computed by discrete integration over $\chi$, and the corresponding correlation functions are obtained by performing Hankel transforms~\eqref{eq:correlations_as_Hankel} with the \code[{https://hankel.readthedocs.io/en/latest/demos/getting_started.html}]{hankel} package~\cite{Murray_2019}. We make extensive use of the \code[https://numpy.org]{numpy}~\cite{Harris_2020} and \code[https://scipy.org]{scipy}~\cite{scipy_2020} functions throughout. Figures are plotted with \code[https://matplotlib.org]{matplotlib}~\cite{Hunter_2007}.

As for the covariance matrix, which involves a number of four- and five-dimensional integrals (see \EC), numerical integrators such as \texttt{numpy.nquad} do not converge on reasonable timescales if we use a realistic numbers of angular and redshift bins. We therefore rely on Monte Carlo integration, record the numerical uncertainties associated with each calculation, and ensure that the relative error is sufficiently small -- see appendix~\ref{appendix:monte_carlo} for details. With $10^7$ Monte-Carlo samples, one ``redshift block'', e.g. $\Cov\pac[2]{(\widehat{\lens\shape_1})^+_\alpha, (\widehat{\lens^+\position_2})_\beta}$ for all $\alpha, \beta$, takes about one day to compute; parallelising to process all blocks simultaneously allows us to evaluate the full covariance matrix on the same timescale.

\section{Forecasted detectability}
\label{sec:forecasted_detectability}

Let us now apply our analytical results to forecast the detectability of the $\lens\lens$, $\lens\shape$, and $\lens\position$ correlations in a stage-IV galaxy survey. Specifically, we consider in this section an \emph{optimistic scenario} (\cref{subsec:optimistic}) and a \emph{conservative scenario} (\cref{subsec:conservative}), which differ by: (i) the number~$L$ of strong-lensing systems with a measurement of the LOS shear; and (ii) the mean uncertainty~$\sigma_{\nlLOS}$ on these measurements. The galaxy sample is the same in both scenarios, and the survey footprint is assumed to cover $\Omega=\SI{15000}{\deg\squared}$.

\subsection{Optimistic scenario}
\label{subsec:optimistic}

\begin{figure}
\centering
\includegraphics[height=5.6cm]{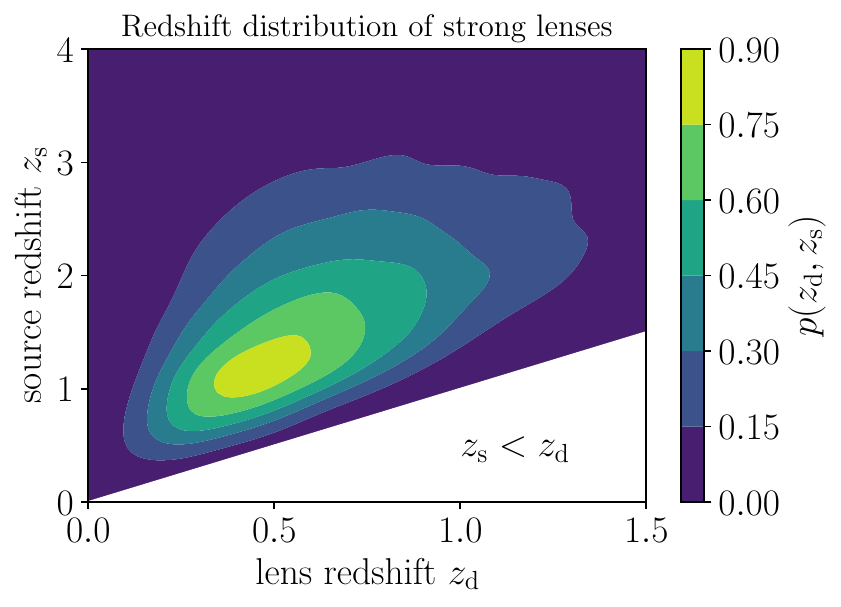}
\hfill
\includegraphics[height=5.6cm]{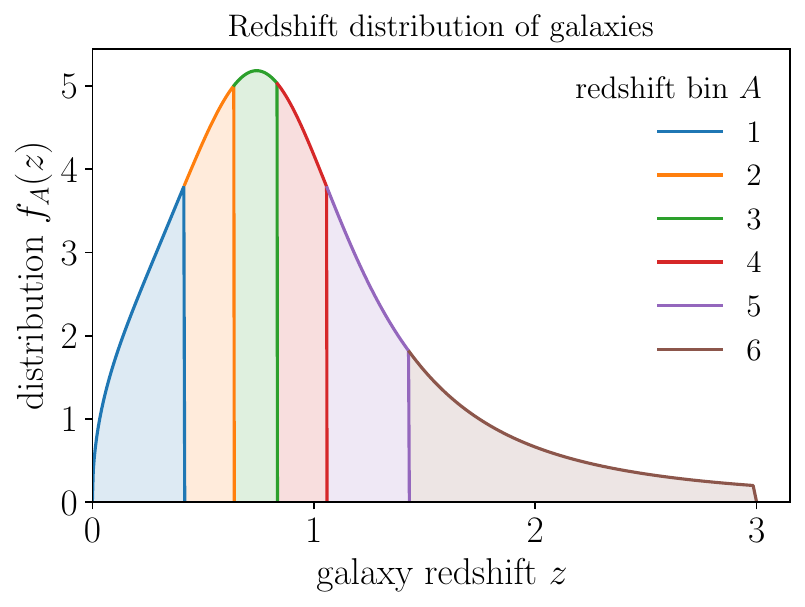}
\caption{Redshift distributions used in our forecasts. \emph{Left:} Expected distribution~$p(\vect{Z})$ of the \Euclid strong lenses from \code[https://github.com/tcollett/LensPop]{LensPop}. \emph{Right:} Expected distribution~$f_A(z)$ of \Euclid galaxies in six bins.}
\label{fig:redshift_distributions}
\end{figure}

\paragraph{Strong-lensing sample} We assume
\begin{empheq}[box=\fbox]{equation}
\label{eq:lens_sample_optimistic}
L = 10^5, \quad
\sigma_{\nlLOS} = 0.05
\qquad \text{(optimistic scenario).}
\end{empheq}
For reference, the complete \Euclid survey is expected to detect $1.7\times 10^5$ strong lenses~\cite{2015ApJ...811...20C}; thus, our optimistic scenario supposes that we shall get reliable measurements of the LOS shear on about 60\,\% of that sample. As for its uncertainty, the proof of concept carried out by Hogg et al. \cite{Hogg:2022ycw} found\footnote{In that reference, the quoted uncertainty of $\bar{\sigma}=0.01$ corresponds to the standard deviation of one component (real or imaginary) of $\nlLOS$. Since $\sigma_{\nlLOS}^2=\lsev{|\nlLOS|^2}=2\times\lsev{\Re(\nlLOS)^2}$, that quoted uncertainty must take on a factor $\sqrt{2}$ to match the definitions considered here.} $\sigma_{\nlLOS}=0.014$ using mock lenses built with analytical profiles. We choose to inflate that uncertainty in order to allow for the many possible systematics associated with lens-mass modelling, source modelling, and lens-light subtraction. The redshift distribution~$p(\vect{Z})=p(z\e{d}, z\e{s})$ of the strong-lensing systems is adapted from the output of \code[https://github.com/tcollett/LensPop]{LensPop}~\cite{2015ApJ...811...20C} for the \Euclid survey, and is depicted in the left panel of \cref{fig:redshift_distributions}.

\paragraph{Galaxy sample} We assume a total number of observed galaxies $G=2\times 10^9$, divided into six equally populated redshift bins, so that $G_A=G/6$. The total redshift distribution is adapted from eq.~(10) of ref.~\cite{2021A&A...646A..62M}, which we artificially divide into six; the resulting distributions~$f_A(z)$ in each bin are depicted on the right panel of \cref{fig:redshift_distributions}. Following ref.~\cite{Euclid:2021osj}, we model the galaxy bias as $b(z)=0.9+1.1\,z^{2.4}/(1+z)$. The amplitude of shape noise is assumed to be identical in all tomographic bins, $\sigma_{\eps_0, A}/\sqrt{2}=0.3$. This is slightly larger than the value reported by KiDS--Legacy, where the same quantity ranges between 0.26 and 0.3 depending on the tomographic bin, with a mean value of 0.28~(see table~1 of ref.~\cite{Wright:2025xka}).

\paragraph{Iso-SNR angular binning} The expectation value of the correlation functions, $(\lens\lens)^\pm_\alpha$, $(\lens\shape_A)^\pm_\alpha$, $(\lens^+\position)_\alpha$, and the covariance matrix all depend on the angular binning of the pairs of lenses and galaxies. Let us denote with~$[\thet_\alpha, \thet_{\alpha+1})$ the limits of the $N$ angular bins. In this work, we choose the $\thet_\alpha$ so that all angular bins roughly have the same signal-to-noise ratio~(SNR) for a given correlation function~$(\observable_1\observable_2)_\alpha$,
\begin{equation}
\label{eq:constant_SNR}
\SNR(\observable_1\observable_2; \thet_\alpha, \thet_{\alpha+1})
\define
\frac{(\observable_1\observable_2)_\alpha}{\sqrt{\Var(\observable_1\observable_2)_\alpha}}
\approx
R \ ,
\end{equation}
where $R$ is the target SNR, which we fix to be at least $R\e{min}=6$ ($8$), for $\lens\lens$ ($\lens\shape$ and $\lens\position$) in the optimistic scenario. In addition, we fix the maximum number of bins to $N\e{max}\approx 20$, so that the actual target SNR may exceed $R\e{min}$.\footnote{Imposing both $R\e{min}$ and $N\e{max}$ aims to make a compromise between precise measurements of the correlation function on the one hand (large SNR), and resolving their evolution with angular separation~$\thet$ on the other hand (large number of bins). At this stage, the values $R\e{min}=8$ and $N\e{max}=20$ are fixed arbitrarily and do not rely on any proper optimisation procedure, which is left for future work.}
See appendix~\ref{appendix:angular_binscheme} for details. In this framework, \emph{the binning scheme generally differs from one correlation function to another}.

\begin{figure}[p]
    \centering
    \includegraphics[width=\columnwidth]{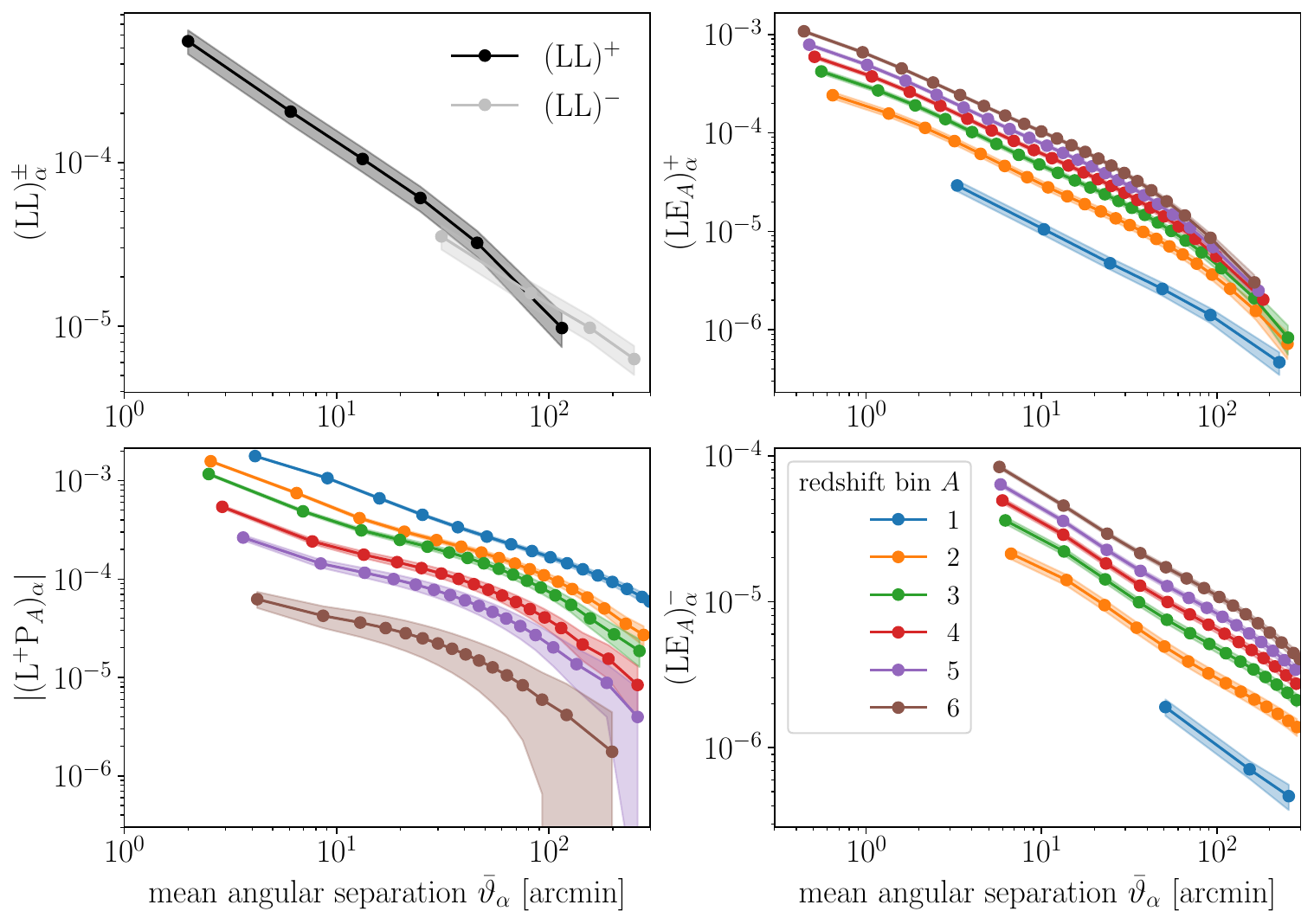}
    \caption{Detection prospects for the $\lens\lens$, $\lens\shape$ and $\lens\position$ correlations in the \emph{optimistic scenario}~\eqref{eq:lens_sample_optimistic}. In each panel, the points indicate the expected value of $(\observable_1\observable_2)_\alpha$ as a function of the mean angular separation~$\bar{\thet}_\alpha$ in the $\alpha$\textsuperscript{th} bin. Shaded regions delimit the $1\sigma$ ($68~\%$) uncertainty about each value, i.e. $(\observable_1\observable_2)_\alpha\pm [\Var\widehat{(\observable_1\observable_2)_\alpha}]^{1/2}$.}
    \label{fig:optimistic_all}
    \vspace{1cm}
    \includegraphics[width=\columnwidth]{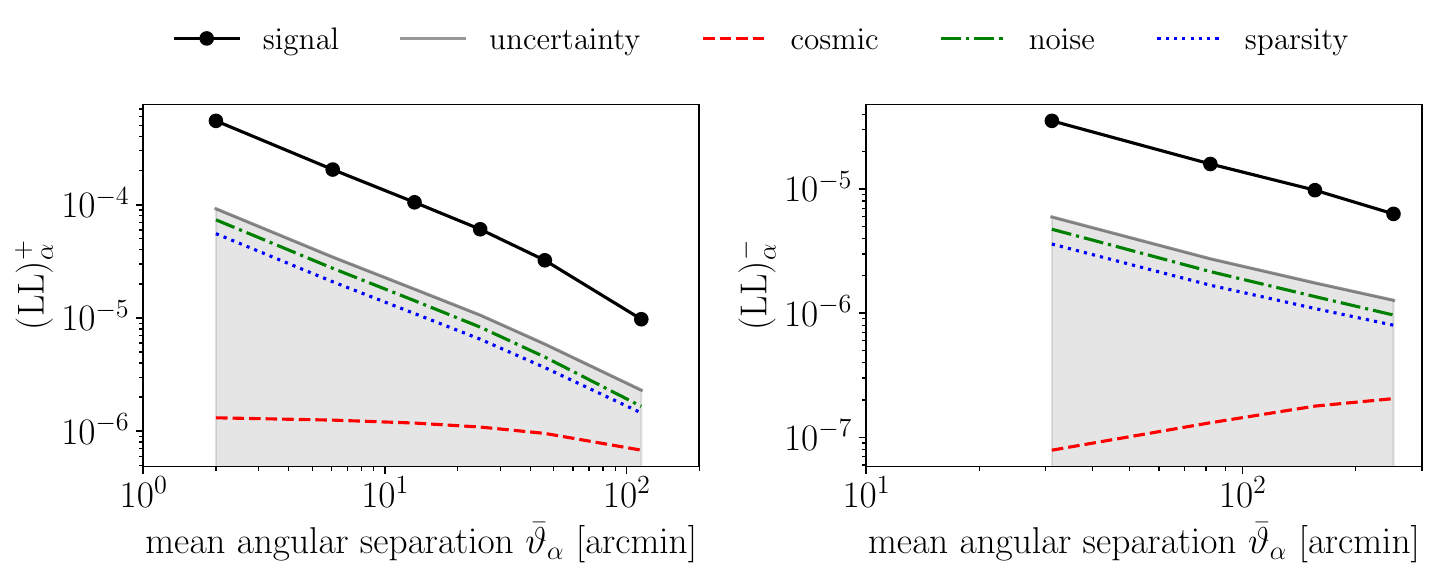}
    \caption{The expected signal and uncertainty on the autocorrelation of the LOS shear, $\lens\lens^\pm$, in the \emph{optimistic scenario}. The contributions of cosmic, noise and sparsity variance are plotted in dashed red, dot-dashed green and blue dotted lines, respectively. The larger uncertainty on the right-most point, $(\lens\lens)^-_7$, is an artefact of the angular binning procedure described in appendix~\ref{appendix:angular_binscheme}, whereby the last bin may not reach the target SNR, since it is bounded from above by $\thet\e{max}=300'$.}
    \label{fig:optimistic_LL}
\end{figure}

\paragraph{All correlations are detectable} \Cref{fig:optimistic_all} shows the expected signal and $68\,\%$ uncertainties for all the correlation functions $(\observable_1\observable_2)_\alpha$ defined in this article, as a function of angular separation.\footnote{%
We use the mean angular separation in the $\alpha$\textsuperscript{th} bin, which is defined as
\begin{equation*}
\bar{\thet}_\alpha
=
\frac{1}{\Omega_\alpha}
\int_{\thet_\alpha}^{\thet_{\alpha+1}}
2\pi\thet\dd\thet \; \thet
=
\frac{2}{3} \frac{\thet_{\alpha+1}^3 - \thet_\alpha^3}{\thet_{\alpha+1}^2 - \thet_\alpha^2} \ .
\end{equation*}
}
These results demonstrate that all correlation functions will be detectable across a range of angular scales and redshifts in the optimistic scenario, which is compatible with stage-IV surveys like \Euclid. We note that, as expected from \cref{fig:kernels_LEP}, detection prospects for $\lens\position_A$ ($\lens\shape_A$) increase for lower (higher) redshift bins~$A$. The expected SNR exceeds $100$ for $\lens\position_1$ and $40$ for $\lens\shape_6$ on angular scales of a few arcminutes.

\paragraph{Full covariance matrix and uncertainty budget} It is instructive to examine how the total uncertainty on, e.g., $\lens\lens$, breaks down into its cosmic, noise and sparsity contributions. In \cref{fig:optimistic_LL}, we can see that the uncertainty budget is dominated by noise and sparsity, while cosmic variance remains subdominant. Here, we actually have $\sigma_{\nlLOS}\approx\sigma_{\lLOS}$ by coincidence; the contribution of noise appears larger than that of sparsity due to our conventional choice to count the cross-term~$\propto2\sigma_{\nlLOS}\sigma_{\lLOS}$ as a noise term. More generally, \cref{fig:optimistic_LL} also emphasises that there is not so much to gain by increasing the precision on $\lLOS$ measurements, since one would quickly hit the sparsity floor. In other words, as far as $\lens\lens$ is concerned, \emph{many lenses with mediocre precision are better than few lenses with good precision.}

\begin{figure}[t]
    \centering
    \includegraphics[width=1\columnwidth]{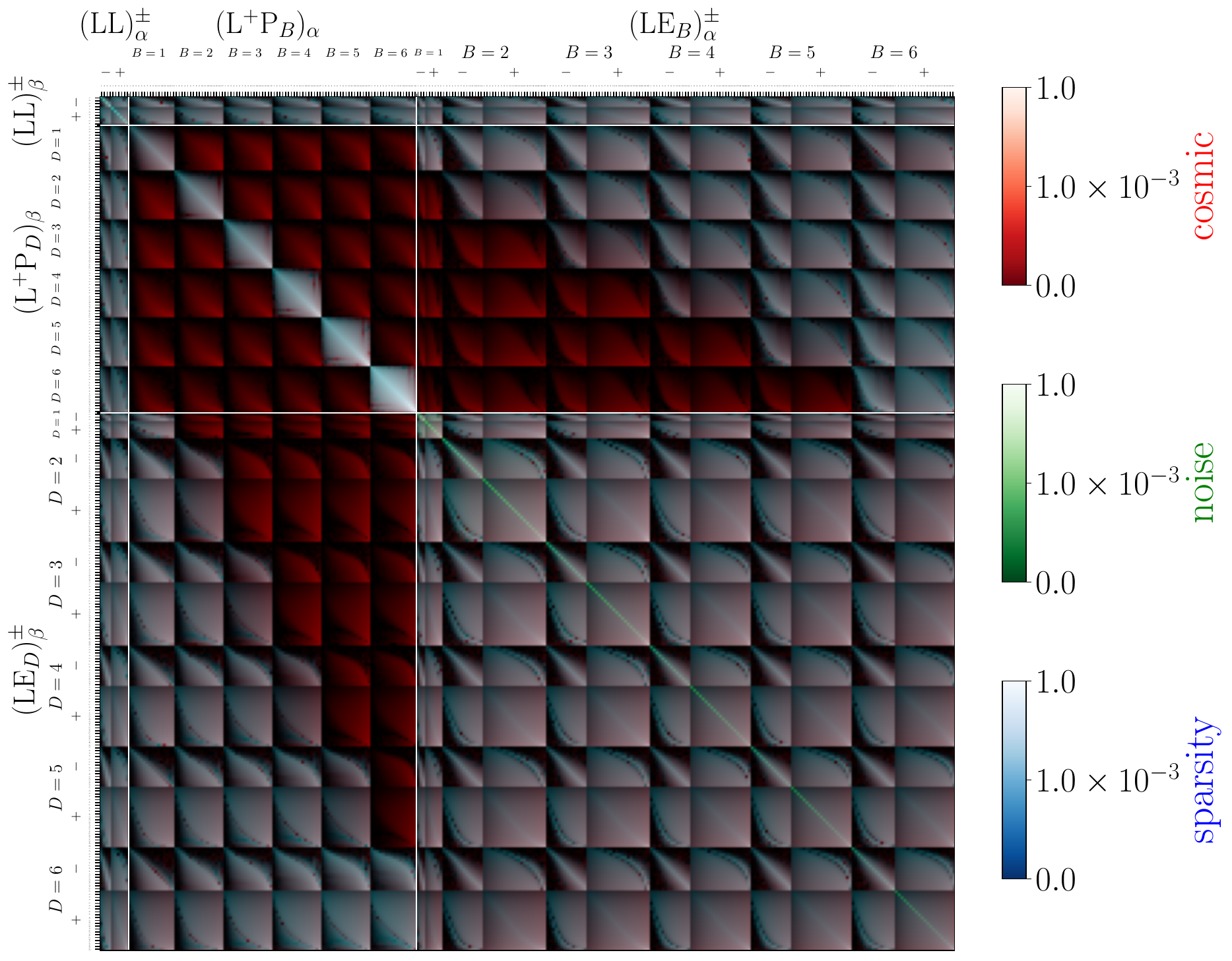}
    \caption{Full covariance matrix of the $\lens\lens$, $\lens\position$ and $\lens\shape$ correlation functions in the \emph{optimistic scenario}, normalised by the corresponding signals:
    $\Cov\pac[2]{
    \widehat{(\observable_1\observable_2)}_\alpha,
    \widehat{(\observable_3\observable_4)}_\beta}
    /\pac[2]{
    (\observable_1\observable_2)_\alpha (\observable_3\observable_4)_\beta}
    $.
    Each matrix element is an RGB combination of the absolute value of the cosmic, noise and sparsity contributions. Brighter colours indicate a stronger covariance relative to signal, \modified{according to a $\log(1+x)$ scale.}
    }
    \label{fig:optimistic_covmat}
\end{figure}

The fact that cosmic \emph{variance} is generally subdominant compared to its noise and sparsity counterparts does not imply that cosmic \emph{covariance} is always negligible. This appears clearly in \cref{fig:optimistic_covmat}, which depicts the full covariance matrix for $\lens\lens$, $\lens\shape$ and $\lens\position$, normalised by the corresponding signal, that is
$
\Cov\pac[2]{
    \widehat{(\observable_1\observable_2)}_\alpha,
    \widehat{(\observable_3\observable_4)}_\beta
    }
/
\pac[2]{
(\observable_1\observable_2)_\alpha
(\observable_3\observable_4)_\beta
}
$.
The diagonal of that matrix thus corresponds to the squared SNR. We use a red-green-blue (RGB) colour coding in order to identify the relative contributions of cosmic, noise, and sparsity, respectively.

The diagonal of \cref{fig:optimistic_covmat} shows that the uncertainty on $\lens\lens$ and $\lens\position$ is a fair mix of noise and sparsity, while $\lens\shape$ is (shape)-noise limited. As for cosmic covariance, it is only non-negligible in off-diagonal elements of the covariance matrix. While negligible in detectability considerations, this contribution will therefore play an important role in Fisher-matrix forecasts. 

\subsection{Conservative scenario}
\label{subsec:conservative}

\begin{figure}[p]
\centering
\includegraphics[width=\columnwidth]{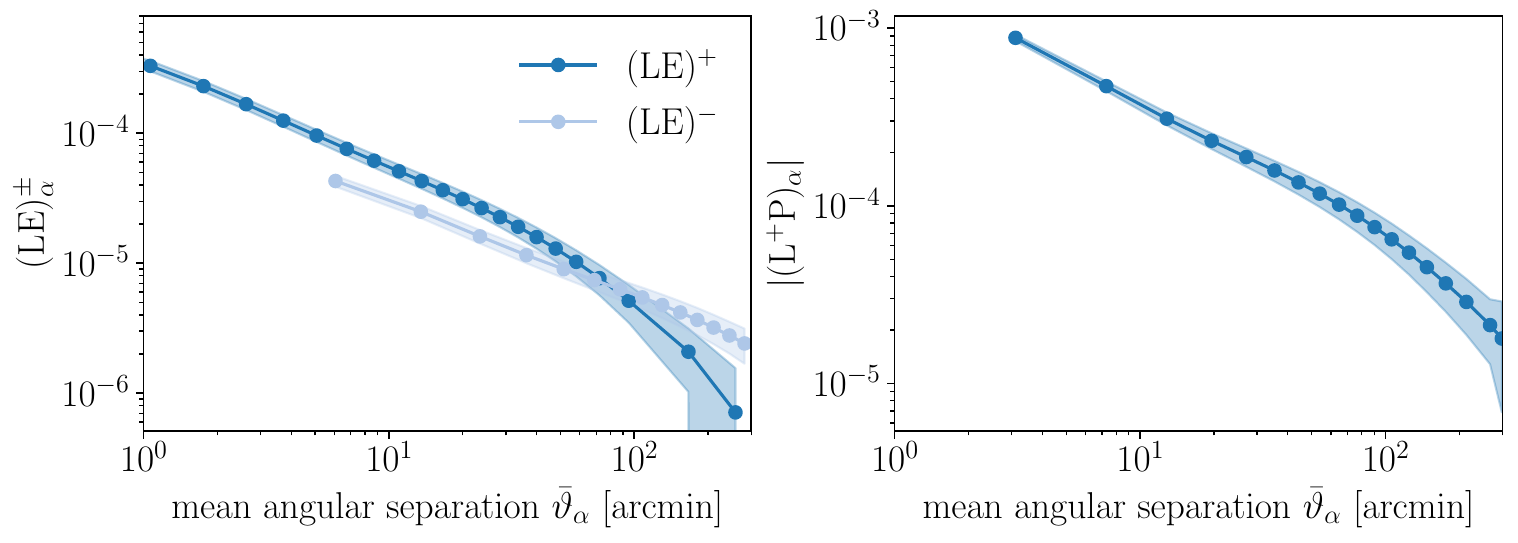}
\caption{Detection prospects for the $\lens\shape$ and $\lens\position$ correlations in the \emph{conservative scenario}~\eqref{eq:lens_sample_conservative}, where there is no redshift binning of the galaxy sample. In each panel, the points indicate the expected value of $(\observable_1\observable_2)_\alpha$ as a function of the mean angular separation~$\bar{\thet}_\alpha$ in the $\alpha$\textsuperscript{th} bin. Shaded regions delimit the $1\sigma$ ($68~\%$) uncertainty about each value.}
\label{fig:conservative_LE_LP}
\vspace{1cm}
\centering
\includegraphics[width=1\columnwidth]{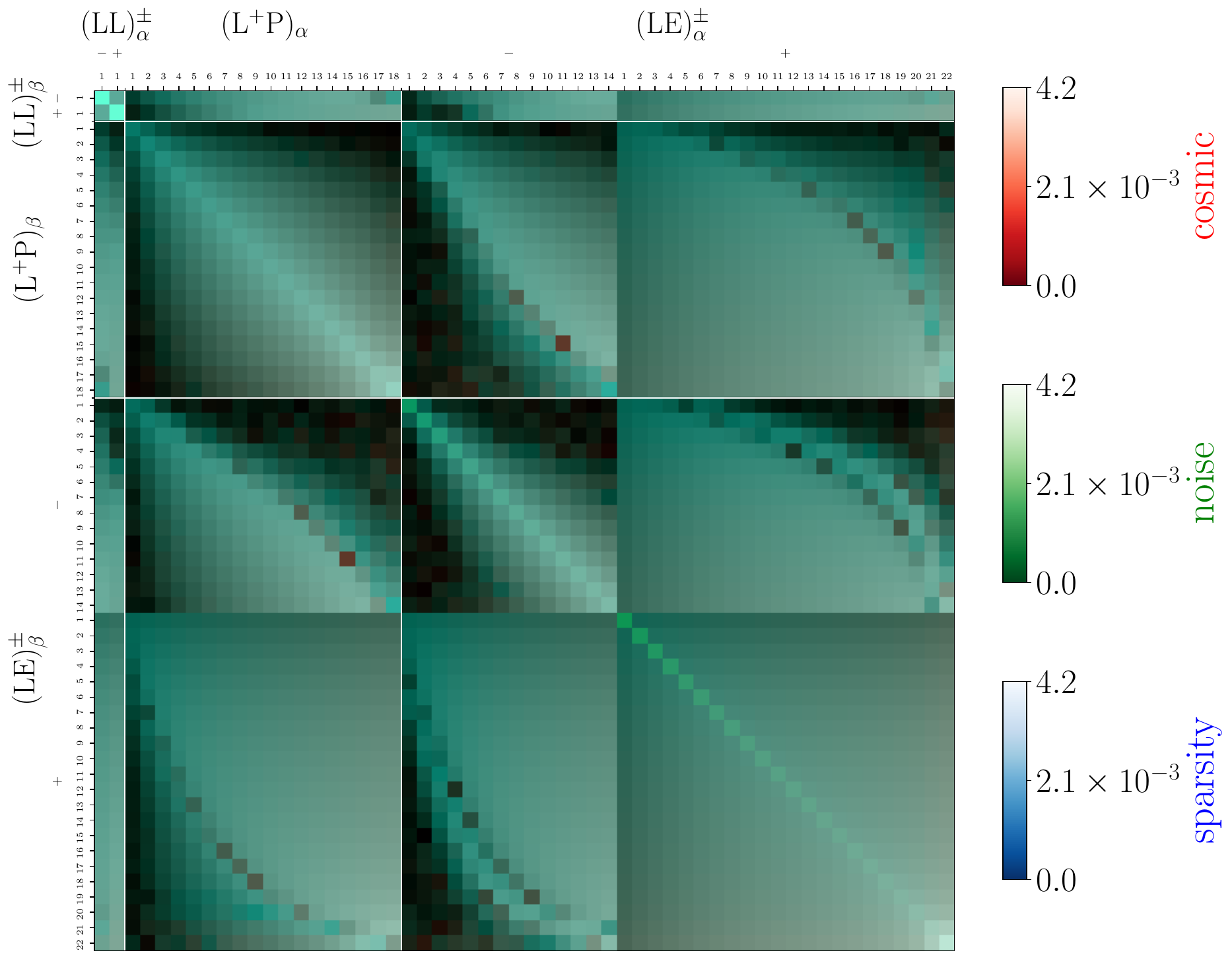}
\caption{Same as \cref{fig:optimistic_covmat} but in the \emph{conservative scenario.}}
\label{fig:conservative_covmat}
\end{figure}

In the conservative scenario, we are more pessimistic about the strong-lensing sample: we divide by ten the number of lenses and multiply by two the uncertainty on LOS shear measurements, relative to the optimistic scenario,
\begin{empheq}[box=\fbox]{equation}
\label{eq:lens_sample_conservative}
L = 10^4, \quad
\sigma_{\nlLOS} = 0.1
\qquad \text{(conservative scenario).}
\end{empheq}
The strong-lensing redshift distribution~$p(\vect{Z})$ is unchanged (\cref{fig:redshift_distributions}). Our assumptions on the galaxy sample are the same as in the optimistic scenario, except that we do not divide it into tomographic redshift bins, so as to boost the SNR for $\lens\shape$ and $\lens\position$.

In this scenario, the autocorrelation of the LOS shear, $\lens\lens$, is not detectable (total SNR of about 0.5). On the other hand, thanks to the impressive size of the galaxy sample, the cross-correlations with galaxy ellipticities and positions, $\lens\shape$, $\lens\position$, remain detectable with very high SNR, as seen in \cref{fig:conservative_LE_LP}. The full covariance matrix of $\lens\lens$, $\lens\shape$, $\lens\position$ is depicted in \cref{fig:conservative_covmat}. Compared to that of the optimistic scenario (\cref{fig:optimistic_covmat}), we notice the brighter colours, which mean a higher covariance relative to signal, as well as a prominence of the green and blue colours, compatible with higher noise and sparsity.

\subsection{Detectability assessment}

\begin{figure}[p]
    \vspace*{-1cm}
    \includegraphics[width=1\columnwidth]{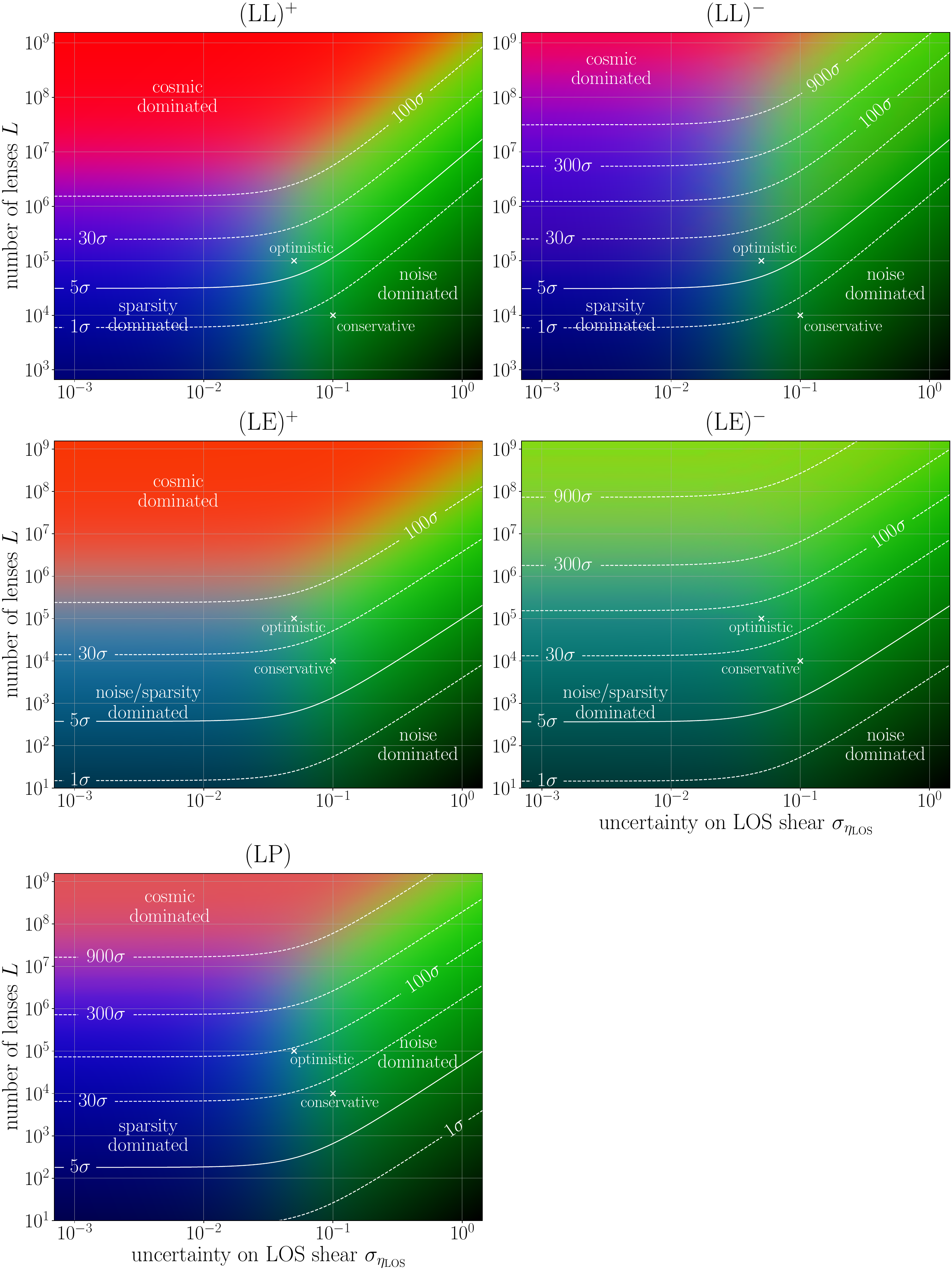}
    \caption{Detectability of the $\lens\lens$, $\lens\shape$, and $\lens\position$ correlations depending on the number of observed strong lenses, $L$, and the uncertainty on individual LOS shear measurements, $\sigma_{\nlLOS}$, assuming a \Euclid-like galaxy sample. The optimistic and conservative scenarios considered in \cref{subsec:optimistic,subsec:conservative} are indicted with crosses for reference. Lines labelled with ``$N\sigma$'' correspond to level lines of the total SNR function, $\SNR(\observable_1\observable_2)=N$, if all pairs of lenses and galaxies are placed in a single, optimised angular bin~\eqref{eq:SNR_max}, and if the galaxy sample consists of a single redshift bin. Red, green and blue colours represent the respective contributions of cosmic variance~($\CCov$), noise~($\NCov$) and sparsity variance~($\SCov$) to the total variance of a correlation function~$\observable_1\observable_2$; hence, they indicate the main limitation to precise measurements of $\observable_1\observable_2$.}
    \label{fig:SNR_grid}
\end{figure}

Let us conclude this section by assessing the detectability of $\lens\lens$, $\lens\shape$ and $\lens\position$ as a function of the parameters $L$ and $\sigma_{\nlLOS}$. To this end, we define a \emph{single quantity} summarising the SNR for each correlation function. We fix the redshift distribution of the strong-lens sample and the properties of the galaxy sample, which we keep in a single redshift bin as in the conservative scenario. We then define the SNR of $\observable_1\observable_2$ as the maximum of the SNR function of \cref{eq:constant_SNR} that can be reached in a single angular bin~$[0,\thet)$,
\begin{equation}
\label{eq:SNR_max}
\SNR(\observable_1\observable_2)
\define
\max_{\thet\in[2', 300']} \SNR(\observable_1\observable_2; 0, \thet) \ .
\end{equation}
Unlike the binning procedure applied in \cref{subsec:optimistic,subsec:conservative}, we use here the full expression of the variance of $\observable_1\observable_2$ without relying on the approximations in \cref{eq:Var_LL_diag,eq:Var_LE_diag,eq:Var_LP_diag}.

\Cref{fig:SNR_grid} shows the behaviour of $\SNR(\observable_1\observable_2)$ for each $\observable_1\observable_2\in\{(\lens\lens)^\pm, (\lens\shape)^\pm, \lens\position\}$, as a function of $L$ and $\sigma_{\nlLOS}$. Level lines labelled with ``$N\sigma$'' correspond to $\SNR(\observable_1\observable_2)=N$. We used the same RGB colour coding as in \cref{fig:optimistic_covmat,fig:conservative_covmat} so as to highlight which component ($\CCov, \NCov, \SCov$) dominates the uncertainty budget. The positions of the optimistic and conservative scenarios of \cref{subsec:optimistic,subsec:conservative} are indicated with crosses for reference.

\paragraph{The detection threshold} For $\lens\lens$, a 5$\sigma$ detection for the autocorrelation signal lies between the optimistic and conservative scenarios; no matter the precision in LOS shear measurements, no autocorrelation signal can be conclusively detected without at least $\approx 2\times 10^4$ lenses. The cross-correlation signal is, however, much more accessible: with $\sigma_{\nlLOS} = 0.05$, $\lens\position$ could be detected to $5\sigma$ with fewer than 200 lenses, and $\lens\shape$ with $\approx 300$ lenses. Alternatively, even with $\sigma_{\nlLOS} = 1$ -- an uncertainty two orders of magnitude larger than the typical value of $\lLOS$ and its uncertainty estimated with mock lenses -- a \Euclid-like sample of $10^5$ lenses is still sufficient for a $5\sigma$ detection of the cross-correlation signals.

\paragraph{The limiting factor} For a realistic number of lenses, $L < 10^6$, detectability is limited by noise or sparsity. In every case, we observe the same pattern: when precision in $\lLOS$ measurements is poor, $\NCov$ tends to dominate the uncertainty budget, and the SNR can be improved upon both by improving this precision or increasing the number of lenses. However, once $\sigma_{\nlLOS}$ is sufficiently low ($\lesssim 5\times10^{-2}$ in the case of $\lens\lens$ and $\lens\position$), $\SCov$ begins to dominate, and meaningful gains to the SNR can only be made by observing more lenses. In the case of $\lens\shape$, the noise always remains significant because of the galaxy shape noise~$\sigma_{\eps_0}$, which is independent of $\sigma_{\nlLOS}$ -- this is why $(\lens\shape)^-$ is nowhere cosmic-variance limited. In the other cases, the region of parameter space where $\CCov$ dominates corresponds to extreme values of the SNR;\footnote{In \cref{fig:SNR_grid}, the SNR of $(\lens\lens)^+$ and $(\lens\shape)^+$ saturates above $100\sigma$, in the top-left corner of the plots, whereas it seems to grow indefinitely for the other correlation functions. This behaviour is due to the fact that we imposed $\thet>2'$ in the definition~\eqref{eq:SNR_max} of the SNR, which ensures that our numerical results are well behaved. In principle, however, since both $(\lens\lens)^+(\thet)$ and $(\lens\shape)^+(\thet)$ go to infinity as $\thet\to 0$, higher values of their SNR could be reached if we allowed for smaller values of the lower bound of $\thet$.} this is only achieved for $L > 10^7$, which is beyond the reach of stage-IV surveys. 

\section{Conclusion}
\label{sec:conclusion}

The line-of-sight (LOS) shear is a physical quantity which summarises the perturbations to a strong-lens system caused by cosmological inhomogeneities between the source, the main lens, and the observer. This quantity is expected to be directly measurable in strong-lensing images, independently of the properties of the main lens. As stage-IV galaxy surveys, such as \Euclid and LSST, are expected to detect $\order(10^5)$ strong lenses in the near future, the LOS shear becomes a viable cosmological probe, alongside galaxy shapes and positions.

In this article, we extended the standard $3\times 2$pt correlation method, built from galaxy ellipticities ($\shape$) and positions ($\position$), into a $6\times 2$pt correlation scheme comprising the LOS shear of strong lenses ($\lens$): $\lens\lens$, $\lens\shape$, $\lens\position$, $\shape\shape$, $\shape\position$, $\position\position$. Besides the added information, $\lens$ will be subject to different systematics than $\shape$ and $\position$, thereby mitigating known cosmic-shear limitations (e.g. intrinsic alignments).

We defined estimators for the three new correlation functions involving the LOS shear, $\lens\lens$, $\lens\shape$, $\lens\position$, and computed their expectation value (i.e. the expected signal) in a $\Lambda$CDM cosmology. We then established and applied a general method to analytically calculate the full covariance matrix of the $6\times 2$pt correlation scheme. The results are explicitly presented in the \EC, and are implemented numerically in a publicly available Python code, \code[https://github.com/ELROND-project/loscov]{loscov}. Analytical results have the double advantage of being flexible and easy to interpret. In particular, they allow us to understand the relative contribution of cosmic, sparsity and noise covariances in the final result.

For concreteness, we considered two observational scenarios. In the optimistic (but realistic) scenario, the LOS shear would be measured on $L=10^5$ lenses, i.e. $60\,\%$ of the expected \Euclid strong-lensing sample, with a mean uncertainty of $\sigma_{\nlLOS}=0.05$, which is about $3.5$ times larger than in the proof of concept~\cite{Hogg:2022ycw}. In this scenario, all three new correlation functions, $\lens\lens$, $\lens\shape$ and $\lens\position$, are detectable with very high SNR. In the conservative scenario, with ten times fewer lenses $(L=10^4)$ and an uncertainty twice as large ($\sigma_{\nlLOS}=0.1$), only $\lens\shape$ and $\lens\position$ are detectable, still with high SNR. More generally, a definitive detection of the $\lens\position$ and $\lens\shape$ cross-correlation functions could be achieved with only a few hundred lenses if $\sigma_{\nlLOS}=0.05$. Even if $\sigma_{\nlLOS}\approx 1$, i.e. two orders of magnitude larger than the expected uncertainty, the cross-correlations will still be detectable with the expected \Euclid lens sample.

This work was focused on theory and methodology. Our main goal was to introduce the new cosmological probes involving the LOS shear and assess their detectability. Hence, it must be considered as the first step towards cosmological forecasts. In particular, we did not yet take into account redshift uncertainties, because they do not affect the detectability of correlation functions, but they will be crucial for cosmological interpretations.

In a subsequent article, we shall quantify the constraining power of our $6\times 2$pt method on cosmological parameters~$\Omega\e{m}$ and $\sigma_8$, and assess the gain from adding the LOS shear to the usual $3\times 2$pt scheme. Particular attention will be paid to the mitigation of weak-lensing systematics such as galaxy intrinsic alignments.

The impact of selection effects, source and lens clustering, reduced-shear corrections and magnification bias will also be carefully investigated in future work, as well as non-Gaussian corrections to the covariance matrix.

\section*{Acknowledgements}

We thank the informatics team at LUPM, in particular St\'ephane Nou, for their support for and facilitation of our use of the lab's computing cluster for the numerical aspects of this work. We also thank Tom Collett, Raphaël Gavazzi, Fabien Lacasa, Nicolas Martinet, Calum Murray, Giacomo Queirolo, and Simone Vinciguerra\modified{, and the anonymous referee} for discussions and comments on the draft.  PF acknowledges support from the French \emph{Agence Nationale de la Recherche} through the ELROND project (ANR-23-CE31-0002). DJ acknowledges support by the First Rand Foundation, South Africa, and the Centre National de la Recherche Scientifique of France. NBH is supported by the research environment and infrastructure of the Handley Lab at the University of Cambridge.

\section*{\href{https://www.elsevier.com/authors/policies-and-guidelines/credit-author-statement}{CRediT} authorship contribution statement}

\textbf{Pierre Fleury:} Conceptualisation, Methodology, Formal analysis, Writing -- Original Draft, Visualisation, Supervision, Project administration, Funding acquisition.
\textbf{Daniel Johnson:} Software, Formal analysis, Writing -- Original Draft, Visualisation.
\textbf{Théo Duboscq:} Methodology, Formal analysis, Writing -- Review \& Editing.
\textbf{Natalie B. Hogg:} Validation, Writing -- Review \& Editing.
\textbf{Julien Larena:} Conceptualisation, Methodology, Supervision, Writing -- Review \& Editing.

\appendix

\section{Detailed calculation of $\Cov\pac[1]{(\widehat{\lens^+\position_B})_\alpha,(\widehat{\lens^+\position_D})_\beta}$}
\label{appendix:detailed_calculation_LPLP}

In this appendix, we present a complete and detailed calculation of
\begin{multline}
\label{eq:covariance_example_detailed}
\Cov\pac{(\widehat{\lens^+\position_B})_\alpha,(\widehat{\lens^+\position_D})_\beta}
\\
=
\frac{1}{L^2 G_B G_D}
\sum_{i=1}^L \sum_{b\in\galaxies_B} \sum_{k=1}^L \sum_{d\in\galaxies_D}
\cev{\lsev{\gsev{%
W^\alpha_{ib} W^\beta_{kd} \mhLOS_{+ib} \mhLOS_{+kd}
}}}
-
(\lens^+\position_B)_\alpha (\lens^+\position_D)_\beta \ .
\end{multline}
Besides serving as an example of how such a derivation is performed, it should also demonstrate why the recipe presented in \cref{subsec:recipe_covariance_matrix} works. The core of the calculation consists in evaluating the expectation value $\cev[2]{\lsev[2]{\gsev[2]{S_{ibkd}^{\alpha\beta}}}}$ of the summand $S_{ibkd}^{\alpha\beta}\define W^\alpha_{ib} W^\beta_{kd} \mhLOS_{+ib} \mhLOS_{+kd}$.

\subsection{Galaxy-sample average $\gsev{\ldots}$}

We first apply the galaxy-sample averaging operator,
\begin{align}
\gsev{S_{ibkd}^{\alpha\beta}}
=
\pac{\prod_{A=1}^T \prod_{a\in\galaxies_A} \dd^5\vect{\Gamma}_a \; p_A(\vect{\Gamma}_a)}
S_{ibkd}^{\alpha\beta} \ .
\end{align}
Here, contrary to \cref{subsec:expectation_values_estimators}, the galaxy indices $b$ and $d$ can be equal. If they are, only one integral over $\vect{\Gamma}_b=\vect{\Gamma}_d$ remains; this is only possible if the bins are identical, $B=D$. If the indices differ, then we must keep two separate integrals over $\vect{\Gamma}_b$ and $\vect{\Gamma}_d$. In other words:
\begin{equation}
\gsev{S_{ibkd}^{\alpha\beta}}
=
(1-\Delta_{bd})
\int\dd^5\vect{\Gamma}_2 \; p_B(\vect{\Gamma}_2)
\int\dd^5\vect{\Gamma}_4 \; p_D(\vect{\Gamma}_4) \,
S_{i2k4}^{\alpha\beta}
+
\Delta_{bd}
\int\dd^5\vect{\Gamma}_2 \; p_B(\vect{\Gamma}_2) \,
S_{i2k2}^{\alpha\beta} \ ,
\end{equation}
where $\Delta_{bd}$ is the Krönecker symbol.

When summing over the galaxies in the bins, this yields
\begin{align}
\sum_{b\in\galaxies_B} \sum_{d\in\galaxies_D} 
\gsev{S_{ibkd}^{\alpha\beta}}
&=
\pa{G_B G_D - \delta_{BD} G_D}
\int \dd^5\vect{\Gamma}_2 \; p_B(\vect{\Gamma}_2)
\int \dd^5\vect{\Gamma}_4 \; p_D(\vect{\Gamma}_4) \,
S_{i2k4}^{\alpha\beta}
\nonumber\\
&\quad
+ \delta_{BD} G_B
\int \dd^5\vect{\Gamma}_2 \; p_B(\vect{\Gamma}_2) \,
S_{i2k2}^{\alpha\beta}
\\
&=
G_B G_D \int \dd^5\vect{\Gamma}_2 \int \dd^5\vect{\Gamma}_4 \;
\tilde{p}_{BD}(\vect{\Gamma}_2, \vect{\Gamma}_4) \, 
S_{i2k4}^{\alpha\beta} \ ,
\end{align}
where we have introduced the effective PDF
\begin{equation}
\tilde{p}_{BD}(\vect{\Gamma}_2, \vect{\Gamma}_4)
=
\pa{1 - \frac{\delta_{BD}}{G_B}} p_B(\vect{\Gamma}_2) p_D(\vect{\Gamma}_4)
+ \frac{\delta_{BD}}{G_B} \, p_B(\vect{\Gamma}_2) \Dirac(\vect{\Gamma}_2 - \vect{\Gamma}_4) \ ,
\end{equation}
which is indeed normalised to unity. The $\delta_{BD}/G_B\ll 1$ term in the first prefactor may be neglected in practice.

\subsection{Lens sample average $\lsev{\ldots}$}

Similarly, taking the lens-sample average of $\gsev[2]{S_{ibkd}^{\alpha\beta}}$ yields
\begin{align}
\lsev{\gsev{S_{ibkd}^{\alpha\beta}}}
&=
\pac{\prod_{i'=1}^{L} \int \dd^6\vect{\Lambda}_{i'} \; p(\vect{\Lambda}_{i'}) }
\gsev{S_{ibkd}^{\alpha\beta}}
\\
&=
(1-\Delta_{ik})
\int\dd^6\vect{\Lambda}_1 \; p(\vect{\Lambda}_1)
\int\dd^6\vect{\Lambda}_3 \; p(\vect{\Lambda}_3) \,
\gsev{S_{1b3d}^{\alpha\beta}}
+
\Delta_{ik}
\int\dd^6\vect{\Lambda}_1 \; p(\vect{\Lambda}_1) \,
\gsev{S_{1b1d}^{\alpha\beta}} ,
\end{align}
where $\Delta_{ik}$ accounts for the possibility of equal indices. When summing over all lenses in the sample, we thus get
\begin{gather}
\sum_{i=1}^L \sum_{k=1}^L \lsev{\gsev{S_{ibkd}^{\alpha\beta}}}
=
L^2 \int\dd^6\vect{\Lambda}_1 \int\dd^6\vect{\Lambda}_3 \;
\tilde{p}(\vect{\Lambda}_1, \vect{\Lambda}_3) \,
\lsev{\gsev{S_{1b3d}^{\alpha\beta}}} \ ,
\\
\text{with} \qquad
\tilde{p}(\vect{\Lambda}_1, \vect{\Lambda}_3)
\define
\pa{1 - \frac{1}{L}} p(\vect{\Lambda}_1) p(\vect{\Lambda}_3)
+ \frac{1}{L} \, p(\vect{\Lambda}_1) \Dirac(\vect{\Lambda}_1 - \vect{\Lambda}_3) \ .
\end{gather}
We may take $1-1/L\approx 1$  in the first prefactor.

\subsection{Cosmic average $\cev{\ldots}$}

Putting together the results of the previous two subsections, and applying the cosmic-averaging operator, we end up with the following expression
\begin{multline}
\frac{1}{L^2 G_B G_D}
\sum_{i=1}^L \sum_{b\in\galaxies_B} \sum_{k=1}^L \sum_{d\in\galaxies_D}
\cev{\lsev{\gsev{S_{ibkd}^{\alpha\beta}}}}
\\
=
\int \dd^6\vect{\Lambda}_1
\int \dd^5\vect{\Gamma}_2
\int \dd^6\vect{\Lambda}_3
\int \dd^5\vect{\Gamma}_4 \;
\tilde{p}(\vect{\Lambda}_1, \vect{\Lambda}_3)
\cev{
\tilde{p}_{BD}(\vect{\Gamma}_2, \vect{\Gamma}_4) \,
S_{1234}^{\alpha\beta}
} .
\label{eq:ev_quadruple_sum_formal}
\end{multline}

\paragraph{Extracting the density contrast} Note that, in \cref{eq:ev_quadruple_sum_formal}, we have kept $\tilde{p}_{BD}(\vect{\Gamma}_2, \vect{\Gamma}_4)$ inside the cosmic-averaging operator because we cannot neglect the density contrast that appears in the definition~\eqref{eq:decomposition_p_A_Gamma} of $p_A(\vect{\Gamma})$.
To be more explicit,
\begin{align}
&\cev{
\tilde{p}_{BD}(\vect{\Gamma}_2, \vect{\Gamma}_4) \,
S_{1234}^{\alpha\beta}
}
\nonumber\\
&=
\frac{1}{\Omega^2} \, f_B(z_2) f_D(z_4) \,
p(\eps_{0, 2}; z_2)
p(\eps_{0, 4}; z_4) \,
W_{12}^\alpha W_{34}^\beta
\cev{(1+\deltag_2)(1+\deltag_4)\,\mhLOS_{+12}\mhLOS_{+34}}
\nonumber\\
&\quad +
\frac{\delta_{BD}}{G_B} \, \Dirac(\vect{\Gamma}_2-\vect{\Gamma}_4) \,
\frac{1}{\Omega} \, f_B(z_2) \, p(\eps_{0, 2}; z_2) \,
W_{12}^\alpha W_{32}^\beta
\cev{(1+\deltag_2)\,\mhLOS_{+12}\mhLOS_{+32}} ,
\end{align}
where $\deltag_a\define b(z_a)\delta(z_a, \uvect{u}_a)$ for short.

If we assume that the gravitational potential, and hence all the other quantities that derive from it, is a \emph{Gaussian random field}, then the three-point function $\cev[2]{\deltag_2\mhLOS_{+12}\mhLOS_{+34}}$ vanishes, and the above becomes
\begin{align}
&\cev{
\tilde{p}_{BD}(\vect{\Gamma}_2, \vect{\Gamma}_4) \,
S_{1234}^{\alpha\beta}
}
\nonumber\\
&=
\frac{1}{\Omega^2} \, f_B(z_2) f_D(z_4) \,
p(\eps_{0, 2}; z_2)
p(\eps_{0, 4}; z_4) \,
W_{12}^\alpha W_{34}^\beta
\pac{
\cev{\mhLOS_{+12}\mhLOS_{+34}}
+
\cev{\mhLOS_{+12}\deltag_2\mhLOS_{+34}\deltag_4}
}
\nonumber\\
&\quad +
\frac{\delta_{BD}}{G_B} \, \Dirac(\vect{\Gamma}_2-\vect{\Gamma}_4) \,
\frac{1}{\Omega} \, f_B(z_2) \, p(\eps_{0, 2}; z_2) \,
W_{12}^\alpha W_{32}^\beta
\cev{\mhLOS_{+12}\mhLOS_{+32}}
\\
&=
\cev{p_B(\vect{\Gamma}_2)} \cev{p_D(\vect{\Gamma}_4)}
W_{12}^\alpha W_{32}^\beta
\pac{
\vanishes{
\cev{\mhLOS_{+12}\mhLOS_{+34}}
}
+
\cev{\mhLOS_{+12}\deltag_2\mhLOS_{+34}\deltag_4}
}
\nonumber\\
&\quad +
\frac{\delta_{BD}}{G_B} \, \Dirac(\vect{\Gamma}_2-\vect{\Gamma}_4) \,
\cev{p_B(\vect{\Gamma}_2)} \,
W_{12}^\alpha W_{32}^\beta
\cev{\mhLOS_{+12}\mhLOS_{+32}} ,
\end{align}
with $\cev{p_A(\vect{\Gamma})}=\Omega^{-1} f_A(z) p(\eps_0; z)$ the cosmic-averaged PDF of the galaxy parameters, from which $\delta$ has disappeared. The term indicated in \vanishes{red} in the above will vanish after integrating over, e.g., $\vect{\Gamma}_2$, because $\mhLOS_{+12}\mhLOS_{+34}\propto \cos(2\psi_{12})$, which averages to zero as the galaxy position~$\uvect{u}_2$ moves around the lens at $\uvect{u}_1$. Strictly speaking, this is only true if $\footprint=\mathcal{S}^2$; if not, there is a residual contribution when $\uvect{u}_1$ lies near the boundary of $\footprint$, since $\uvect{u}_2$ may have to exit $\footprint$ to span a complete annulus around $\uvect{u}_1$. We neglect such boundary effects here (\emph{pseudo-full-sky approximation}).

\paragraph{Isserlis's theorem} Since we assume that we are dealing with Gaussian random fields, we may apply Isserlis's theorem~\cite{Isserlis_1918} and decompose four-point functions in terms of two-point functions. Here, this leads to
\begin{equation}
\cev{\mhLOS_{+12}\deltag_2\mhLOS_{+34}\delta_4}
=
\cev{\mhLOS_{+12}\deltag_2}\cev{\mhLOS_{+34}\deltag_4}
+
\cev{\mhLOS_{+12}\mhLOS_{+34}}\cev{\deltag_2\deltag_4}
+
\cev{\mhLOS_{+12}\deltag_4}\cev{\deltag_2\mhLOS_{+34}} .
\end{equation}

\subsection{Putting everything together}

Substituting, in \cref{eq:ev_quadruple_sum_formal}, the expression of $\tilde{p}(\vect{\Lambda}_1, \vect{\Lambda}_2)$ and $\cev{\tilde{p}_{BD}(\vect{\Gamma}_2, \vect{\Gamma}_4) S_{1234}}$, we get
\begin{align}
&\frac{1}{L^2 G_B G_D}
\sum_{i=1}^L \sum_{b\in\galaxies_B} \sum_{k=1}^L \sum_{d\in\galaxies_D}
\cev{\lsev{\gsev{%
S_{ibkd}^{\alpha\beta}
}}}
\nonumber\\
&=
\int \dd^6\vect{\Lambda}_1 \; p(\vect{\Lambda}_1)
\int \dd^5\vect{\Gamma}_2 \cev{p_B(\vect{\Gamma}_2)}
\int \dd^6\vect{\Lambda}_3 \; p(\vect{\Lambda}_3)
\int \dd^5\vect{\Gamma}_4 \cev{p_D(\vect{\Gamma}_4)}
W^\alpha_{12} W^\beta_{34}
\nonumber \\
&\qquad\times
\pa{%
{\color{blue}
\cev{\hLOS_{+12}\deltag_2}\cev{\hLOS_{+34}\deltag_4}
}
+
\cev{\hLOS_{+12}\hLOS_{+34}}\cev{\deltag_2\deltag_4}
+
\cev{\hLOS_{+12}\deltag_4}\cev{\deltag_2\hLOS_{+34}} 
}
\nonumber \\
&\quad +
\frac{1}{L}
\int \dd^6\vect{\Lambda}_1 \; p(\vect{\Lambda}_1)
\int \dd^5\vect{\Gamma}_2 \cev{p_B(\vect{\Gamma}_2)}
\int \dd^5\vect{\Gamma}_4 \cev{p_D(\vect{\Gamma}_4)}
W^\alpha_{12} W^\beta_{14}
\nonumber \\
&\qquad \times
\pa{%
\negligible{%
\cev{\mhLOS_{+12}\deltag_2}\cev{\mhLOS_{+14}\deltag_4}
}
+
\cev{\mhLOS_{+12}\mhLOS_{+14}}\cev{\deltag_2\deltag_4}
+
\negligible{%
\cev{\mhLOS_{+12}\deltag_4}\cev{\deltag_2\mhLOS_{+14}}
}
}
\nonumber \\
&\quad +
\frac{\delta_{BD}}{G_B}
\int \dd^6\vect{\Lambda}_1 \; p(\vect{\Lambda}_1)
\int \dd^5\vect{\Gamma}_2 \cev{p_B(\vect{\Gamma}_2)}
\int \dd^6\vect{\Lambda}_3 \; p(\vect{\Lambda}_3) \,
W^\alpha_{12} W^\beta_{32} \,
\cev{\hLOS_{+12}\hLOS_{+32}}
\nonumber \\
&\quad +
\frac{\delta_{BD}}{L G_B}
\int \dd^6\vect{\Lambda}_1 \; p(\vect{\Lambda}_1)
\int \dd^5\vect{\Gamma}_2 \cev{p_B(\vect{\Gamma}_2)}
W^\alpha_{12} W^\beta_{12} \,
\cev{(\mhLOS_{+12})^2} ,
\end{align}
where we have replaced $\mlLOS$ with $\lLOS$ in terms where it is evaluated at two different points, since the LOS measurement noise~$\nlLOS$ is assumed to be uncorrelated for different lenses. In the above expression, we recognise in \textcolor{blue}{blue} the product of expectation values $(\lens^+\position_B)_\alpha (\lens^+\position_D)_\beta$, which will be subtracted in the definition~\eqref{eq:covariance_example_detailed} of the covariance. Besides, we note that the two terms in \negligible{grey} can be neglected, because they are of the same order of magnitude as the preceding line, but suppressed by the $1/L\ll 1$ prefactor.

The full covariance of \cref{eq:covariance_example_detailed} thus takes the form
\begin{align}
&\Cov\pac{(\widehat{\lens^+\position_B})_\alpha,(\widehat{\lens^+\position_D})_\beta}
\nonumber\\
&=
\int \dd^6\vect{\Lambda}_1 \; p(\vect{\Lambda}_1)
\int \dd^5\vect{\Gamma}_2 \cev{p_B(\vect{\Gamma}_2)}
\int \dd^6\vect{\Lambda}_3 \; p(\vect{\Lambda}_3)
\int \dd^5\vect{\Gamma}_4 \cev{p_D(\vect{\Gamma}_4)}
W^\alpha_{12} W^\beta_{34}
\nonumber\\
&\quad\times
\bigg\{
\pac{%
\cev{\hLOS_{+12}\hLOS_{+34}}
+ \frac{1}{L} \cev{\mhLOS_{+12}\mhLOS_{+14}} \Dirac(\vect{\Lambda}_1-\vect{\Lambda}_3)
}
\pac{%
\cev{\deltag_2\deltag_4}
+ \frac{\delta_{BD}}{G_B} \,  \Dirac(\vect{\Gamma}_2-\vect{\Gamma}_4)
}
\nonumber\\
&\qquad
+ \cev{\hLOS_{+12}\deltag_4}\cev{\hLOS_{+34}\deltag_2}
\bigg\} .
\label{eq:covariance_example_penultimate}
\end{align}
We recognise, in the third line, similar expressions to the tilded correlations $\wnoise{\lens^+\lens^+}$, $\wnoise{\position_B\position_D}$ introduced in \cref{subsec:recipe_covariance_matrix}, which contain the contributions of noise and sparsity.

\subsection{Polarisation angles}

The last step of the calculation consists in extracting the misaligned polarisation angles from the integrand of \cref{eq:covariance_example_penultimate}. The $\lens^+\lens^+$ correlation function, for example, is indeed defined as $\cev[2]{\hLOS_{+ij}\hLOS_{+ji}}$, while \cref{eq:covariance_example_penultimate} features terms like $\cev[2]{\hLOS_{+12}\hLOS_{+34}}$.

In order to recover our building blocks, the trick is to introduce by hand the angles that cause the expressions to depart from the relevant axis -- here the axis $13$ between the points where the LOS shear is evaluated. From the definition~\eqref{eq:plus_cross_polarisations_definition} of the $+/\times$ polarisations,
\begin{align}
\hLOS_{+12}
&\define \Re\pac{\hLOS_1 \ex{-2\ii\psi_{12}}}
\\
&= \Re\pac{(\hLOS_{+13}+\ii\hLOS_{\times 13}) \, \ex{2\ii(\psi_{13}-\psi_{12})}}
\\
&= \hLOS_{+13} \cos2(\psi_{13}-\psi_{12}) - \hLOS_{\times 13} \sin2(\psi_{13}-\psi_{12}) \ ,
\end{align}
and similarly for $\hLOS_{+34}$, so that
%
\begin{multline}
\cev{\hLOS_{+12}\hLOS_{+34}}
=
\overbrace{%
\cev{\hLOS_{+13}\hLOS_{+31}} 
}^{\lens^+\lens^+(\thet_{13}; \vect{Z}_1, \vect{Z}_3)}
\cos2(\psi_{13}-\psi_{12}) \cos2(\psi_{31}-\psi_{34})
\\
+
\underbrace{%
\cev{\hLOS_{\times 13}\hLOS_{\times 31}}
}_{\lens^\times\lens^\times(\thet_{13}; \vect{Z}_1, \vect{Z}_3)}
\sin2(\psi_{13}-\psi_{12}) \sin2(\psi_{31}-\psi_{34}) \ ,
\end{multline}
where we used $\cev[2]{\hLOS_{+ 13}\hLOS_{\times 31}}=0$. The same applies to hatted quantities. As for the $\lens\position$ terms, a similar calculation yields
\begin{align}
\cev{\hLOS_{+12}\deltag_4}
&= \cev{\hLOS_{+14}\deltag_4} \cos2(\psi_{14}-\psi_{12})
= \lens^+\position(\thet_{14}; \vect{Z}_1, z_4) \, \cos2(\psi_{14}-\psi_{12}) \ ,
\\
\cev{\hLOS_{+34}\deltag_2}
&= \lens^+\position(\thet_{32}; \vect{Z}_3, z_2) \, \cos2(\psi_{32}-\psi_{34}) \ .
\end{align}
The general rules for such conversions are summarised in \textbf{Step 2} of \cref{subsubsec:using_covariance_formula}.

\subsection{Final expression}

Substituting the relevant trigonometric terms, as derived above, into \cref{eq:covariance_example_penultimate}, and introducing the redshift-averaged correlation functions, we conclude that
\begin{align}
&\Cov\pac{(\widehat{\lens^+\position_B})_\alpha,(\widehat{\lens^+\position_D})_\beta}
\nonumber\\
&=
\int_\footprint \frac{\dd^2\uvect{u}_1}{\Omega}
\int_\footprint \frac{\dd^2\uvect{u}_2}{\Omega} \; W^\alpha_{12}
\int_\footprint \frac{\dd^2\uvect{u}_3}{\Omega}
\int_\footprint \frac{\dd^2\uvect{u}_4}{\Omega} \; W^\beta_{34}
\nonumber\\
&\quad
\times
\bigg\{
\bigg[
\wnoise{\lens^+\lens^+}(\thet_{13}) \cos2(\psi_{13}-\psi_{12}) \cos2(\psi_{31}-\psi_{34})
\nonumber\\
&\hspace{1.3cm}
+ \wnoise{\lens^\times\lens^\times}(\thet_{13}) \sin2(\psi_{13}-\psi_{12}) \sin2(\psi_{31}-\psi_{34})
\bigg]
\wnoise{\position_B\position_D}(\thet_{24})
\nonumber\\
&\hspace{1cm}
+ \wnoise{\lens^+\position_D}(\thet_{14}) \cos2(\psi_{14}-\psi_{12}) \,
    \wnoise{\lens^+\position_B}(\thet_{32}) \cos2(\psi_{32}-\psi_{34})
\bigg\} ,
\end{align}
in agreement with the general formula~\eqref{eq:covariance_matrix_recipe} proposed in \cref{subsec:recipe_covariance_matrix}.

The calculation discussed here illustrates in particular how the summands with coincident indices in the definition~\eqref{eq:covariance_example_detailed} of the covariance translate into effective PDFs for the lens and galaxy parameters, and can be eventually absorbed in effective (tilded) correlation functions.

\section{Angular binning procedure}
\label{appendix:angular_binscheme}

Let $\observable_1\observable_2(\thet)$ be one of the correlation functions considered in this work. The SNR of that correlation function in the angular bin~$[\thet_\alpha, \thet_{\alpha+1})$ is defined as
\begin{equation}
\SNR(\observable_1\observable_2; \thet_\alpha, \thet_{\alpha+1})
\define
\frac{
\displaystyle
\frac{2\pi}{\Omega_{\alpha}} \int_{\thet_\alpha}^{\thet_{\alpha+1}}
\dd\thet \; \thet \, \observable_1\observable_2(\thet)
}
{
\displaystyle
\sqrt{\Var\pac[2]{(\widehat{\observable_1\observable_2})_\alpha}}
}
\ ,
\label{eq:SNR_definition}
\end{equation}
with $\Omega_\alpha=\pi(\thet_{\alpha+1}^2-\thet_\alpha^2)$. Since the full expression of $\Var\pac[2]{(\widehat{\observable_1\observable_2})_\alpha}$ is computationally expensive to evaluate, we choose to crudely approximate it by its strictly diagonal element (see \EC),
\begin{align}
\label{eq:Var_LL_diag}
\Var\pac{(\widehat{\lens\lens})^\pm_\alpha}
&\approx
\frac{\sigma\e{LOS}^4}{L^2} \frac{\Omega}{\Omega_\alpha} \ ,
\\
\label{eq:Var_LE_diag}
\Var\pac{(\widehat{\lens\shape_B})^\pm_\alpha}
&\approx
\frac{1}{2} \frac{\sigma\e{LOS}^2}{L} \frac{\sigma_B^2}{G_B} \frac{\Omega}{\Omega_\alpha} \ ,
\\
\label{eq:Var_LP_diag}
\Var\pac{(\widehat{\lens^+\position_B})_\alpha}
&\approx
\frac{1}{2} \frac{\sigma\e{LOS}^2}{L} \frac{1}{G_B} \frac{\Omega}{\Omega_\alpha} \ .
\end{align}
This approximation allows for a quick evaluation of the $\SNR$ function, because it does not require us to evaluate any integral. In practice, the approximation is acceptable in the case of $\lens\lens$, but it is poor for $\lens\shape$ and $\lens\position$, especially for large values of $\thet$. \emph{Hence, \cref{eq:Var_LL_diag,eq:Var_LE_diag,eq:Var_LP_diag} should not be used if one really cares about the uncertainty on $\lens\shape$ and $\lens\position$}. It should be considered to be a first step towards more refined binning procedures in the future.

\subsection{Binning algorithm given a target SNR}

Suppose that we want all the angular bins of $\observable_1\observable_2$ to have the same SNR, $R$. Starting from $\thet_1=0$, we may proceed iteratively and determine~$\thet_{\alpha+1}$ from $\thet_\alpha$ as follows:
\begin{enumerate}
\item Evaluate $M_\alpha\define\max_{\thet>\thet_\alpha}\SNR(\observable_1\observable_2; \thet_\alpha, \thet)$.
\item If $M_\alpha\geq R$, then numerically solve $\SNR(\observable_1\observable_2; \thet_\alpha, \thet_{\alpha+1})=R$ for $\theta_{\alpha+1}$. The limits of the $\alpha$\textsuperscript{th} bin are then fixed and we may consider the next one.
\end{enumerate}
Repeat steps~1 and 2 as long as $M_\alpha\geq R$. If $M_\alpha<R$, then no bin starting from $\thet_\alpha$ would have enough signal to fulfil the target requirement. We may then either stop the procedure here, or add a last bin $[\thet_\alpha, \thet\e{max}]$ with $\SNR(\observable_1\observable_2; \thet_\alpha, \thet\e{max})=M_\alpha$, if $M_\alpha\geq M\e{min}=2.5$.

\subsection{Determining the target SNR}

Our baseline choice is $R=R\e{min}=8$. However, for high-SNR observables such as $\lens\shape$ and $\lens\position$ in the optimistic scenario (\cref{subsec:optimistic}), this choice  results in a very large number of angular bins. We choose to avoid this by imposing a maximum number of bins, $N\e{max}=20$, and increasing the target SNR accordingly.

In practice, we found that for all the observables considered here ($\lens\lens$, $\lens\shape$, $\lens\position$), the number of angular bins, $N$, and the SNR in each bin, $R$, roughly follow the  empirical relationship
\begin{equation}
N(R) \approx 2 \pa{\frac{M_1}{R}}^2 ,
\end{equation}
where $M_1=\max_{\thet>0}\SNR(\observable_1\observable_2; 0, \thet)$ as defined above. In order to ensure both conditions $R\geq R\e{min}$ and $N\lesssim N\e{max}$, we thus fix the target SNR as
\begin{equation}
R = \max\pac{R\e{min}, M_1\pa{0.5\,N\e{max}}^{-1/2}} .
\end{equation}
Since this whole procedure relies on the approximations of \cref{eq:Var_LL_diag,eq:Var_LE_diag,eq:Var_LP_diag}, the actual SNR in each bin can turn out to be quite far from the target.

\section{Monte Carlo integration and error analysis}
\label{appendix:monte_carlo}

Consider the definite integral
\begin{equation}
    I = \int_{\mathcal{D}} \dd^m\vect{x} \; f(\vect{x}),
\end{equation}
of a real-valued function $f$ over $m$ variables $\vect{x}=(x_1, x_2, ..., x_m)$, within the integration domain $\mathcal{D} \subset \mathbb{R}^{m}$. We denote with $V = \int_{\mathcal{D}} \dd^m \vect{x}$ the volume of $\mathcal{D}$.

$I$ can be efficiently estimated with a Monte Carlo integral. Drawing a random sample of $N$ points $\vect{x}_1,\vect{x}_2, \ldots, \vect{x}_N$ uniformly within the domain $\mathcal{D}$, we can approximate the mean of the function over this sample as 
\begin{equation}
\langle f \rangle_N \define \frac{1}{N}\sum^N_{i=1}f(\vect{x}_i) \ , 
\end{equation}
from which our integral can be approximated as 
\begin{equation}
I \approx I_N = V\langle f \rangle_N \ .
\end{equation}
The error in $I_N$ can be estimated from the sample variance of the $N$ evaluations of the function, $f(\vect{x}_1), f(\vect{x}_2),...,f(\vect{x}_N)$, as
\begin{equation}
\mathrm{err}(I_N)
= \sqrt{\frac{V^2}{N} \, s^2} \ ,
\qquad
s^2
= \frac{1}{N-1}
\sum^N_{i=1}\pac{f(\vect{x}_i)-\langle f \rangle_N}^2.
\end{equation}

Each element in the covariance matrices presented in \cref{fig:optimistic_covmat,fig:conservative_covmat} takes the form of a sum of a cosmic, sparsity and noise contribution, each of which involving one or more integrals over multiple dimensions. The total numerical uncertainty of each element is therefore propagated from the uncertainty arising from each integral. Because the cosmic variance generally involves integrals over five dimensions, rather than four in the case of sparsity and noise, larger samples are generally needed to achieve a similar fractional uncertainty. 

We use samples with $10^6$ points for noise/sparsity-covariance integrals, and $10^7$ for cosmic-covariance integrals. The slowest single redshift ``block'' in the matrices in \cref{fig:optimistic_covmat,fig:conservative_covmat} then takes about one day to compute; parallelising to process each block simultaneously means the full covariance matrix can be evaluated numerically on the same timescale. Further improvements with more sophisticated integrators and parallelisation schemes are possible, but we leave this for a future project. 

\begin{figure}[t]
\includegraphics[width=\columnwidth]{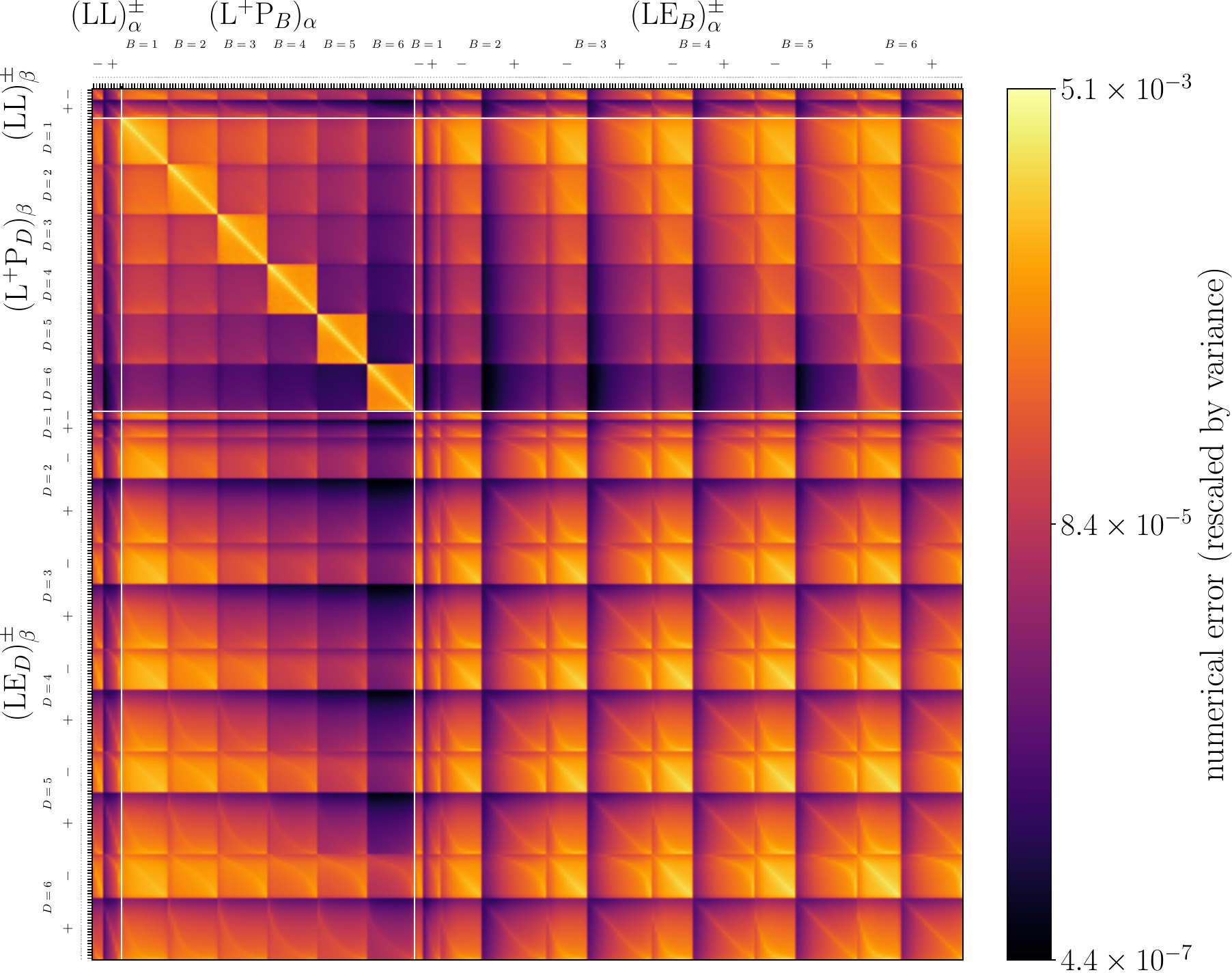}
\caption{Numerical uncertainty, arising from the Monte Carlo integration, on the covariance matrix of $\lens\lens$, $\lens\position$ and $\lens\shape$ in the \emph{optimistic scenario}, normalised by variance according to
$
\mathrm{err}
\paac[2]{
\Cov\pac[2]{
\widehat{(\observable_1\observable_2)}_\alpha,
\widehat{(\observable_3\observable_4)}_\beta}
}
/\pac[2]{
\Var\widehat{(\observable_1\observable_2)}_\alpha
\Var\widehat{(\observable_3\observable_4)}_\beta
}^{1/2}
$.}
\label{fig:optimistic_matrix_errors}
\end{figure}

\begin{figure}[t]
\centering
\includegraphics[width=1\columnwidth]{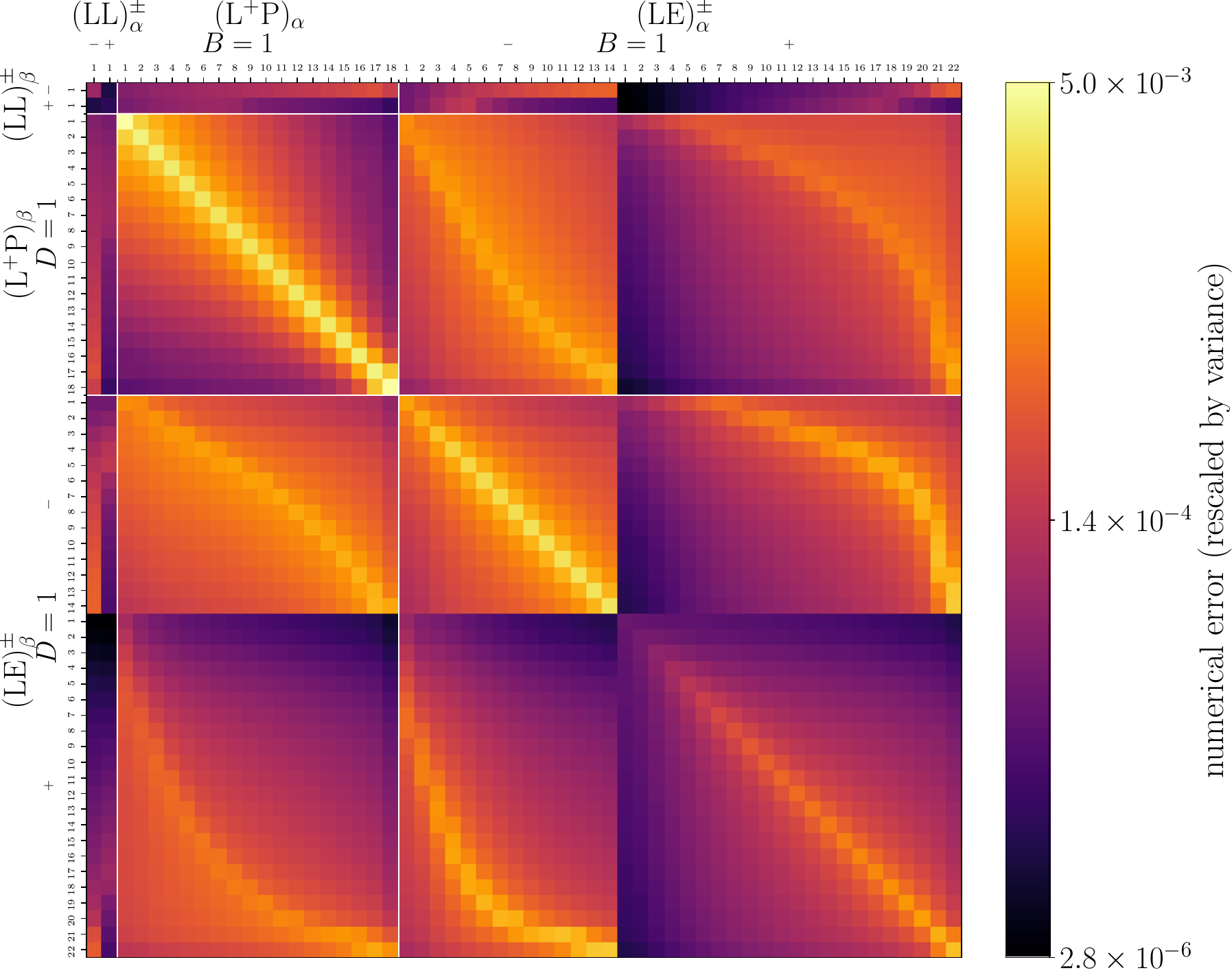}
\caption{The same as in \cref{fig:optimistic_matrix_errors}, but for the \emph{conservative scenario}.}
\label{fig:conservative_matrix_errors}
\end{figure}

We plot the errors corresponding to \cref{fig:optimistic_covmat} in \cref{fig:optimistic_matrix_errors}, and those corresponding to \cref{fig:conservative_covmat} in \cref{fig:conservative_matrix_errors}. From these figures, we see that, while there are elements for which the numerical uncertainty may be comparable to or even larger than the covariance in that block, it is always orders of magnitude smaller than the corresponding variance, and so these errors are not expected to have a perceivable impact on the relative strength of these elements, or in applications such as Fisher forecasting.

\bibliographystyle{JHEP.bst}
\bibliography{bibliography.bib}

\end{document}